\documentclass[preprint,aps,12pt,notitlepage,nofootinbib,tightenlines]{revtex4}
\usepackage{amsmath}
\usepackage{enumerate}
\usepackage{bm}
\usepackage{times}
\usepackage{braket}
\usepackage{color}
\usepackage{epsfig}
\usepackage{slashed}
\usepackage{hyperref}
\usepackage{multirow}
\usepackage{booktabs}
\usepackage{array}
\usepackage{float}
\usepackage{graphicx}

\usepackage{booktabs}
\usepackage{multirow}

\usepackage{booktabs}    
\usepackage{amsmath}     
\usepackage{float}       

\newcommand{\beq}{\begin{eqnarray}}
\newcommand{\eeq}{\end{eqnarray}}
\newcommand{\be}{\begin{equation}\begin{aligned}}
\newcommand{\ee}{\end{aligned}\end{equation}}

\definecolor{Red}{rgb}{1.,0.,0.}

\definecolor{Blue}{rgb}{0.,0.,1.}

\definecolor{nicered}{rgb}{0.7,0.1,0.1}
\definecolor{nicegreen}{rgb}{0.1,0.5,0.1}
\def\lsim{ {\ \lower-1.2pt\vbox{\hbox{\rlap{$<$}\lower6pt\vbox{\hbox{$\sim$}}}}\ } }
\def\gsim{ {\ \lower-1.2pt\vbox{\hbox{\rlap{$>$}\lower6pt\vbox{\hbox{$\sim$}}}}\ } }

\hypersetup{colorlinks,citecolor=nicegreen,linkcolor=nicered}
\begin{document}

\title{Probing the vector-like $X$ quark via the $tW$ channel at future muon-proton colliders}

\author{Liangliang Shang$^{1}$\footnote{Email: shangliangliang@htu.edu.cn; liangliang.shang@physics.uu.se}, Shuting Zhao$^1$\footnote{Email: zhaoshuting@stu.htu.edu.cn}, Bingfang Yang$^1$\footnote{Email: yangbingfang@htu.edu.cn}, Stefano Moretti$^{2,3}$\footnote{Email: s.moretti@soton.ac.uk; stefano.moretti@physics.uu.se}}

\affiliation{\footnotesize
$^1$School of Physics, Henan Normal University, Xinxiang 453007,
PR China\\
$^2$Department of Physics and Astronomy, University of Southampton, Southampton, SO17 1BJ, United Kingdom\\
$^3$Department of Physics and Astronomy, Uppsala University, Box 516, SE-751 20 Uppsala, Sweden
}

\begin{abstract}

We investigate the discovery potential for the vector-like $X$-quark (VLX)
at future muon--proton ($\mu p$) colliders through the process
$\mu^+ p \to \bar{\nu}_\mu X \to \bar{\nu}_\mu t W^+$.
A simplified effective model is adopted 
in which the production and decay of the VLX are governed
by the coupling strength $g^{*}$, the generation-mixing parameter $R_{L}$,
and the VLX mass $m_X$.
A comprehensive Monte Carlo analysis is performed at
$\sqrt{s}=5.29$, $6.48$, and $9.16\ \mathrm{TeV}$,
considering four complementary decay channels:
the Fully Leptonic (FL),
Fully Hadronic (FH),
and two Semi-Leptonic (SL1 and SL2) modes.
An $80\%$ polarized muon beam together with boosted-object
reconstruction based on fat-jet techniques is employed
to improve the signal sensitivity.
The expected exclusion and discovery reaches are 
evaluated using the Asimov significance.
We find that the sensitivity can be improved substantially with
increasing center-of-mass energy and larger values of $R_L$.
Among the four channels, the FH mode provides the strongest sensitivity,
reaching a $2\sigma$ exclusion limit of $m_X\simeq8.3\ \mathrm{TeV}$
with  $g^* = 0.009 $ 
for $R_L=0.1$ at $\sqrt{s}=9.16\ \mathrm{TeV}$,
whereas the FL mode gives the weakest reach because of its smallest branch ratio.
These results demonstrate that future $\mu p$ colliders can offer 
significant sensitivity to heavy VLX over a broad
region of parameter space.
    
\end{abstract}
\maketitle  
\newpage
\section{Introduction}

The Standard Model (SM), based on the gauge symmetry $\mathrm{SU}(3)_C\otimes\mathrm{SU}(2)_L\otimes\mathrm{U}(1)_Y$, provides a remarkably successful description of the strong, electromagnetic, and weak interactions~\cite{Glashow:1961tr, Weinberg:1967tq, Salam:1968rm}. It successfully predicted the $W^{\pm}$ and $Z$ bosons, the top quark, and the Higgs boson, all of which have been confirmed experimentally~\cite{UA1:1983crd,UA2:1983tsx,CDF:1995wbb,D0:1995jca,ATLAS:2012yve,CMS:2012qbp}. 
Nevertheless, the SM remains an incomplete theory. It does not incorporate quantum gravity~\cite{DeWitt:1967yk, tHooft:1974toh}, or provide a viable Dark Matter (DM) candidate~\cite{Zwicky:1933gu, Rubin:1970zza}, or account for non-zero neutrino masses~\cite{Super-Kamiokande:1998kpq, SNO:2001kpb}, or explain the baryon asymmetry of the Universe~\cite{Sakharov:1967dj}, or resolve the Higgs naturalness problem~\cite{Gildener:1976ih, Susskind:1978ms, tHooft:1979rat}. These questions motivate the exploration of physics Beyond the SM (BSM).

Vector-like quarks (VLQs) are well-motivated BSM fermions that can address several of these SM limitations~\cite{Aguilar_Saavedra_2013,Buchkremer:2013bha,Branco:2022gja,Alves_2024,Banerjee_2024}. 
In contrast to SM chiral quarks, 
the left- and right-handed components of a VLQ transform in the same representation of $\mathrm{SU}(2)_L$,
leading to identical gauge-coupling structures~\cite{Aguilar_Saavedra_2013}. 
Furthermore, their masses are not solely generated by the Higgs Yukawa mechanism~\cite{PeraltaCano:2017nih}.
This allows for the existence of TeV-scale states whose loop effects naturally alleviate the fine-tuning problem of Higgs mass,
while also offering a potential mechanism for generating the fermion mass hierarchy at the TeV scale~\cite{Arkani-Hamed:2026wwy}. 
VLQs frequently emerge in little Higgs~\cite{Arkani-Hamed:2002ikv,PhysRevD.67.095004,Cao:2007pv}, 
extra-dimensional~\cite{Agashe:2006wa}, and composite Higgs models~\cite{Agashe:2004rs,Bellazzini:2014yua,Low:2015nqa,Bian:2015ota,He:2001fz}. 
Consequently, their collider phenomenology has been extensively investigated in the literatures (see, e.g., \cite{DELAGUILA19901, Cacciapaglia:2011fx, Liu:2015kmo, Fuks_2017, Cacciapaglia:2018lld, Liu:2019jgp, Yang:2019uea, Wang:2020ips, Deandrea:2021vje, Yang:2022wfa, SHANG2022115977, Han:2023ied, Liu_2024, Moretti:2025ckw, Han:2025itd}).
Typical VLQ multiplets contain the $T$, $B$, $X$, and $Y$ quarks, with electric 
charges $+2/3$, $-1/3$, $+5/3$, and $-4/3$, respectively. They can be organized into singlets ($T$, $B$), 
doublets ($(X,T)$, $(T,B)$, $(B,Y)$), or triplets ($(X,T,B)$, $(T,B,Y)$).

While many studies focus on $T$ and $B$ states, the vector-like $X$-quark (VLX) has also attracted phenomenological attention~\cite{SHANG2023116185,LIU2024116667,Zhang:2024nto,Arhrib:2024mbq}. 
In this work, we focus on the $(X,T,B)$ triplet, in which the VLX mixes with SM up-type quarks via Yukawa couplings
to the Higgs field. Experimental searches for both pair and single VLX production have been 
performed at the LHC~\cite{ATLAS:2018mme, PhysRevLett.112.171801, ATLAS:2022tla, CMS:2018dcw, CMS:2018ubm, 2021mku}. 
At $\sqrt{s}=13\ \mathrm{TeV}$ and assuming SM decay channels, CMS has excluded $X$ masses up to 1.33 TeV via pair production searches~\cite{CMS:2018ubm},
while ATLAS has set {a lower mass limit} of 1.46 TeV 
\cite{ATLAS:2022tla}. 
For single production, CMS has excluded the coupling parameter $\kappa>0.16$ for $m_X=0.8$--$1.6\ \mathrm{TeV}$~\cite{CMS:2018dcw} 
and $\kappa>0.2$ for $m_X=1.6$--$1.8\ \mathrm{TeV}$~\cite{2021mku}.

Furthermore, indirect bounds from Electro-Weak Precision Observables (EWPOs)
impose stringent constraints on the left-handed mixing parameter, $\sin\theta_L^t$.
Specifically, tree-level shifts in the $Zb\bar{b}$ coupling yields a 95\% Confidence Level (CL) upper bound of $\sin\theta_L^u\lesssim0.06$~\cite{Aguilar_Saavedra_2013} in the $(X,T,B)$ triplet extension. 
Combined with constraints from the oblique parameters
($S$, $T$, and $U$)~\cite{Peskin:1990zt}, 
these limits highly restrict the parameter space, particularly in regions that could otherwise explain anomalies such as the $W$-boson
mass measurement~\cite{deBlas:2022hdk, Cao:2022mif}.

To probe higher mass scales and weaker couplings that are challenging for the
Large Hadron Collider (LHC) due to complex QCD backgrounds and kinematic limits,
further collider programmes are essential. The proposed muon-proton ($\mu p$) collider represents a novel asymmetric machine that combines
the distinct advantages of both lepton and hadron colliders~\cite{Cheung_2021,Dagli:2022idi}.
Due to the large muon mass (as opposed to electrons), synchrotron radiation is heavily suppressed, allowing for highly efficient acceleration to reach larger $\sqrt{s}$ values.
Simultaneously, the cleaner initial state mitigates the combinatorial backgrounds typically encountered in $pp$ collisions.
With target integrated luminosities ranging from 10 to 100 fb$^{-1}$, a $\mu p$ collider offers an ideal
environment for BSM discoveries.

Motivated by these considerations, this paper presents a comprehensive phenomenological study of the discovery potential for the VLX
via $\mu^+ p \to \bar{\nu}_\mu X \to \bar{\nu}_\mu t W^+$ at a future $\mu p$ collider with $\sqrt{s}=5.29$, $6.48$, $9.16~\mathrm{TeV}$.
Utilizing a simplified model, we formulate the effective Lagrangian and interaction vertices governing the VLX dynamics.
To effectively reconstruct the hadronic decays of the $t$ quark and $W$ boson in the ensuing highly boosted regime, 
we employ jet substructure techniques~\cite{Thaler:2008ju}. 

This paper is organized as follows. In Section II, we introduce the simplified model framework for the VLX and detail the effective Lagrangian. In Section III, we describe the event generation and analysis procedures, including the simulation of signal and background events as well as the design of selection criteria for each decay channel. In Section IV, we 
present the phenomenological results, detailing the exclusion limits and discovery potential across different analysis modes and $\sqrt{s}$ values.
Finally, we summarize the main conclusions in Section V.

\section{Simplified Model}

VLXs typically arise in extended representations of the EW symmetry. For instance, the VLX can manifest within an $(X, T, B)$ triplet with
quantum numbers $\mathrm{SU}(3)_C\otimes\mathrm{SU}(2)_L\otimes\mathrm{U}(1)_Y = (3, 3, 2/3)$. 
Throughout this section, primed fermion fields denote weak-interaction eigenstates, while unprimed fields denote the corresponding mass eigenstates unless otherwise specified.
Its interaction Lagrangian with the SM Higgs field is given by
\begin{equation}
\begin{aligned}
-\mathcal{L}_{(X,T,B),H} &= \xi \bar{q}'_L \tau \cdot \psi \tilde{H} + \text{h.c.} \\
&= \xi \bar{t}'_L T' H^0 - \xi \sqrt{2} \bar{t}'_L X H^- + \xi \sqrt{2} \bar{b}'_L B' H^0 - \xi \bar{b}'_L T' H^- + \text{h.c.},
\end{aligned}
\end{equation}
where $q'_L = (t'_L,\, b'_L)^T$, $H = (H^+,\, H^0)$, with the Vacuum Expectation Value (VEV) 
$\langle H^0 \rangle = v/\sqrt{2}$, $v \simeq 246\ \text{GeV}$, and $\tilde{H} = i\tau_2 H^*$.
The symbols $\tau_i\ (i=1,2,3)$ denote the Pauli matrices. The multiplet $\Psi$ in the adjoint representation
is constructed as\footnote{The quantum number $T_3$ of the component operators is obtained via $[\tau/2,\,\Psi]_{ij} = T_3 \Psi_{ij}$.}
\begin{equation}
  \tau \cdot \psi \equiv
\begin{pmatrix}
T' & \sqrt{2}X \\
\sqrt{2}B' & -T'
\end{pmatrix}
\equiv \Psi.
\end{equation}
The bare mass term for the triplet is defined as
\begin{equation}
-\mathcal{L}_{\text{bare mass}} = \frac{M_0}{2} \mathrm{Tr}(\bar{\Psi}\Psi) = M_0\left(\bar{T}'T' + \bar{B}'B' + \bar{X}X\right).
\end{equation}
Such bare mass terms can be dynamically generated via 
interactions with a dilaton field~\cite{Cao:2013cfa},
which elegantly resolves the vacuum stability problem associated with VLQs, 
provided the field VEV is significantly larger than the EW scale~\cite{Arsenault:2022xty}.
Alternatively, the VLX can reside in a $Q=(X,T)$ doublet with quantum numbers $(3, 2, 7/6)$.
The corresponding Yukawa interactions and bare mass terms are parameterized as
\begin{equation}
-\mathcal{L}_{(X,T),H} = \xi_Q \bar{Q} H t'_R + \text{h.c.} = \xi_Q \bar{X} H^+ t'_R + \xi_Q \bar{T}' H^0 t'_R + \text{h.c.}
\end{equation}
and $M_0 (\bar{T}' T + \bar{X} X)$, respectively.

After Electro-Weak symmetry breaking (EWSB), the mass matrices are diagonalized via bi-unitary transformations.
Since the VLX possesses an exotic electric charge of $+5/3$, it does not mix with any SM quarks at the tree-level
and is therefore a mass eigenstate. 
Nevertheless, it still participates in charged-current interactions through the gauge couplings $X-T'-W$.
After rotating the weak eigenstates into the mass basis, the mixing between $T'$ and the SM top quark induces the
effective $X-t-W$ interaction, with a coupling strength proportional to the corresponding mixing angle.
The top quark and its vector-like partner $T$ undergo mixing, 
parameterised by the left- and right-handed mixing angles $\theta_{L/R}^t$:
\begin{equation}
\begin{aligned}
\begin{pmatrix}
t_{L/R} \\
T_{L/R}
\end{pmatrix}
&=
\begin{pmatrix}
\cos\theta_{L/R}^t & \sin\theta_{L/R}^t \\
-\sin\theta_{L/R}^t & \cos\theta_{L/R}^t
\end{pmatrix}
\begin{pmatrix}
t'_{L/R} \\
T'_{L/R}
\end{pmatrix} \\
\end{aligned}
\end{equation}

The mixing of a VLQ with more than one SM quark flavour induces flavour-changing neutral currents (FCNCs) among the SM quarks~\cite{Branco:1986my, delAguila:1998tp, Aguilar_Saavedra_2013, Buchkremer:2013bha}, however, these constraints can be evaded in non-minimal models with more than one VLQ~\cite{Atre:2008iu, Aguilar-Saavedra:2013wba}. 
Mixing with the first generation implies a substantial enhancement of the production rates due to the large parton distributions of the valence quarks, therefore, we assume the VLX couples to both the first generation and the third-generation SM quarks.  
For our phenomenological simulations, we adopt an effective Lagrangian
approach within a simplified model. This effective Lagrangian respects 
$\mathrm{SU}(2)_L \otimes \mathrm{U}(1)_Y$ gauge symmetry, and the relevant Lagrangian describing the interactions of the VLX with SM quarks is given by
\begin{equation}
\mathcal{L}_X = g^* \left\{
\sqrt{\frac{R_L}{1+R_L}} \frac{g}{\sqrt{2}}
\left[ \bar{X}_L W_\mu^+ \gamma^\mu u_L \right]
+
\sqrt{\frac{1}{1+R_L}} \frac{g}{\sqrt{2}}
\left[ \bar{X}_L W_\mu^+ \gamma^\mu t_L \right]
\right\} + \text{h.c.} 
\end{equation}
where $g$ is the SM weak coupling constant, 
$g^*$ represents the new coupling strength parameter, and 
$R_L$ dictates the branch ratio (BR) between  the first and
third generation.

Based on this effective Lagrangian, the decay width of the VLX is analytically calculated as
\begin{equation}
\begin{aligned}
\Gamma(X \to Wq)
&= \frac{C^2 \left( m_q^4 - 2 m_W^4 + m_W^2 m_X^2 + m_X^4 + m_q^2 \left( m_W^2 - 2 m_X^2 \right) \right)}
{32 m_W^2 \pi \left| m_X \right|^3} \\
&\times \sqrt{m_q^4 + \left( m_W^2 - m_X^2 \right)^2 - 2 m_q^2 \left( m_W^2 + m_X^2 \right)},
\end{aligned}
\end{equation}
where $q = u, t$. The effective coupling coefficient $C$ is defined as $C = \frac{gg^*}{\sqrt{2}} \sqrt{\frac{R_L}{1 + R_L}}$ 
for the $q=u$ channel, and $C = \frac{gg^*}{\sqrt{2}} \sqrt{\frac{1}{1 + R_L}}$ for the $q=t$ channel.

\begin{figure}[htbp]
  \centering
  \includegraphics[width=0.6\textwidth]{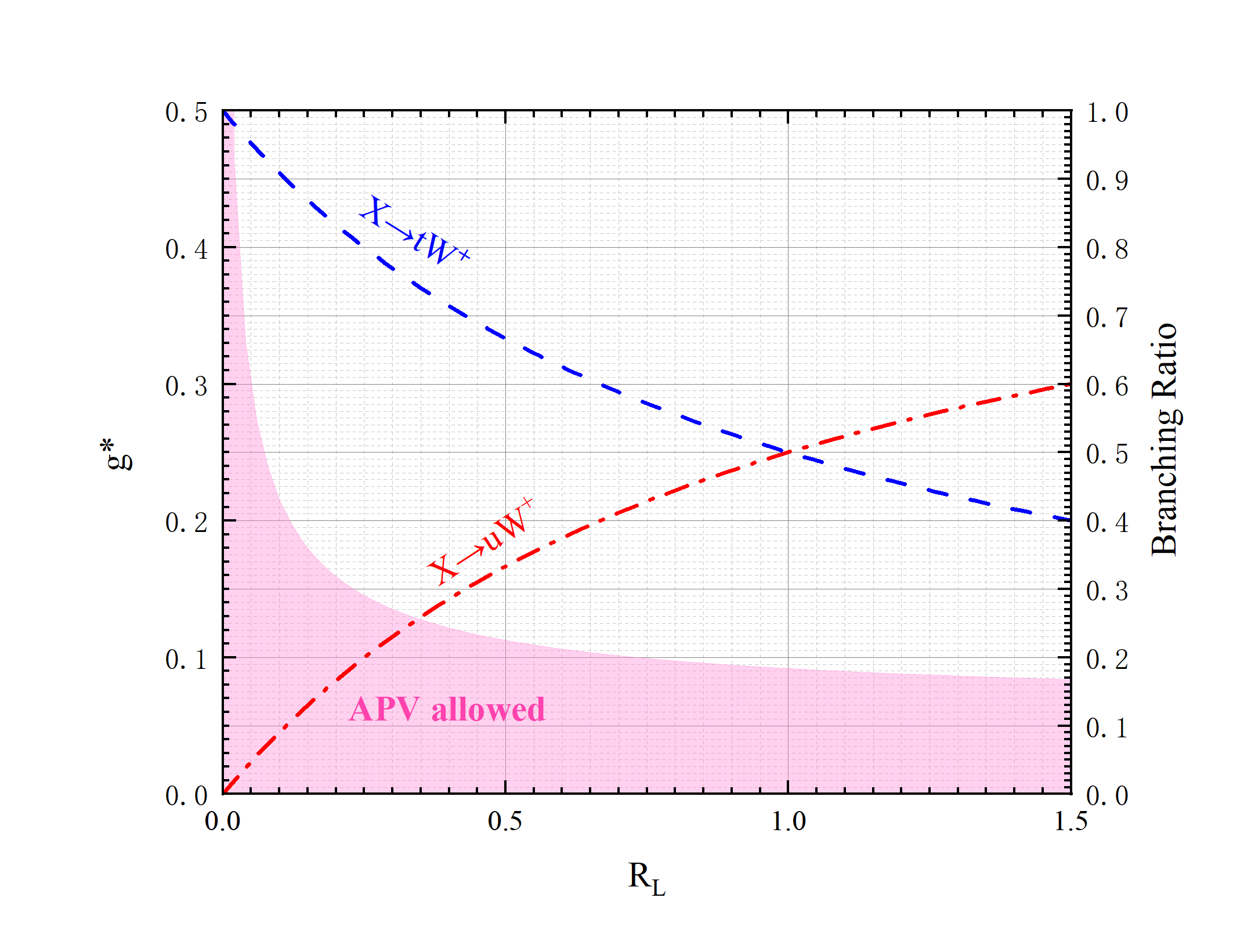}
   \vspace{-1cm}
  \caption{VLX BRs and the upper limit on the coupling parameter $g^*$ as functions of $R_L$.}
  \label{fig1}
\end{figure}

Utilizing this model, we analyze the dependence of the VLX BRs and the coupling constraints on $R_L$ 
in Figure~\ref{fig1}.
The blue dashed and red dot-dashed lines represent the BRs for $X\to tW^+$ and $X\to uW^+$, respectively.
For $R_L \geq 1$, the VLX couples predominantly to the $u$ quark, whereas for $R_L < 1$, it couples primarily to the
$t$ quark. In this work, we focus on the $R_L \leq 1$ regime to maximize the $X \to tW^+$ experimental sensitivity.
It is worth noting that the effective couplings of the VLX to first-generation quarks face stringent constraints from 
Atomic Parity Violation (APV) experiments, as shown by the pink region in this figure. The region denotes the
constraint $g^* \sqrt{\frac{R_L}{1+R_L}} < 6.5\times10^{-2}$~\cite{Okada:2012gy}.
Although the upper limit on $g^*$ is tightly constrained to approximately $0.1$ for most regions where $R_L\leq 1$,
we relax this strict APV bound in our collider simulations for the sake of exploring full phenomenological projections,
and scan the broader parameter space of $g^* \leq 0.5$ and $R_L \leq 1$.

\begin{figure}[htbp]
  \centering
  \includegraphics[width=0.35\textwidth]{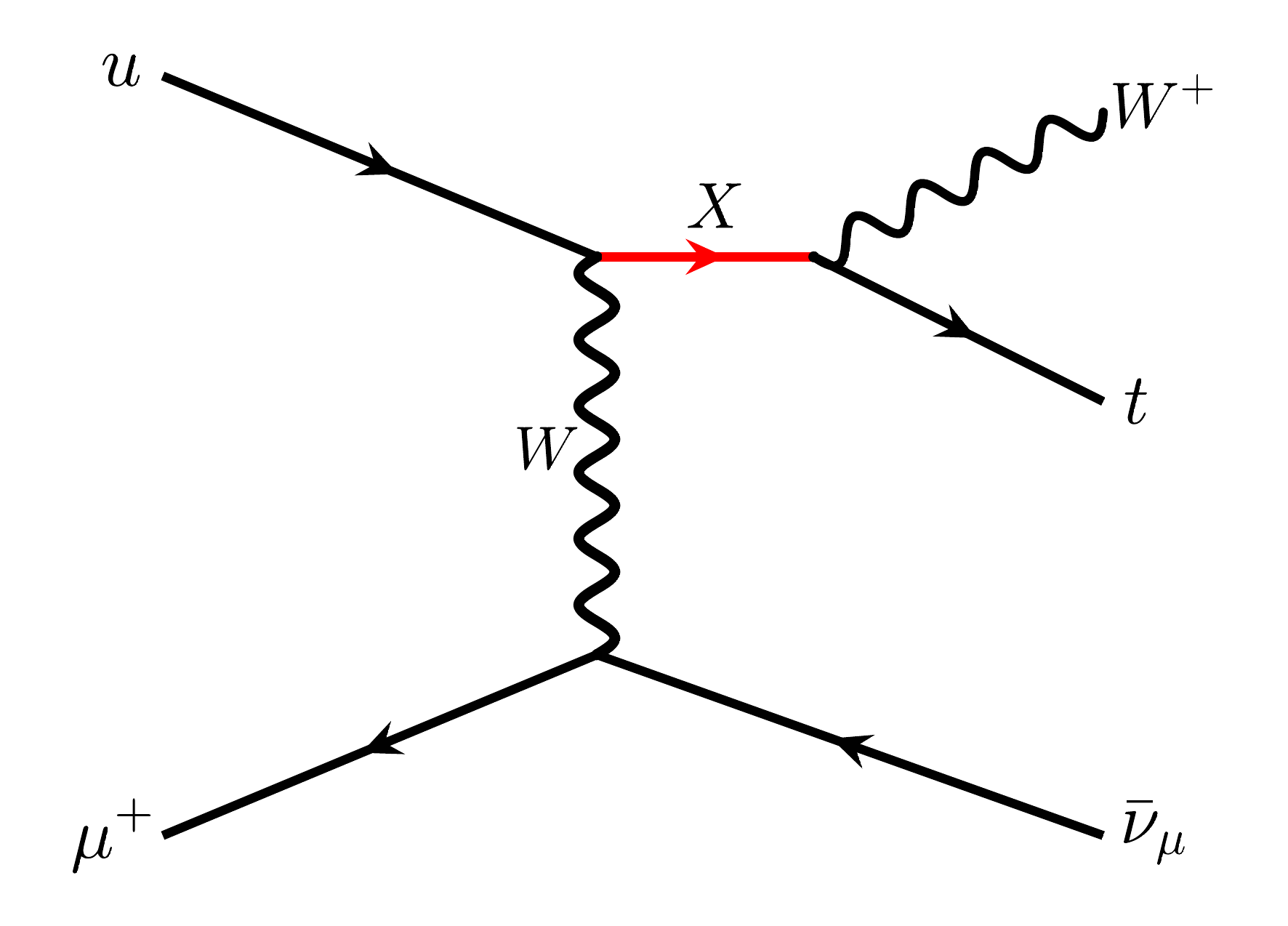} 
  \vspace{-0.5cm}
  \caption{Feynman diagram for the single VLX production with decay $X\to tW^+$ at $\mu p$ collider.}
  \label{fig2}
\end{figure}

\begin{figure}[htbp]
  \centering
  \includegraphics[width=0.6\textwidth]{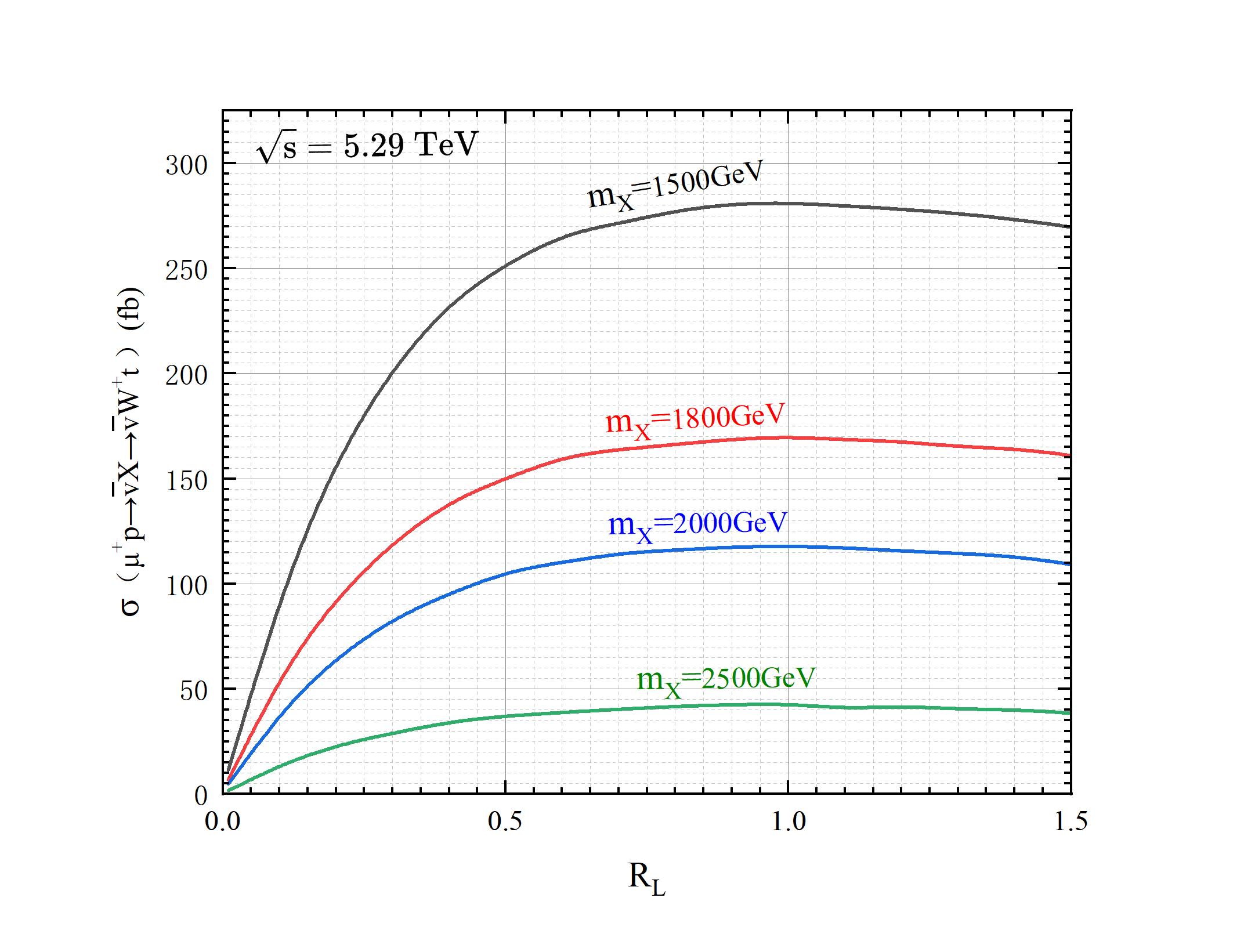}
  \vspace{-1cm}
  \caption{The signal production cross section as a function of $R_L$ at $\mu p$ collider with $\sqrt{s}=5.29\ \text{TeV}$.}
  \label{fig3}
\end{figure}

The tree-level Feynman diagram of signal process $\mu^+ p \to \bar{\nu}_\mu X \to \bar{\nu}_\mu t W^+$ 
is shown in Figure~\ref{fig2}.
The dependence of the cross section on $R_L$ and $m_X$ is shown in Figure~\ref{fig3}, where the
black, red, blue, and green lines correspond to $m_X = 1500$, $1800$, $2000$, and $2500\ \mathrm{GeV}$, respectively.
For $m_X = 1500\ \mathrm{GeV}$, the cross section peaks at approximately $275\ \mathrm{fb}$,
and, even for a heavier mass of $m_X = 2500\ \mathrm{GeV}$, it can reach up to $40\ \mathrm{fb}$.
The signal cross section increases with $R_L$ in the low-$R_L$ region, a trend that is more 
pronounced for lighter $m_X$. 
The cross section subsequently drops in the $R_L > 1$ regime due to the suppression of BR$(X \to tW^+$).

\begin{figure}[htbp]
  \centering
  \includegraphics[width=0.5\textwidth]{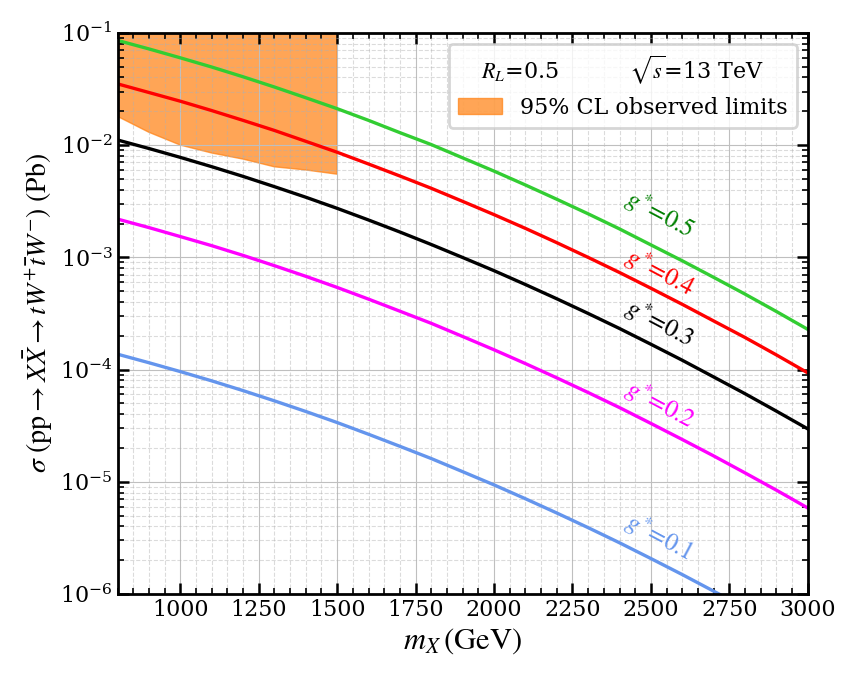}
  \vspace{-0.5cm}
  \caption{VLX pair production cross section at the $13\mathrm{TeV}$ LHC as a function of $m_X$ and $g^*$. The orange band 
  represents the 95\% CL exclusion region obtained from the CMS search~\cite{CMS:2018ubm}.}
  \label{fig4}
\end{figure}

Finally, in Figure~\ref{fig4}, we confront our model with the LHC
constraints from searches for pair-produced VLX decaying via 
$X\to tW^+$ in the same-sign dilepton and 
single-lepton final state~\cite{CMS:2018ubm}, and 
map these limits onto our parameter space at a fixed $R_L = 0.5$. 
The orange region delineates the 95\% CL exclusion boundary. 
The exclusion capability strengthens considerably with larger $g^*$: 
for $g^* = 0.5$, the region $m_X < 1500\ \mathrm{GeV}$ is completely excluded.
Conversely, for weaker couplings, corresponding to $g^* \leq 0.3$ in the scenarios shown here, current LHC searches lose sensitivity, 
underscoring the necessity and potential of a future $\mu p$ collider in probing
this high-mass, weak-coupling parameter space.

\section{Event generation}

\begin{figure}[!t]
\centering
{\includegraphics[width=0.4\textwidth]{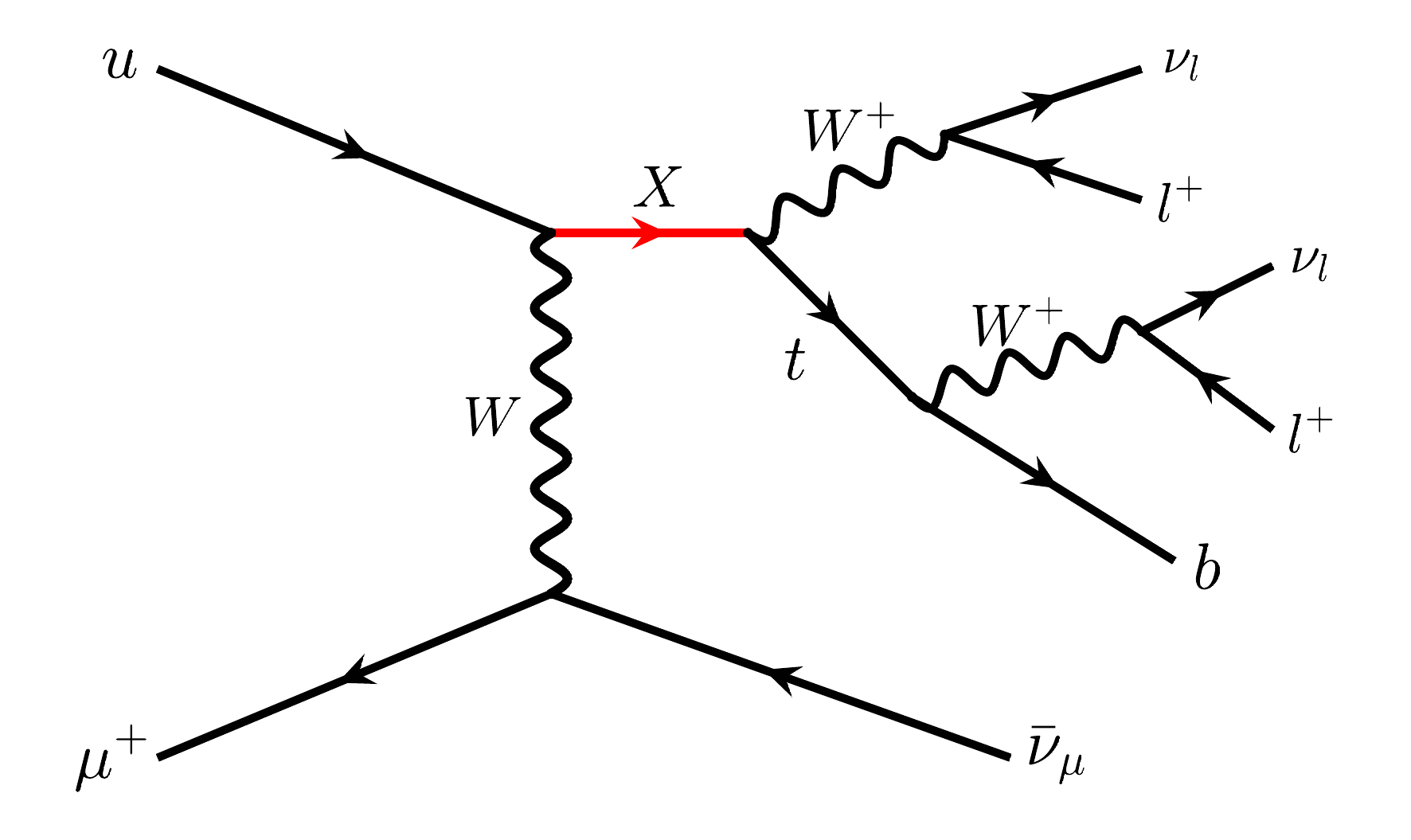}}{\includegraphics[width=0.4\textwidth]{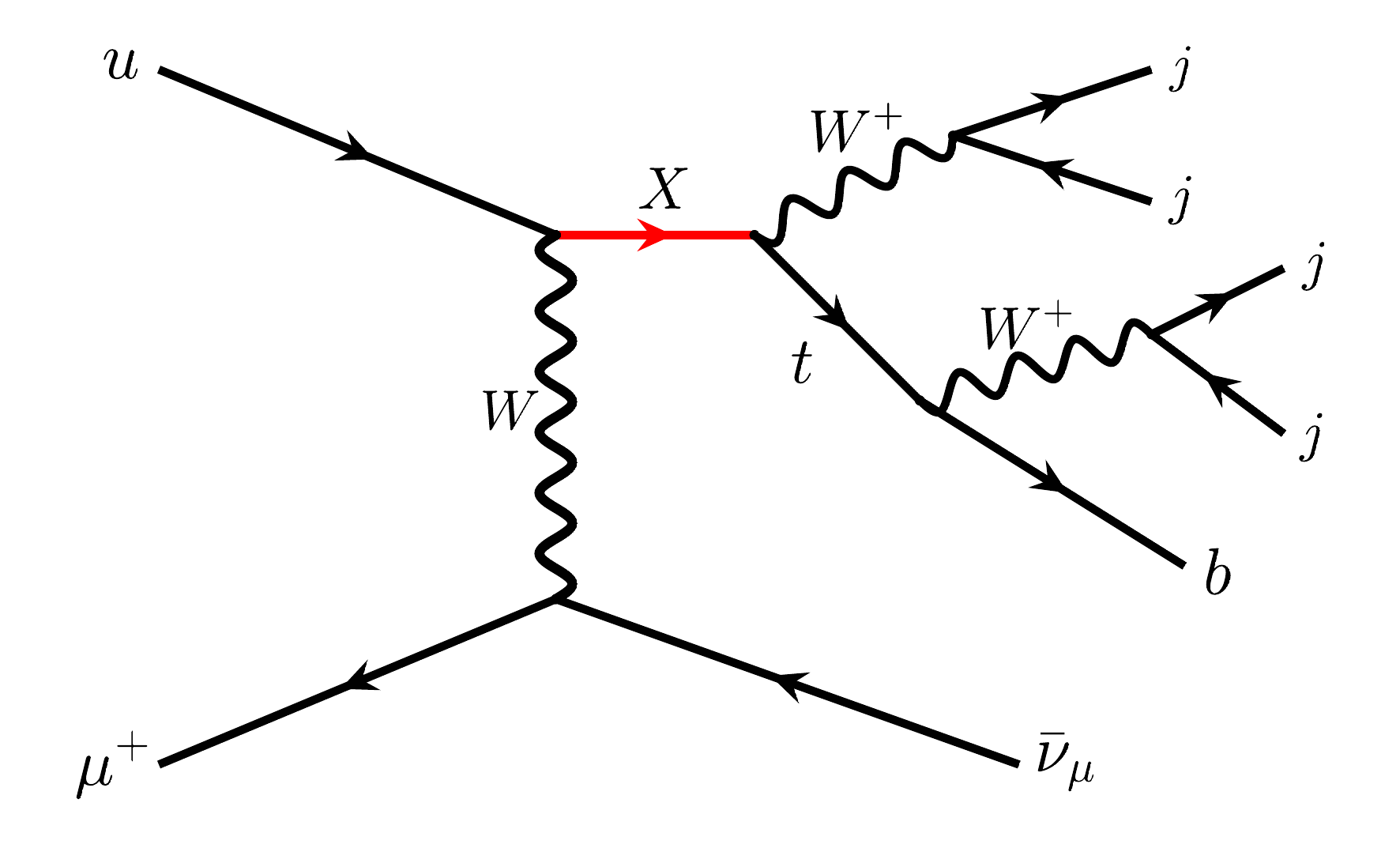}}
(a)~~~~~~~~~~~~~~~~~~~~~~~~~~~~~~~~~~~~~~~~~~~~~~~~~~~~~~~~~~~~(b)
{\includegraphics[width=0.4\textwidth]{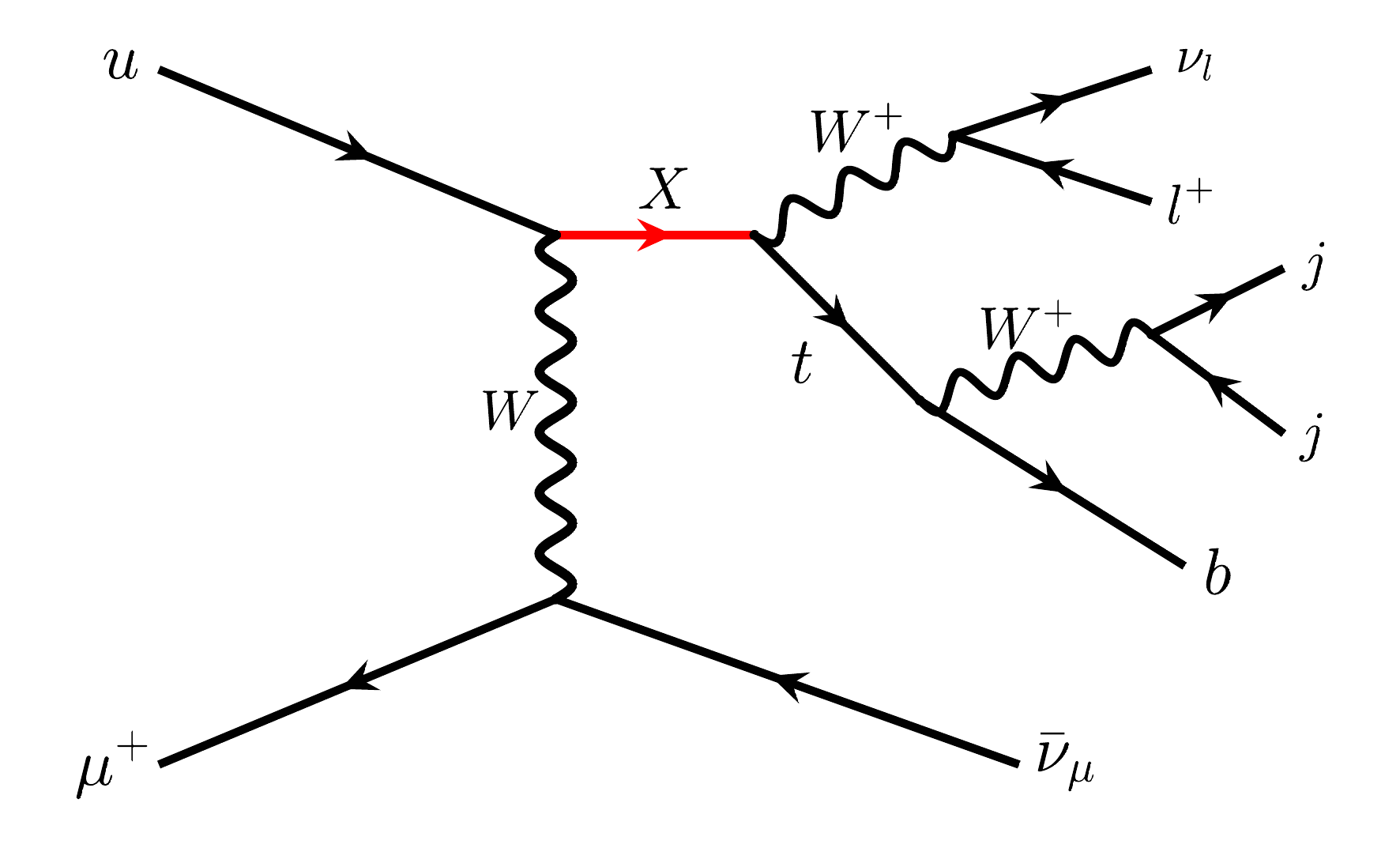}}{\includegraphics[width=0.4\textwidth]{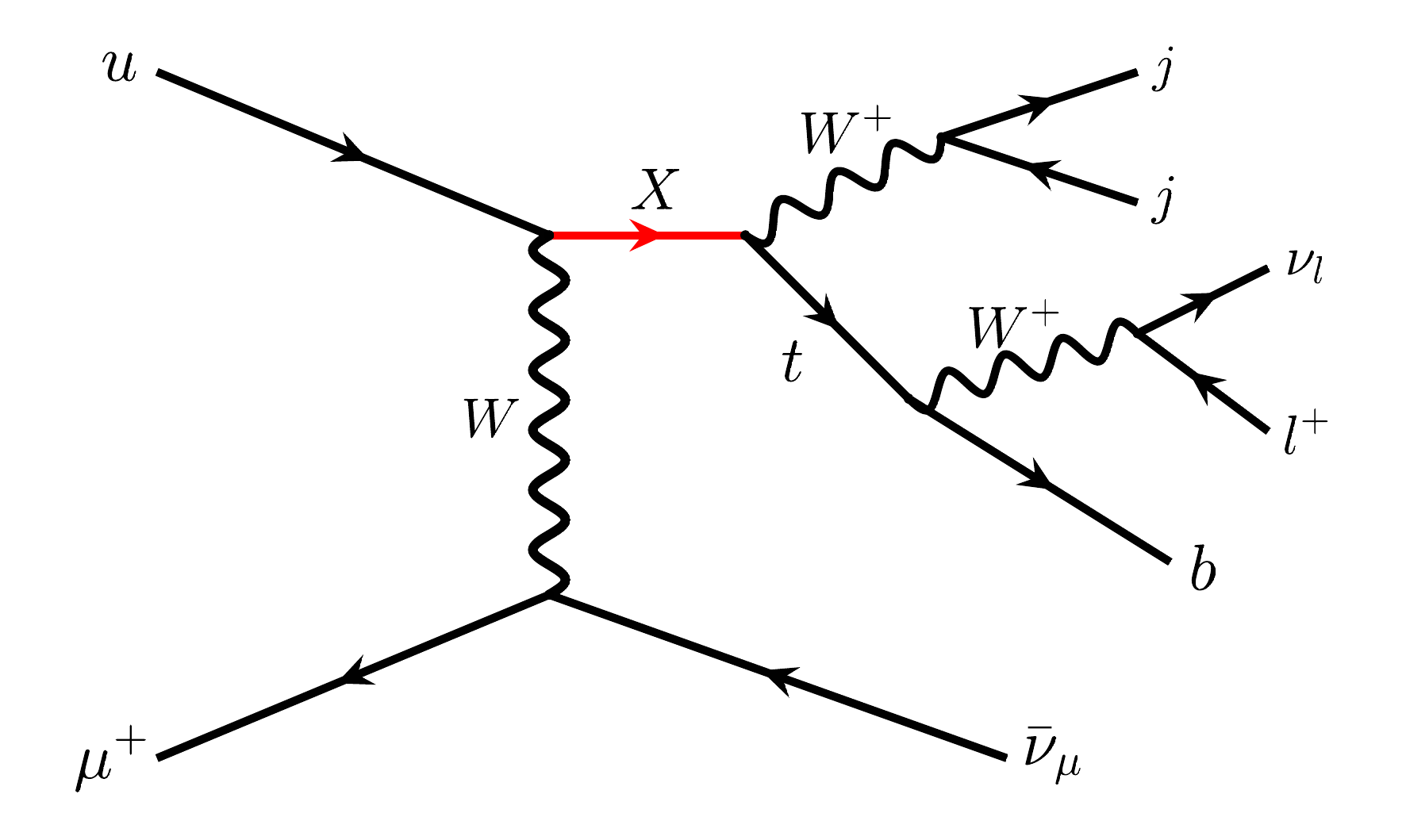}}
(c)~~~~~~~~~~~~~~~~~~~~~~~~~~~~~~~~~~~~~~~~~~~~~~~~~~~~~~~~~~~~(d)
\caption{Representative Feynman diagrams for the single production and decay of the VLX at $\mu p$ collider.
Panels (a), (b), (c), and (d) stand for the FL, FH, SL1, and SL2 modes, respectively.
}
\label{fig5}
\end{figure}

To investigate the discovery potential of the VLX at a future $\mu p$ collider,
we focus on the single production process $\mu^+ p \to \bar{\nu}_\mu X$ followed by the dominant decay $X\to t W^+$ . 
The primary final state is therefore characterized by the presence of a top quark,
a $W$ boson, and significant missing transverse energy ($\not\!\!\mathrm{E}_\mathrm{T}$)
stemming from the neutrino $\bar{\nu}_\mu$.
Depending on the subsequent decay modes of the top quark ($t\to b W^+$) and the two $W$ bosons,
the signal is systematically categorized into four analysis channels.
The representative Feynman diagrams for these channels are illustrated in Figure~\ref{fig5}.
The specific topologies are defined as follows: 
\begin{itemize}
    \item \underbar{Fully Leptonic (FL):} both $W$ bosons decay leptonically. The
    cascade is $X \to t W^+ \to (bW^+) W^+\to b (\ell^+ \nu_\ell) (\ell^+ \nu_\ell)$.
    The final state consists of exactly two charged leptons, one $b$-jet, and a large
    $\not\!\!\mathrm{E}_\mathrm{T}$. (Here and throughout, $\ell = e, \mu$).
    
    \item \underbar{Fully Hadronic (FH):} Both $W$ bosons decay hadronically. 
    The cascade is $X \to t W^+ \to (bW^+) W^+\to b (q \bar{q}') (q \bar{q}')$.
    The final state features one $b$-jet, four light-flavor jets, and 
    $\not\!\!\mathrm{E}_\mathrm{T}$ solely from the primary $\bar{\nu}_\mu$.
    
    \item \underbar{Semi-Leptonic1 (SL1):} The $W$ boson originating directly 
    from the $X$ decays leptonically, while the $W$ from the top quark decays hadronically.
    The cascade is $X \to t W^+(\to \ell^+ \nu_\ell) \to (bW^+(\to q \bar{q}'))  \ell^+ \nu_\ell$.
    The final state comprises one charged lepton, one $b$-jet, two light-flavor jets, and $\not\!\!\mathrm{E}_\mathrm{T}$.

    \item \underbar{Semi-Leptonic2 (SL2):} The topology features identical final-state particles to SL1,
    but with inverted intermediate resonant origins. Here, the $W$ from the $X$ decays hadronically,
    while the $W$ from the top decays leptonically:
    $X \to t W^+(\to q \bar{q}')\to (b W^+(\to \ell^+ \nu_\ell)) q \bar{q}'$.
    
\end{itemize}

Given the signal characteristics described above, the dominant SM background processes
include $\mu^+ W^+ j$, $\bar{\nu} W^+ W^+ j$, $Z \mu^+ W^+ j$, $\bar{\nu} t Z$, $\bar{\nu} Z jj$, and $ \bar{\nu} W^+ jj$ (where $j$ represents a jet).
For the collider configuration, the proton beam energy is fixed at $E_p = 7\ \mathrm{TeV}$. We consider three
typical muon beam energies: $E_\mu = 1,\ 1.5,$ and $3\ \mathrm{TeV}$.
Using the asymmetric collision kinematic relation $\sqrt{s} \simeq 2\sqrt{E_\mu E_p}$,
these correspond to $\sqrt{s}=5.29,\ 6.48,$ and $9.16\ \mathrm{TeV}$, respectively.

\begin{figure}[htbp]  
  \centering
  \includegraphics[width=0.6\textwidth]{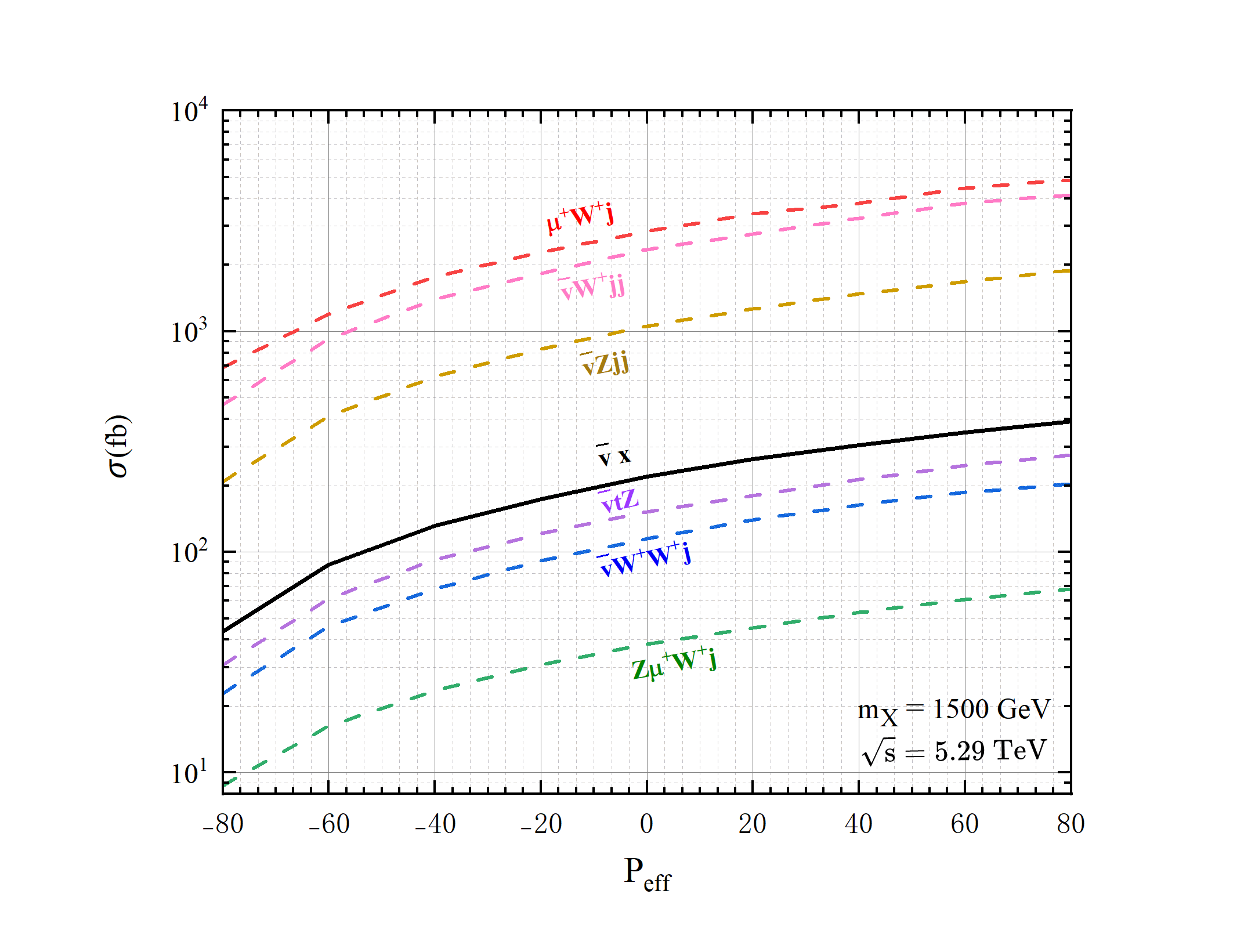}
  \vspace{-0.5cm}
  \caption{Cross sections of the polarized signal process (solid line) and dominant backgrounds (dashed lines) as functions of the effective muon polarization degree $P_{\mathrm{eff}}$.}
  \label{fig6}
\end{figure}

Furthermore, (longitudinal) beam polarization can be effectively utilized to enhance the signal yield.
Figure~\ref{fig6} demonstrates the cross sections of the $\mu^+ p \to \bar{\nu}_\mu X$ signal and the dominant backgrounds
as functions of the effective muon beam polarization degree ($P_{\mathrm{eff}}$).
The numerical values for the unpolarized ($P_{\mathrm{eff}}=0$) and highly polarized ($P_{\mathrm{eff}}=80\%$)
scenarios, along with their corresponding enhancement factors, are summarized in Table~\ref{tab:polarization_xsec_ratio}.
To maximize the signal-to-background ratio, we adopt a polarization degree of $80\%$ in all subsequent simulations.

\begin{table}[htbp]
\centering
\setlength{\tabcolsep}{8pt}
\begin{tabular}{l|l c c c}
\hline
 & Process & $\sigma\left(P_{\mathrm{eff}}=0\right)$ [fb] & $\sigma\left(P_{\mathrm{eff}}=80\%\right)$ [fb] & Factor \\
\hline
\multirow{1}{*}{\centering Signal} 
& $\bar{\nu}_\mu X$ & 219.27 & 391.55 & 1.79 \\
\hline
\multirow{6}{*}{\centering Backgrounds}
& $\bar{\nu} tZ$ & 152.27 & 275.18 & 1.81 \\
& $\bar{\nu} Z jj$ & 1056.10 & 1892.01 & 1.79 \\
& $\bar{\nu} W^+ W^+ j$ & 114.67 & 203.41 & 1.77 \\
& $Z\mu^+ W^+ j$ & 38.24 & 67.75 & 1.77 \\
& $\mu^+ W^+ j$ & 2838.14 & 4855.45 & 1.71 \\
& $\bar{\nu} W^+ jj$ & 2349.53 & 4146.89 & 1.76 \\
\hline
\end{tabular}
\caption{Cross sections of the signal and dominant background processes for  unpolarized $P_{\mathrm{eff}}=0$ and polarized $P_{\mathrm{eff}}=80\%$
muon beam configurations, 
along with their respective enhancement factors.}
\label{tab:polarization_xsec_ratio}
\end{table}

The event generation is structured as follows.
We encode our simplified model using \textsc{FeynRules} 2.0~\cite{Alloul:2013bka} 
and interface it
into \textsc{MadGraph5\_aMC@NLO}~\cite{Alwall_2011,Alwall_2014} to compute 
the cross sections and generate parton-level events for both the signal and backgrounds.
The default Parton Distribution Functions (PDFs) and dynamic renormalization/factorization scales
are utilized. The parton-level events are subsequently passed to \textsc{PYTHIA 8}~\cite{Sj_strand_2015,10.21468/SciPostPhysCodeb.8} 
to perform parton showering and hadronization. 
The detector response is simulated using \textsc{DELPHES 3.4.2}~\cite{deFavereau2014}, 
and kinematic distributions are analyzed via \textsc{MadAnalysis 5}~\cite{Conte:2012fm}.
The simulation workflow, including the integration of different packages and the parameter space scan,
is managed using \textsc{EasyScan\_HEP}~\cite{Shang:2023gfy}.

The relevant SM input parameters are set as follows~\cite{ParticleDataGroup:2024cfk}:
\[
\begin{gathered}
m_b = 4.18\ \mathrm{GeV},\quad m_t = 172.69\ \mathrm{GeV},\quad m_Z = 91.1876\ \mathrm{GeV},\quad m_H = 125.25\ \mathrm{GeV}, \\
\sin^2\theta_W = 0.231,\quad \alpha_{\mathrm{EM}}(m_Z) = 1/128.
\end{gathered}
\]

To suppress soft and collinear regions of phase space, the following cuts are imposed at the parton-level:
\[
\begin{gathered}
\Delta R > 0.4, \quad |\eta_{l}| < 5.0, \quad |\eta_{j/b}| < 5.0, \quad p_T^l > 10\ \mathrm{GeV}, \quad p_T^{j/b} > 20\ \mathrm{GeV}.
\end{gathered}
\]
Here, the angular separation is defined as $\Delta R = \sqrt{(\Delta \phi)^2 + (\Delta \eta)^2}$, where $\Delta \phi$ and $\Delta \eta$ 
are the differences in azimuthal angle and pseudorapidity between any two visible final-state particles, respectively.

\section{Numerical results}

\subsection{FL mode}
\begin{figure}[htbp]
  \centering
  \includegraphics[width=0.45\textwidth]{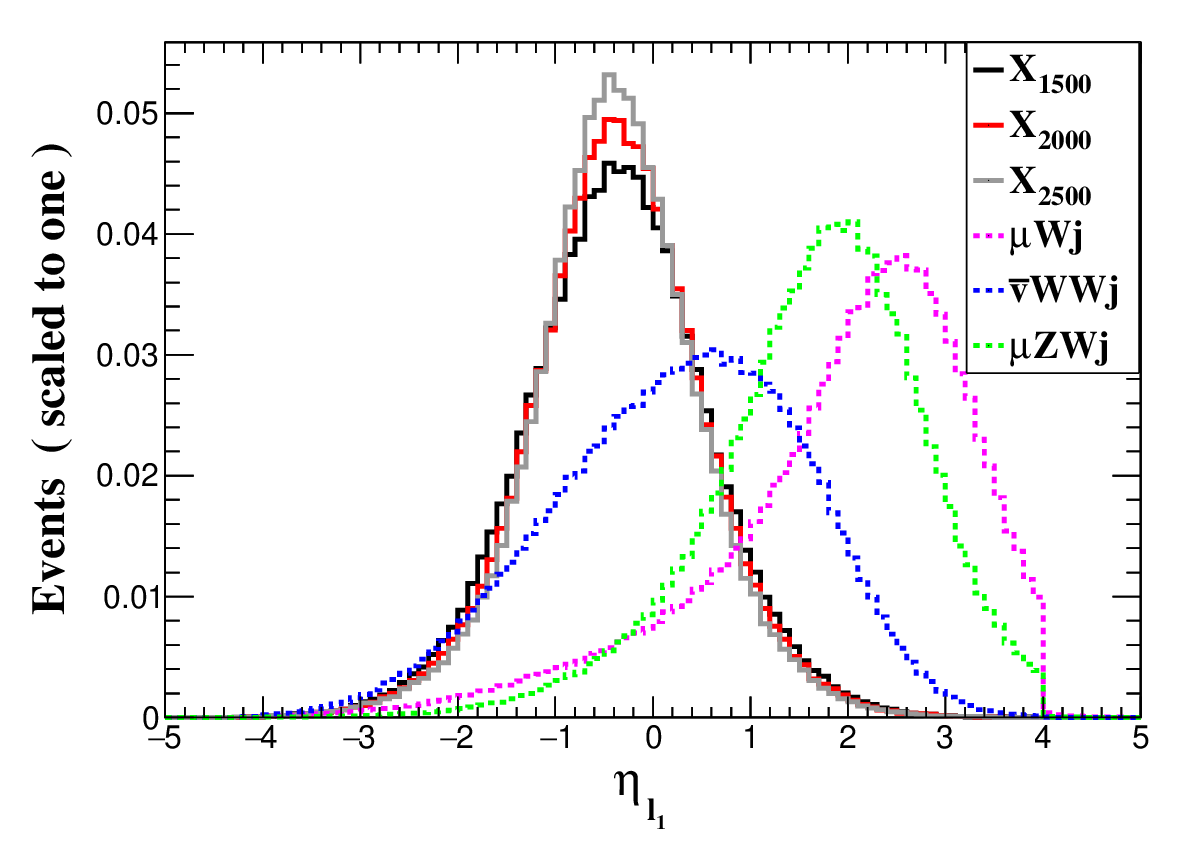} 
  \includegraphics[width=0.45\textwidth]{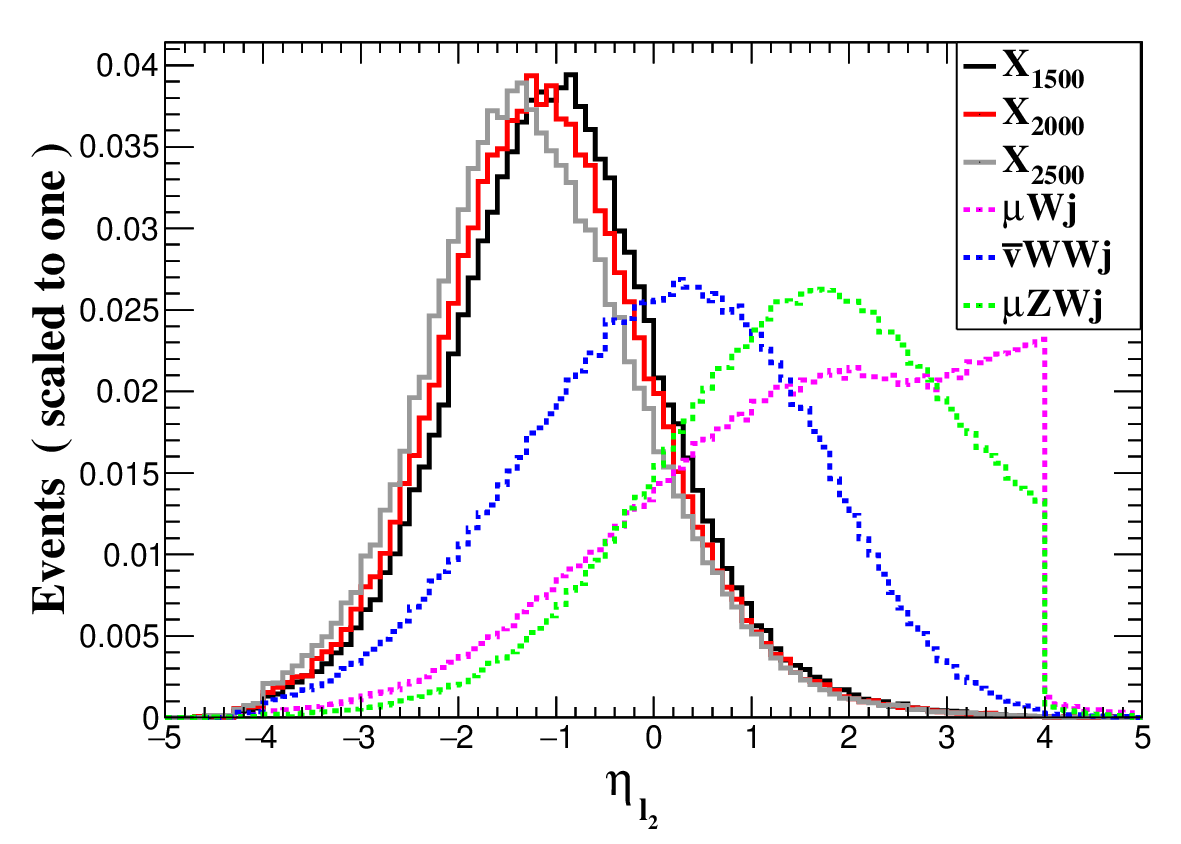}
  \includegraphics[width=0.45\textwidth]{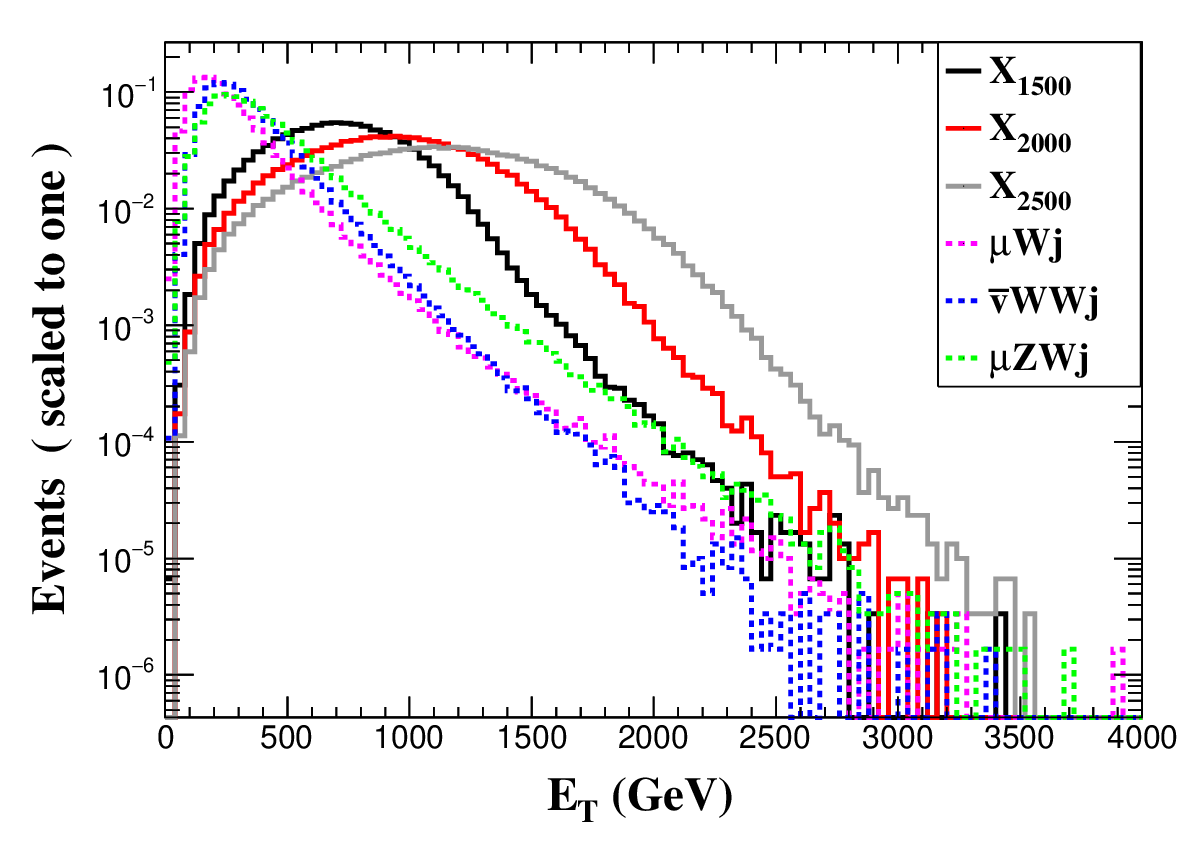}
  \includegraphics[width=0.45\textwidth]{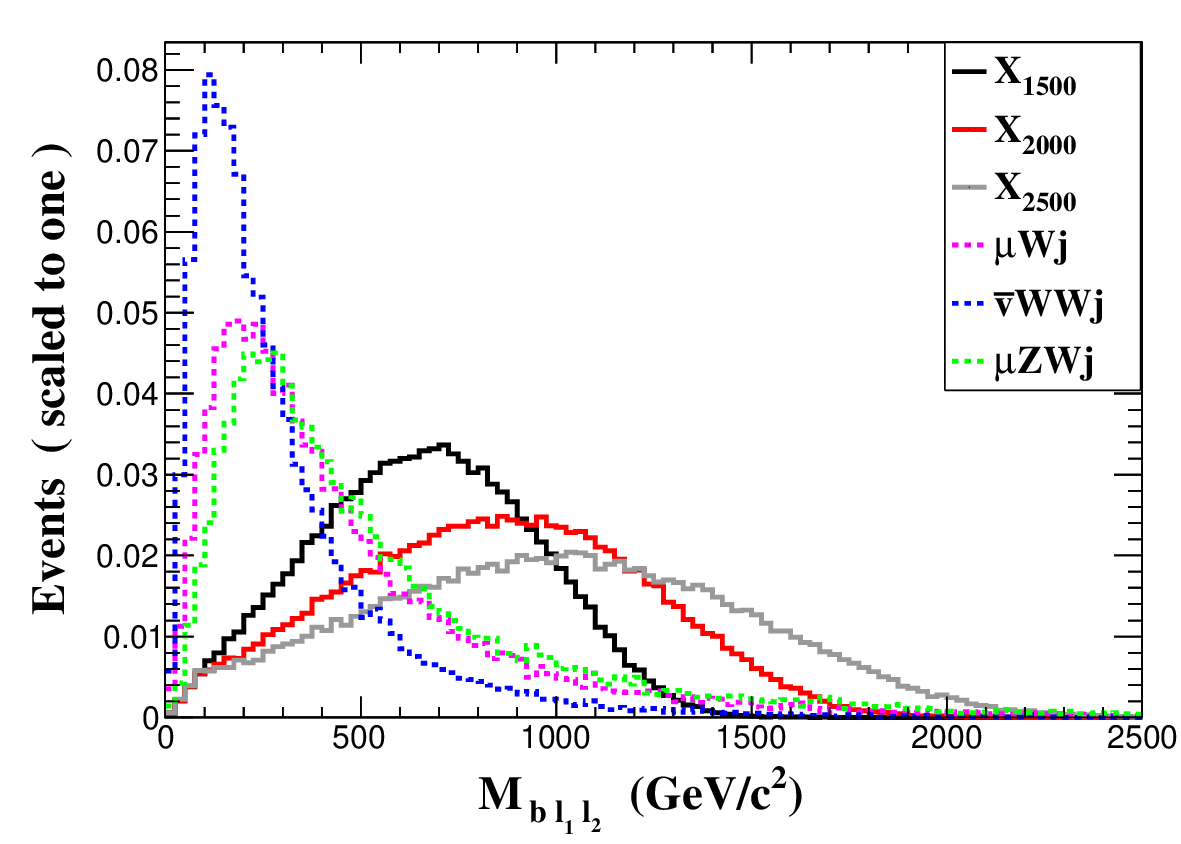}
  \caption{Normalized distributions of representative kinematic observables for the FL signal with $m_X = 1500$, $2000$ and $2500\ \mathrm{GeV}$ and the dominant backgrounds at $\sqrt{s}=5.29\ \mathrm{TeV}$.}
  \label{fig7}
\end{figure}

In the FL mode,
both $W$ bosons decay leptonically, and the dominant backgrounds competing with this topology are:
\[
\begin{aligned}
&\mu^+ p \to \mu^+ W^+ j \quad && \mathrm{with} \quad W^+\to \ell^+ \nu_\ell, \\
&\mu^+ p \to \bar{\nu} W^+ W^+ j \quad && \mathrm{with} \quad W^+\to \ell^+ \nu_\ell, \\
&\mu^+ p \to Z\mu^+ W^+ j \quad && \mathrm{with} \quad Z\to \nu\bar{\nu},\ W^+\to \ell^+ \nu_\ell.
\end{aligned}
\]
To ensure sufficient Monte Carlo (MC) statistics, we generate $3.0\times10^5$ signal events
for each mass point and $6.0\times10^5$ events for each background process.
Figure~\ref{fig7} shows the normalized distribution of the
pseudorapidities ($\eta_{\ell_1}$, $\eta_{\ell_2}$), the
invariant mass reconstructed from the tagged $b$-jet and the two charged leptons 
($M_{b \ell_1 \ell_2}$), and the total transverse energy $E_T$
for three representative signal mass points ($m_X = 1500, 2000,$ and $2500\ \mathrm{GeV}$) alongside the dominant backgrounds
at $\sqrt{s}=5.29\ \mathrm{TeV}$. Here, 
$E_T$ is defined as the scalar sum of the transverse energies of all reconstructed objects in the event.
It is employed because it increases with the VLX mass, reflecting the harder event topology,
and therefore provides a good discriminant between signal and backgrounds.

Guided by these kinematic features, shown in Figure~\ref{fig7}, we implement a sequential cut flow.
We first require exactly two isolated leptons ($N_\ell=2$) and
exactly one $b$-jet ($N_b=1$).
Owing to the asymmetric beam configuration of the $\mu p$ collider, together with the moderate boost
of the heavy VLX, the charged leptons from the VLX decay are preferentially produced in the negative
pseudorapidity region.
As the VLX mass increases, the decay products become increasingly energetic, resulting in larger
values of the total tranverse energy $E_T$.
Moreover, the invariant mass $M_{b\ell_1\ell_2}$, reconstructed from the visible decay products
of the heavy resonance, exhibits a pronounced peak that shifts toward larger values with increasing
VLX mass, whereas the SM backgrounds remain concentrated in the low-mass region.
The corresponding selection criteria are summarized in Table~\ref{table1}.

\begin{table}[htbp]
  \centering
  \vspace{0.1cm}
  \begin{tabular}
  {cccccccc}
  \toprule[1pt]
  \multirow{2}{*}{Cuts} & \multicolumn{3}{c}{Signal(fb)} && \multicolumn{3}{c}{Backgrounds(fb)} \\ 
  \cline{2-4} \cline{6-8}
  & $X_{1500}$ & $X_{2000}$ & $X_{2500}$ && $\mu Wj$ & $\bar{\nu}WWj$ & $\mu ZWj$ \\ 
  \cline{1-8} \midrule[0.8pt]
  Basic cuts & 10.63 & 4.38 & 1.56 && 3314 & 10.19 & 1.79 \\
  $N_l = 2$ & 5.95 & 1.94 & 0.55 && 2355 & 8.10 & 1.43 \\
  $N_b = 1$ & 3.93 & 1.26 & 0.35 && 60.50 & 0.17 & 0.03 \\
  $-2.0 < \eta_{l_1} < 0.2$ & 2.80 & 0.88 & 0.24 && 3.67 & 0.07 & 0.0024 \\
  $-4.0 < \eta_{l_2} < 0.4$ & 2.44 & 0.77 & 0.21 && 0.35 & 0.04 & 0.00026 \\
  $E_T > 500\ \mathrm{GeV}$ & 2.10 & 0.73 & 0.20 && 0.04 & 0.0097 & 0.00011 \\
  $M_{b l_1 l_2} > 500\ \mathrm{GeV}$ & 1.81 & 0.66 & 0.19 && 0.03 & 0.0026 & 0.00006 \\
   \cline{1-8} \midrule[0.8pt]
  Efficiency & 17\% & 15\% & 12\% &&0.0008\% & 0.0252\% & 0.0035\% \\
  \bottomrule[1pt]
  \end{tabular}
\caption{Cut flow for the FL signal ($m_X = 1500$, $2000$ and $2500\ \mathrm{GeV}$) and the backgrounds at $\sqrt{s}=5.29\ \mathrm{TeV}$.}
  \label{table1}
\end{table}

Table~\ref{table1} also displays the signal and background cross sections
after each event selection step, together with the corresponding overall
efficiencies for the FL mode at $\sqrt{s} =5.29\ \mathrm{TeV}$.
The requirement of exactly two isolated leptons ($N_\ell = 2$) exploits
the characteristic dilepton signature of the
signal and substantially suppresses the $\mu W j$ background, which
typically contains only one charged lepton.
The subsequent requirement of exactly one $b$-jet ($N_b=1$) provides the 
strongest background rejection.
Since the signal always contains a $b$-quark originating from the top-quark
decay, whereas
the dominant SM backgrounds do not contain a real top quark, this selection
reduces the background rates by nearly two orders of magnitude while
retaining a large fraction of the signal.
The pseudorapidity selections further suppress the remaining backgrounds
by exploiting the preference of the signal leptons to populate the negative
pseudorapidity region. 
The requirements $E_T > 500\ \mathrm{GeV}$ and $M_{b\ell_1\ell_2}>500\ \mathrm{GeV}$
take advantage of the larger event energy scale associated with the heavy VLX decay,
providing additional discrimination against the residual backgrounds.
After all selections, the dominant $\mu W j$ background is reduced
from $3314$ fb to $0.03$ fb, corresponding to an overall efficiency of $8\times 10^{-6}$,
while the signal efficiencies are between 12\% and 17\%, demonstrating that the
proposed event selection efficiently suppresses the SM backgrounds while
preserving a sizable fraction of the signal.

Since the kinematic distributions at $\sqrt{s} = 6.48$ and $9.16\ \mathrm{TeV}$ 
are similar to those at $\sqrt{s} = 5.29\ \mathrm{TeV}$,
the same event selection is adopted for the higher center-of-mass (CoM) energies.
The corresponding cut-flow tables are presented in Appendix as  Tables~\ref{table2} and \ref{table3}.

\subsection{FH mode}

In the FH mode, all gauge bosons are required to decay hadronically.
To reconstruct the highly boosted $W$
boson while suppressing the combinatorial backgrounds,
we implement jet substructure techniques.
The collimated decay products of the boosted hadronically decaying $W$ boson
are clustered into a single large-radius (fat) jet, denoted by $j_1$,
whose invariant mass is required to
fall within a window around the $W$-boson mass.
The dominant SM background processes considered are:
\[
\begin{aligned}
&\mu^+ p \to \bar{\nu} t Z \quad && \mathrm{with} \quad t\to W^+ b \to jjb,\ Z\to jj, \\
&\mu^+ p \to \bar{\nu} W^+ W^+ j \quad && \mathrm{with} \quad W^+\to jj, \\
&\mu^+ p \to \bar{\nu} Z jj \quad && \mathrm{with} \quad Z\to jj, \\
&\mu^+ p \to \bar{\nu} W^+ jj \quad && \mathrm{with} \quad W^+\to jj.
\end{aligned}
\]

\begin{figure}[htbp]
  \centering
  \includegraphics[width=0.45\textwidth]{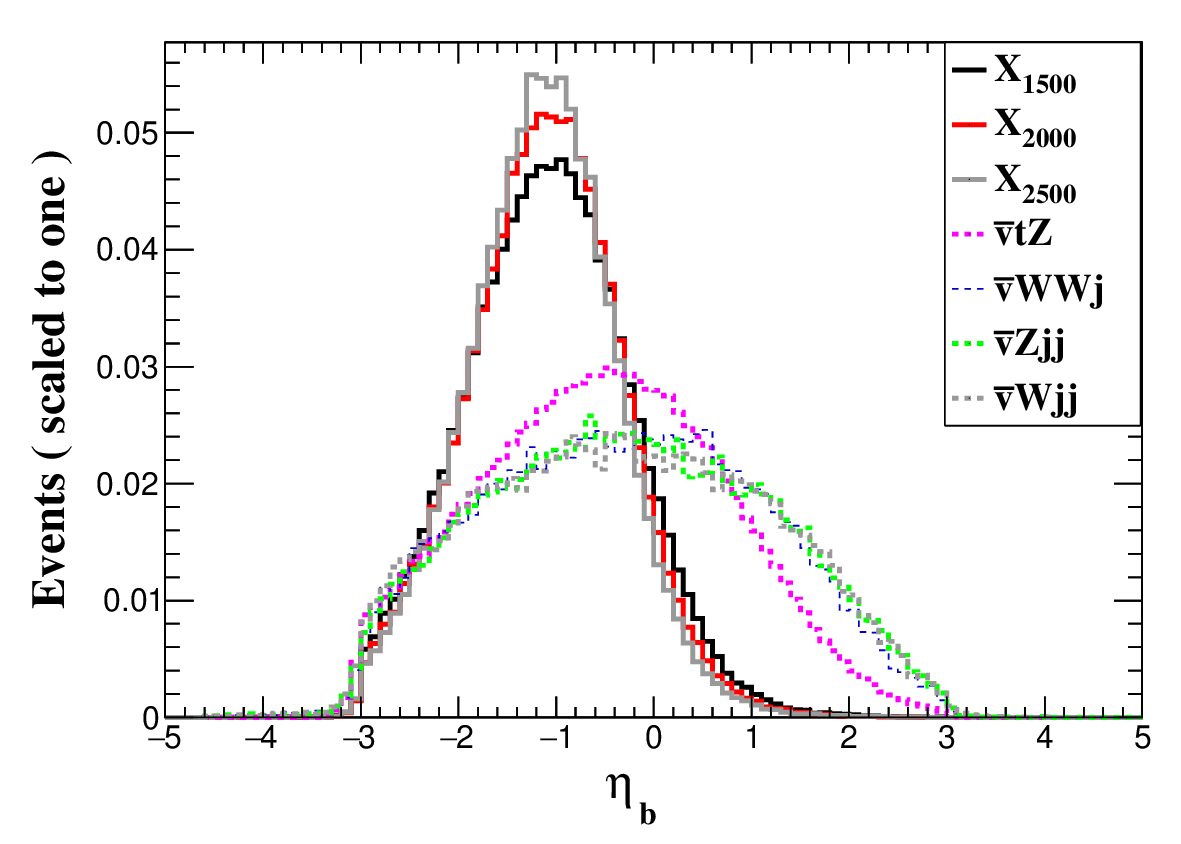}
  \includegraphics[width=0.45\textwidth]{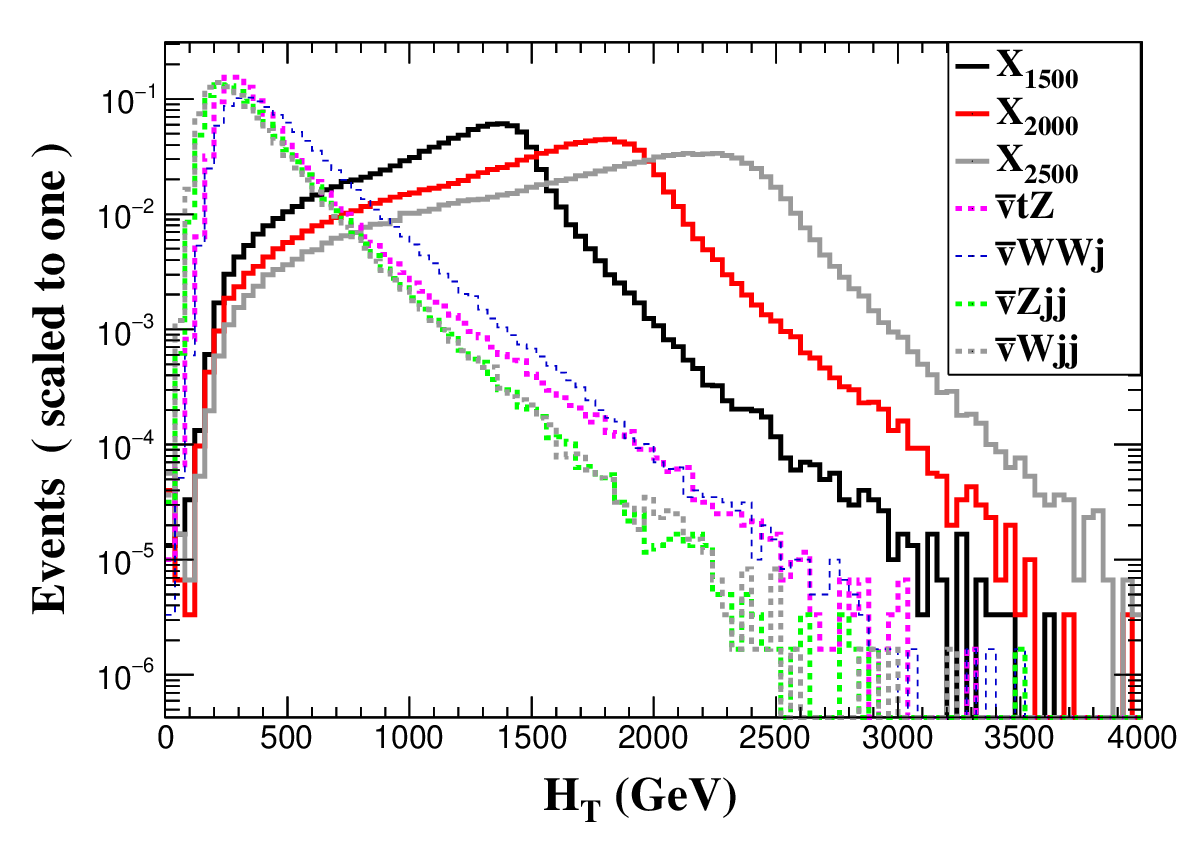}
  \includegraphics[width=0.45\textwidth]{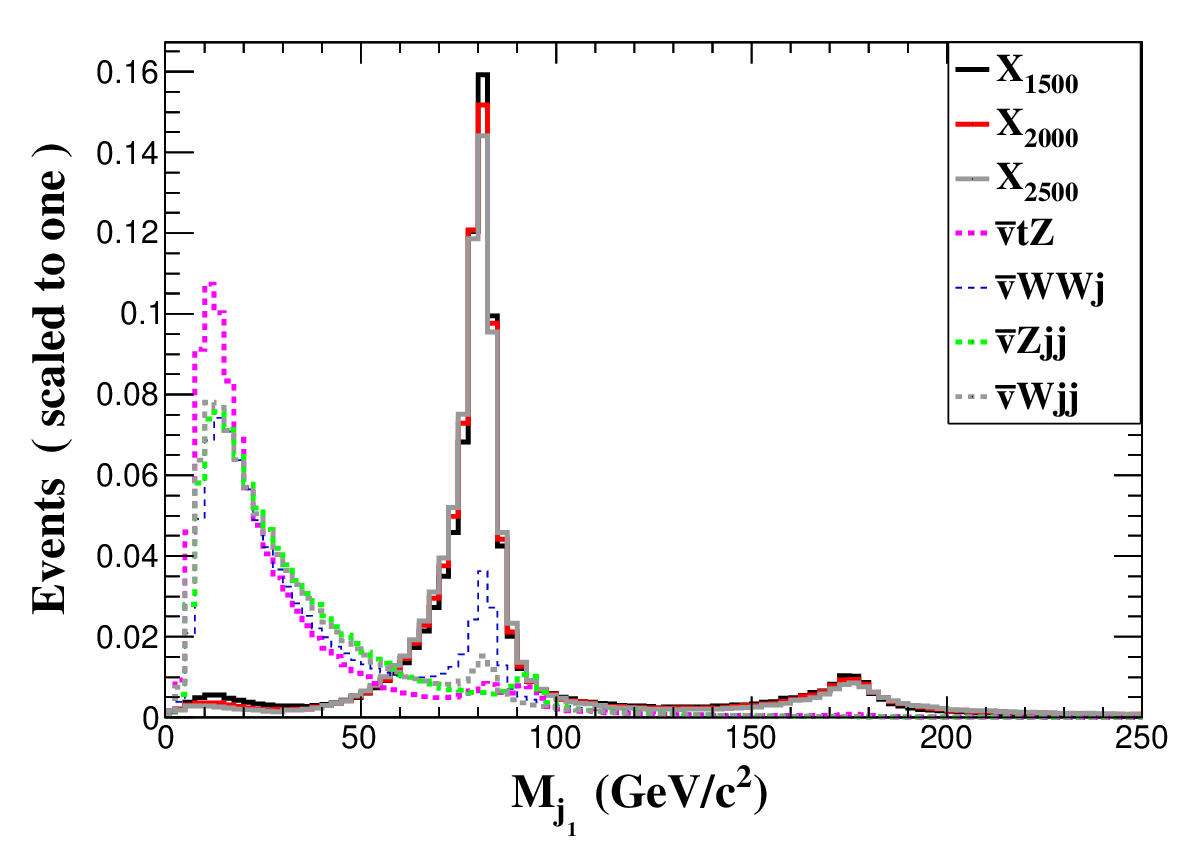}
  \includegraphics[width=0.45\textwidth]{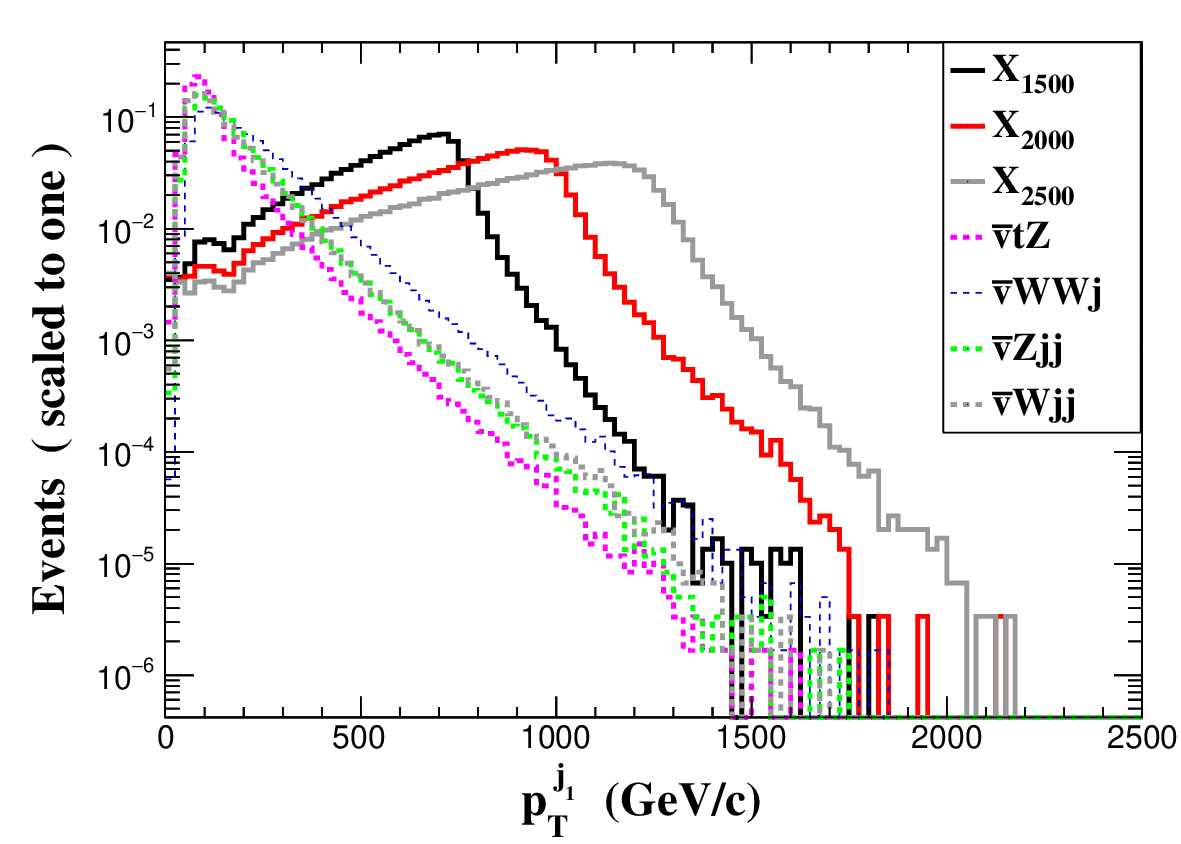}
  \includegraphics[width=0.45\textwidth]{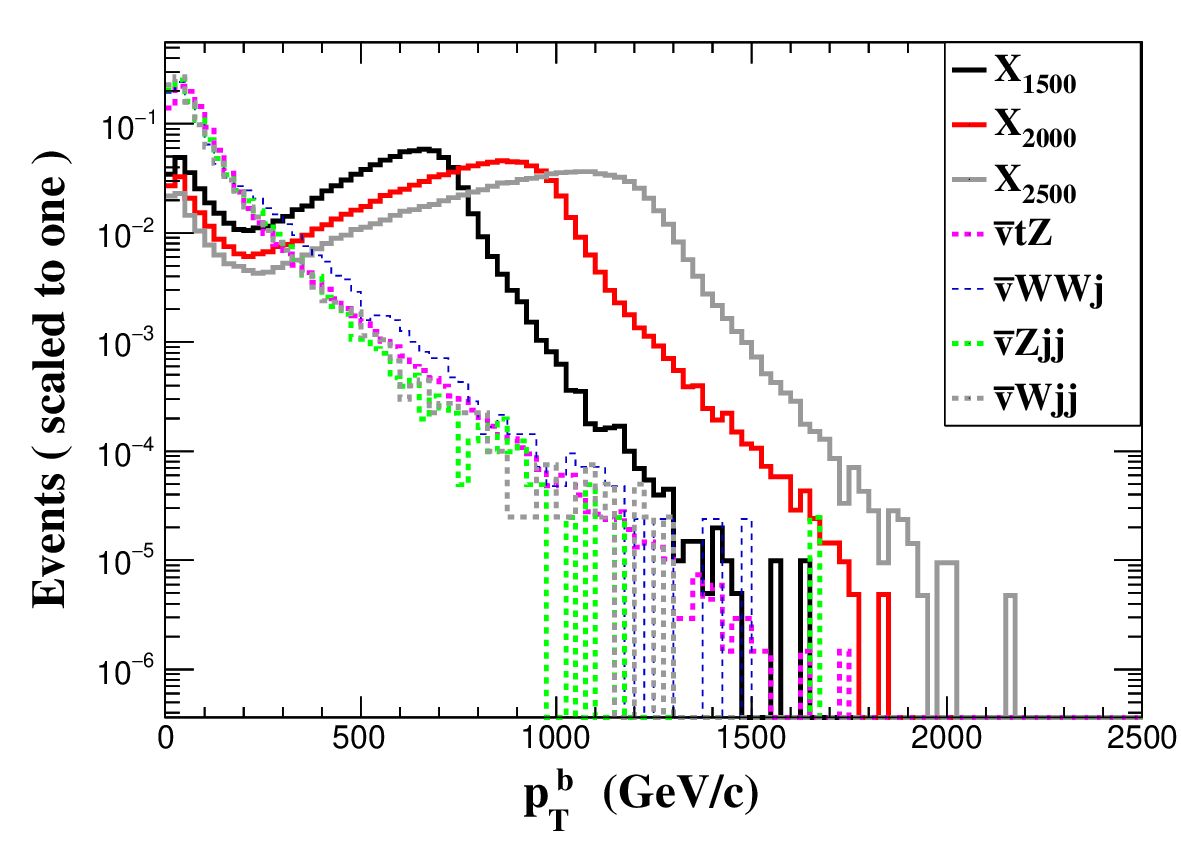}
  \caption{Normalized distributions of representative kinematic observables for the FH signal with $m_X = 1500$, $2000$ and $2500\ \mathrm{GeV}$ and the dominant backgrounds at $\sqrt{s}=5.29\ \mathrm{TeV}$.}
  \label{fig8}
\end{figure}

Guided by the kinematic distributions shown in Figure \ref{fig8},
a sequential cut flow is designed. Owing
to the asymmetric beam configuration of the $\mu p$ collider,
the $b$-jet from the VLX decay is preferentially produced in the negative
pseudorapidity region, providing good discrimination
against the dominant SM backgrounds.
In addition, because of the large VLX mass, the signal events are
characterized by a substantially larger scalar hadronic transverse energy,
$H_T$, defined as the scalar sum of the transverse momenta of all reconstructed jets.
As the VLX mass increases, the $H_T$ distribution shifts toward higher values, 
whereas the SM backgrounds remain concentrated in the low-$H_T$ region.
Furthermore, the boosted $W$ boson from the VLX decay is reconstructed
as a single fat-jet, whose invariant mass exhibits a pronounced peak
around the $W$-boson mass. 
Although backgrounds containing genuine hadronic $W$ bosons also
contribute around $m_W$, a sizable fraction of the background events
populates the low-$M_{j_1}$ region.
Since the parent $W$ boson in the signal carries a large momentum, the reconstructed 
fat-jet also possesses a considerably harder transverse-momentum spectrum than those in the SM
backgrounds.
Finally, the $b$-jet from the heavy VLX decay exhibits a significantly harder $p_T$ distribution,
especially for larger VLX masses. The corresponding selection criteria are summarized in Table~\ref{table4}.

\begin{table}[htbp]
\centering
  \vspace{0.1cm}
  \begin{tabular}
  {ccccccccc}
  \toprule[1pt]
  \multirow{2}{*}{Cuts} & \multicolumn{3}{c}{Signal(fb)} && \multicolumn{3}{c}{Backgrounds(fb)} \\ 
  \cline{2-4} \cline{6-9}
  & $X_{1500}$ & $X_{2000}$ & $X_{2500}$ && $\bar{\nu}tZ$ & $\bar{\nu} WWj$ & $\bar{\nu} Zjj$ & $\bar{\nu}Wjj$ \\
  \cline{1-9} \midrule[0.8pt]
  Basic cuts & 94.13 & 39.50 & 13.99 && 96.03 & 91.78 & 1031 & 2859 \\
  $N_j\ge2$    & 80.43 & 32.00 & 10.94  && 95.56  & 91.18  & 1029 & 2853.91 \\
  $N_b=1$ & 44.31 & 17.72 & 6.09 && 49.90 & 5.77 & 57.90 & 158.90 \\
  $-2.6<\eta_{b}<0$ & 38.56 & 15.84 & 5.52 && 30.55 & 2.96 & 29.41 & 78.98 \\
  $H_T>900\ \mathrm{GeV}$ & 30.47 & 14.32 & 5.24 && 0.83 & 0.21 & 0.94 & 2.47 \\
  $70\ \mathrm{GeV}<M_{j_1}<90\ \mathrm{GeV}$ & 21.23 & 9.82 & 3.53 && 0.13 & 0.08 & 0.15 & 0.62 \\
  $p_{T}^{j_1}>600\ \mathrm{GeV}$ & 13.19 & 8.29 & 3.21 && 0.02 & 0.02 & 0.03 & 0.18 \\
  $p_{T}^{b}>400\ \mathrm{GeV}$ & 11.58 & 7.63 & 3.02 && 0.009 & 0.006 & 0.01 & 0.09 \\
  \cline{1-9} \midrule[0.8pt]
  Efficiency & 12\% & 19\% & 22\% && 0.009\% & 0.007\% & 0.0007\% & 0.001\% \\
\bottomrule[1pt]
\end{tabular}
\caption{Cut flow for the FH signal ($m_X = 1500$, $2000$ and $2500\ \mathrm{GeV}$) and the backgrounds at $\sqrt{s}=5.29\ \mathrm{TeV}$.}
\label{table4}
\end{table}

Table~\ref{table4} also displays the signal and background cross sections after each event selection step,
together with the corresponding overall efficiencies for the FH mode at $\sqrt{s} = 5.29\ \mathrm{TeV}$.
The requirement of at least two reconstructed jets ($N_j \ge 2$) retains most of the signal events while 
having some impact on the dominant backgrounds.
Requiring exactly one $b$-jet ($N_b=1$) provides the first substantial background suppression.
Since the signal always contains a real $b$ quark from the top-quark decay, whereas the
$\bar{\nu}WWj$, $\bar{\nu}Zjj$, and $\bar{\nu}Wjj$ backgrounds do not contain a genuine top quark,
this requirement suppresses these backgrounds by more than one order of magnitude while
preserving about half of the signal events.
The $\eta_b$ selection further improves the signal-to-background ratio by exploiting 
the characteristic backward pseudorapidity distribution of the signal $b$-jet.

Among all kinematic selections, the requirement $H_T>900\ \mathrm{GeV}$ provides
the  strongest rejection of the remaining backgrounds.
Owing to the much larger energy scale associated with the heavy VLX decay,
the signal events populate the high-$H_T$ region,
where the backgrounds are concentrated at lower values of $H_T$.
This requirement reduces all background channels to the fb level or below
while retaining a large fraction of the signal. 
The subsequent $W$-tagging requirement, 
$70\ \mathrm{GeV}<M_{j_1}<90\ \mathrm{GeV}$,
together with the transverse-momentum selections on the leading fat-jet and the $b$-jet,
$p_T^{j_1}>600\ \mathrm{GeV}$ and $p_T^b>400\ \mathrm{GeV}$,
further exploits the boosted topology of the signal and suppresses the residual backgrounds by 
another one to two orders of magnitude.

After all selection criteria are applied, all background processes are reduced to below
$0.1\ \mathrm{fb}$, with overall efficiencies below $0.01\%$, 
while signal efficiencies remain between $12\%$ and $22\%$.
These results demonstrate that the
proposed event selection effectively suppresses the SM backgrounds while
maintaining good signal acceptance.

Since the kinematic distributions at $\sqrt{s} = 6.48$ and $9.16\ \mathrm{TeV}$ are
similar to those at $\sqrt{s} = 5.29\ \mathrm{TeV}$,
the same event selection is adopted for the higher CoM energies.
The corresponding cut-flow tables are presented in Appendix as  Tables~\ref{table5} and \ref{table6}.

\subsection{SL1 and SL2 modes}

The semi-leptonic topology provides a compromise between
signal statistics and background suppression by combining 
the clean experimental signature of a charged lepton
with the relatively large hadronic BR of the $W$ boson.
Depending on which of the two $W$ bosons decays leptonically,
the signal can be classified into two distinct modes,
denoted as SL1 and SL2, as intimated and now illustrated in Figure~\ref{fig2}(c)
and~\ref{fig2}(d). In the SL1 mode, the $W$ boson produced directly
in the decay $X \to tW^+$ decays leptonically,
while the $W$ boson originating from the 
subsequent top-quark decay decays hadronically.
Consequently, the charged lepton directly inherits
the large Lorentz boost of the heavy VLX, whereas
the hadronic activity mainly arises from the boosted top-quark
decay.
By contrast, in the SL2 mode, the $W$ boson emitted directly from
the VLX decays hadronically, producing a highly boosted hadronic 
$W$ that is reconstructed as a fat-jet.
The second $W$ boson, originating from the top-quark decay,
decays leptonically, so that
the charged lepton is produced through the intermediate top quark.
Therefore, these two decay topologies share the same visible final state, 
$\ell^+ + b+ jj+\not\!\!\mathrm{E}_\mathrm{T}$, and the dominant
background processes considered in their analysis are
\[
\begin{aligned}
&\mu^+ p \to \bar{\nu} W^+ W^+ j \quad && \mathrm{with} \quad W^+\to \ell^+ \nu_l, W^+ \to jj, \\
&\mu^+ p \to\bar{\nu} t Z \quad && \mathrm{with} \quad t\to W^+ b \to (\ell^+ \nu_\ell)b,\ Z\to jj, \\
&\mu^+ p \to Z \mu^+ W^+ j \quad && \mathrm{with} \quad Z\to \nu\bar{\nu},\ W^+\to jj, \\
&\mu^+ p \to \mu^+ W^+ j \quad && \mathrm{with} \quad W^+\to \ell^+ \nu_\ell.
\end{aligned}
\]

\subsubsection{SL1 mode}

As illustrated in Figure~\ref{fig2}(c), the charged lepton in the SL1
topology originates from the $W$ boson emitted directly in the decay
$X \to tW^+$, where the second $W$ boson subsequently decays
hadronically through the top quark. As intimated, here, 
since the $W$ boson is produced directly from the decay of the heavy VLX,
it acquires a large fraction of the parent momentum, resulting in an extremely 
energetic charged lepton.
Meanwhile, the accompanying top quark is also significantly boosted,
producing a hard $b$-jet folowing its decay.
Consequently, both the charged lepton and the $b$-jet exhibit
kinematic features that
differ significantly from those of the SM backgrounds.

\begin{figure}[htbp]
  \centering
  \includegraphics[width=0.45\textwidth]{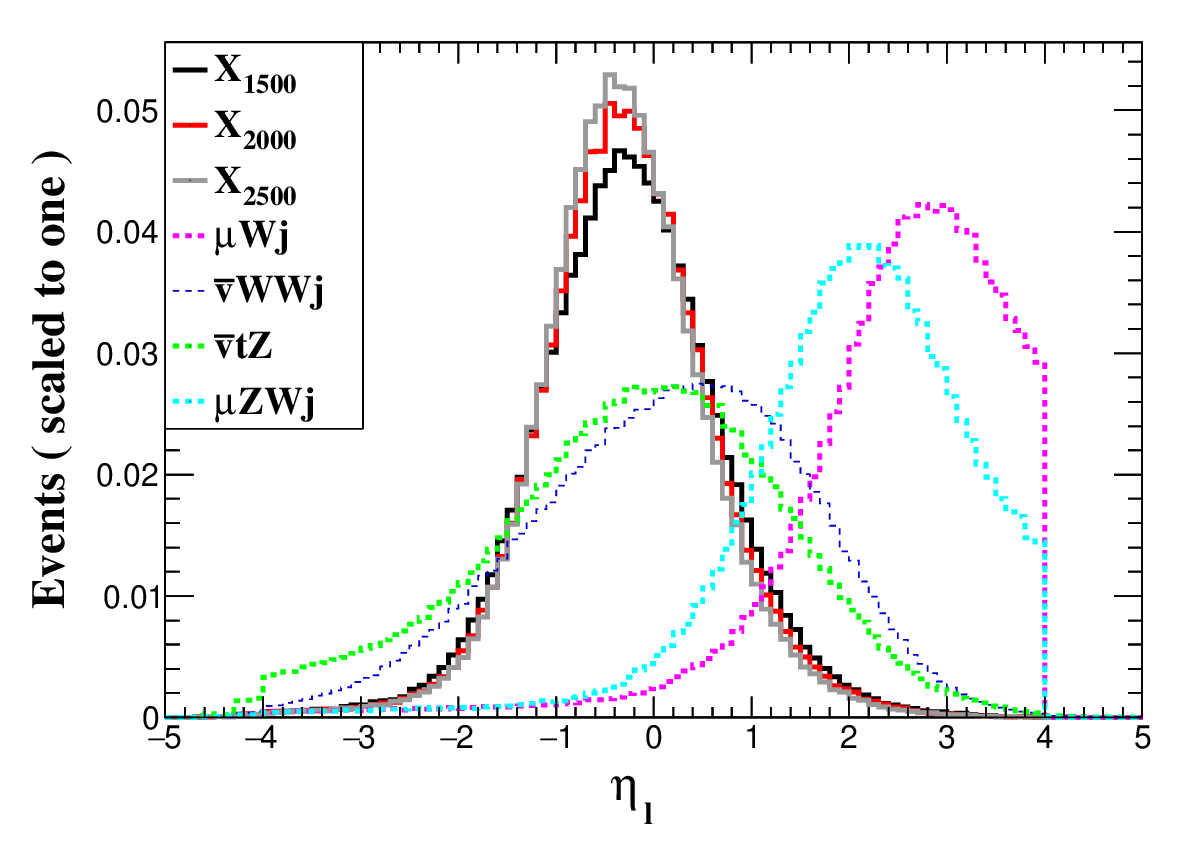}
  \includegraphics[width=0.45\textwidth]{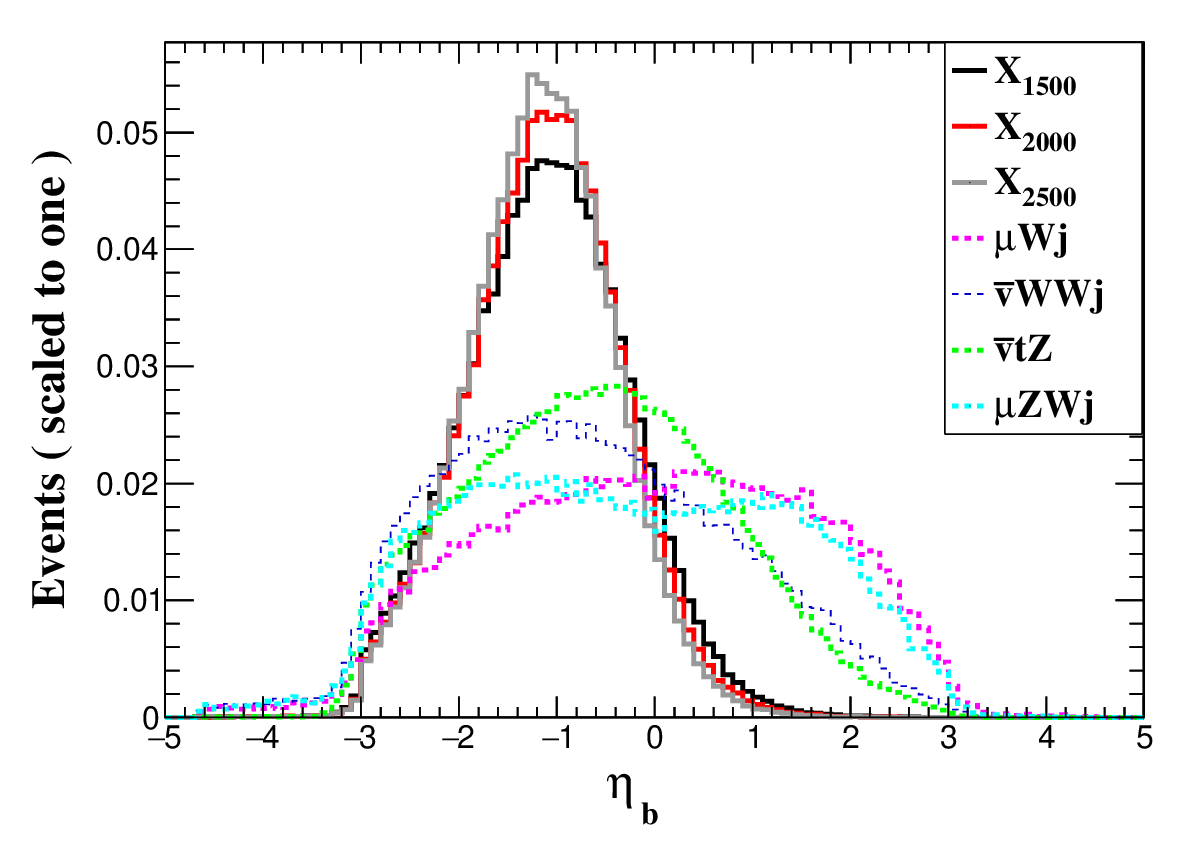}
  \includegraphics[width=0.45\textwidth]{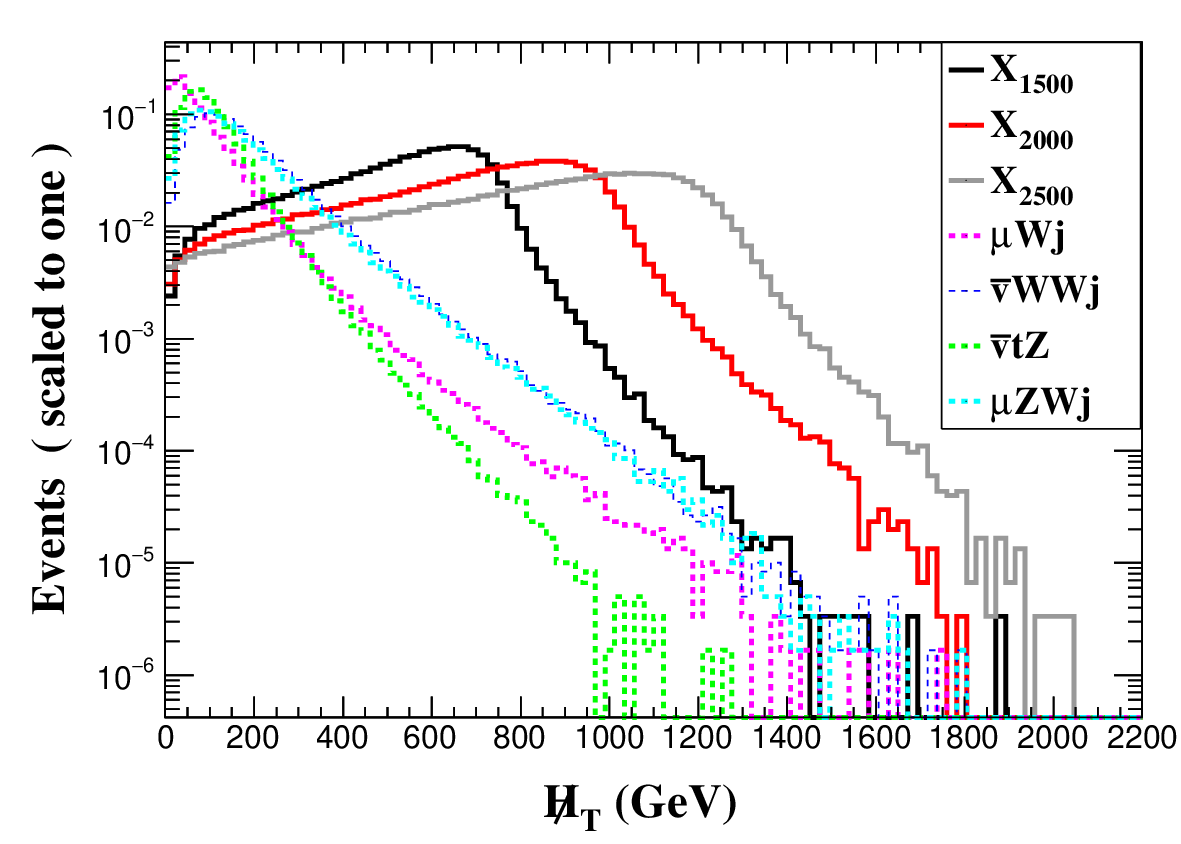}
  \includegraphics[width=0.45\textwidth]{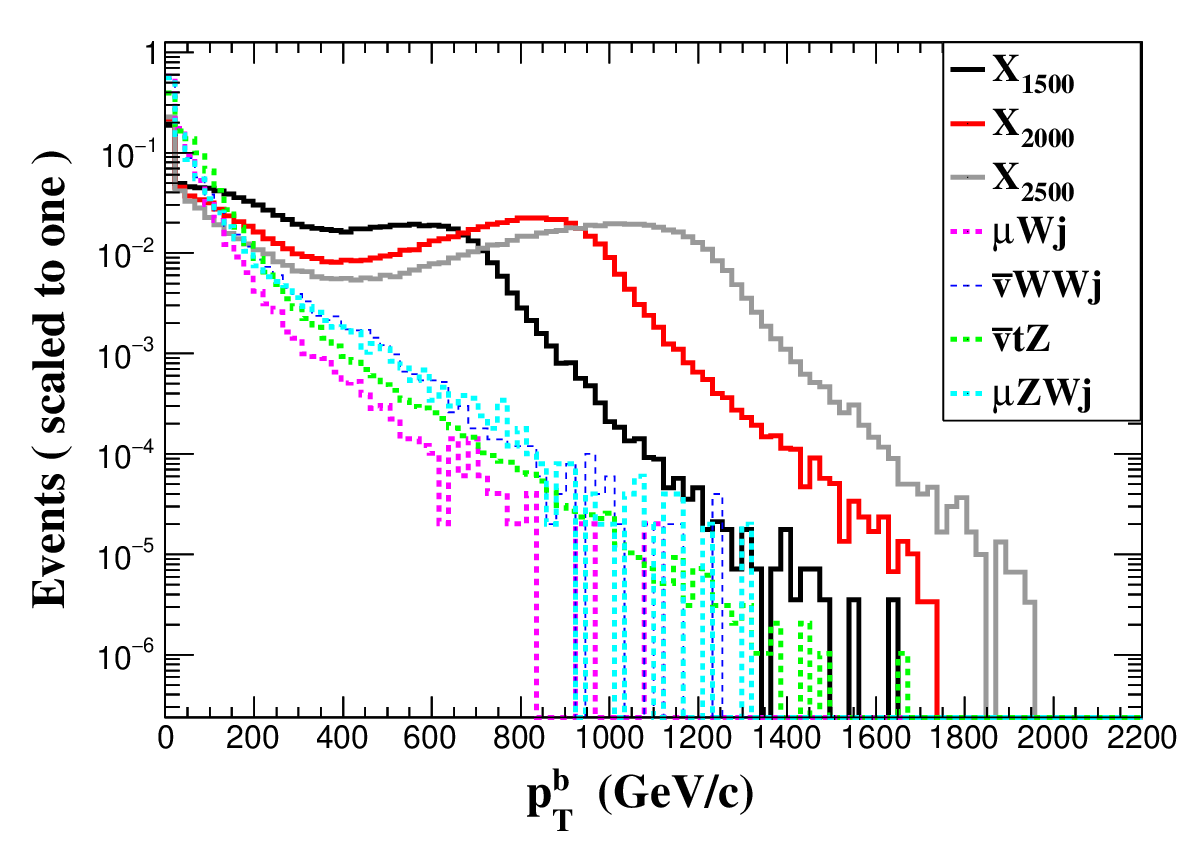}
  \caption{Normalized distributions of representative kinematic observables for the SL1 signal with $m_X = 1500$, $2000$ and $2500\ \mathrm{GeV}$ and the dominant backgrounds at $\sqrt{s}=5.29\ \mathrm{TeV}$.}
  \label{fig9}
\end{figure}

These characteristics are presented in Figure~\ref{fig9}.
The upper-left panel shows that the signal lepton is
preferentially produced in the backward pseudorapidity region,
while the dominant $\mu W j$ background is mainly distributed
toward positive pseudorapidities.
This behavior reflects the production kinematics of the heavy
VLX and motivates the requirement $-1.4<\eta_l<0.6$,
which removes the majority of the $\mu Wj$ background while retaining
most of the signal. A similar feature is observed for the
$b$-jet pseudorapidity shown in the upper-right panel.
Since the top quark originates from the heavy VLX decay, its
decay products are also preferentially emitted in 
the backward direction, leading to a relatively narrow 
signal  distribution centered around $\eta_b \sim~1$.
This motivates the requirement $-2.6 < \eta_b < 0$, 
which suppresses all background processes. 
The lower-right panel shows the transverse momentum of the $b$-jet. 
As expected, due to the large Lorentz boost inherited from the heavy VLX,
the signal exhibits a much harder $p_T^b$ spectrum than the backgrounds,
with the spectrum becoming increasingly energetic as $m_X$
increases. Therefore, the requirement $p_T^b>400\ \mathrm{GeV}$ 
provides one of the effective discriminating variables.

The lower-left panel shows the missing hadronic transverse momentum, $\not\!\!\mathrm{H}_T$,
defined as the magnitude of the vector sum of the transverse momenta of all hadronic particles
in the final state. In the signal topology shown in Figure~\ref{fig2}(c), $\not\!\!\mathrm{H}_T$
serves a proxy for the transverse momentum of the highly boosted top quark originating from
the heavy VLX resonance. Consequently, requiring $\not\!\!\mathrm{H}_T>500~\mathrm{GeV}$
further improves the signal purity.

\begin{table}[htbp]
\centering
  \vspace{0.1cm}
  \begin{tabular}
  {cccccccc}
  \toprule[1pt]
  \multirow{2}{*}{Cuts} & \multicolumn{3}{c}{Signal(fb)} & \multicolumn{4}{c}{Backgrounds(fb)} \\ 
  \cline{2-4} \cline{5-8}
  & $X_{1500}$ & $X_{2000}$ & $X_{2500}$ & $\mu Wj$ & $\bar{\nu}WWj$ & $\bar{\nu}Zt$ & $\mu ZWj$ \\ 
  \cline{1-8} \midrule[0.8pt]
  Basic cuts & 31.89 & 12.97 & 4.67 & 9941 & 61.07 & 32.04 & 5.36 \\
  $N_b=1$ & 19.82 & 7.85 & 2.79 & 592.10 & 3.47 & 12.32 & 0.30 \\
  $N_l=1$ & 17.30 & 6.80 & 2.43 & 447.20 & 2.83 & 8.70 & 0.25 \\
  $-2.6<\eta_b<0$ & 15.19 & 6.20 & 2.22 & 173.40 & 1.53 & 5.58 & 0.09 \\
  $-1.4<\eta_l<0.6$ & 11.48 & 4.98 & 1.86 & 5.37 & 0.71 & 3.26 & 0.006 \\
  $p_{T}^{b}>400\ \mathrm{GeV}$ & 4.65 & 3.14 & 1.35 & 0.15 & 0.02 & 0.02 & 0.0005 \\
  $\not\!\!\mathrm{H}_\mathrm{T}>500\ \mathrm{GeV}$ & 4.36 & 3.08 & 1.34 & 0.07 & 0.008 & 0.003 & 0.0003 \\
  \cline{1-8} \midrule[0.8pt]
  Efficiency & 14\% & 24\% & 29\% & 0.0007\% & 0.01\% & 0.01\% & 0.006\% \\
\bottomrule[1pt]
\end{tabular}
\caption{Cut flow for the SL1 signal ($m_X = 1500$, $2000$ and $2500\ \mathrm{GeV}$) and the backgrounds at $\sqrt{s}=5.29\ \mathrm{TeV}$.}
\label{table7}
\end{table}

The cut flow corresponding to these optimized selections is summarized in Table~\ref{table7}.
The dominant $\mu Wj$ background is reduced from $9941\ \mathrm{fb}$ to only $0.07\ \mathrm{fb}$,
corresponding to an overall efficiency of merely $0.0007\%$.
In particular, the requirement of exactly one tagged $b$-jet ($N_b=1$) and the  one on the lepton pseudorapidity
provide the largest background rejection, while the $\not\!\!\mathrm{H}_T$ requirement
further removes the background events with only a minor loss of signal.
Meanwhile, the signal efficiency increases from $14\%$ to $29\%$ as the VLX mass increases,
reflecting the increasingly boosted decay products of heavier VLXs.

The corresponding cut flow tables at $\sqrt{s}=6.48$ and $9.16~\mathrm{TeV}$ 
are presented in Appendix as Tables~\ref{table8} and~\ref{table9}, respectively.

\subsubsection{SL2 mode}

The SL2 topology, shown in Figure~\ref{fig2}(d), differs from SL1 in the decay pattern of the two $W$ bosons.
In this case, the $W$ boson produced directly from the heavy VLX decays hadronically, 
while the second $W$ boson originating from the top-quark decay subsequently decays leptonically.
Consequently, as explained, the hadronic $W$ receives the full boost from the heavy VLX and
is reconstructed as a highly energetic fat-jet, whereas
the charged lepton acquires only the boost of the intermediate top quark.

\begin{figure}[htbp]
  \centering
  \includegraphics[width=0.45\textwidth]{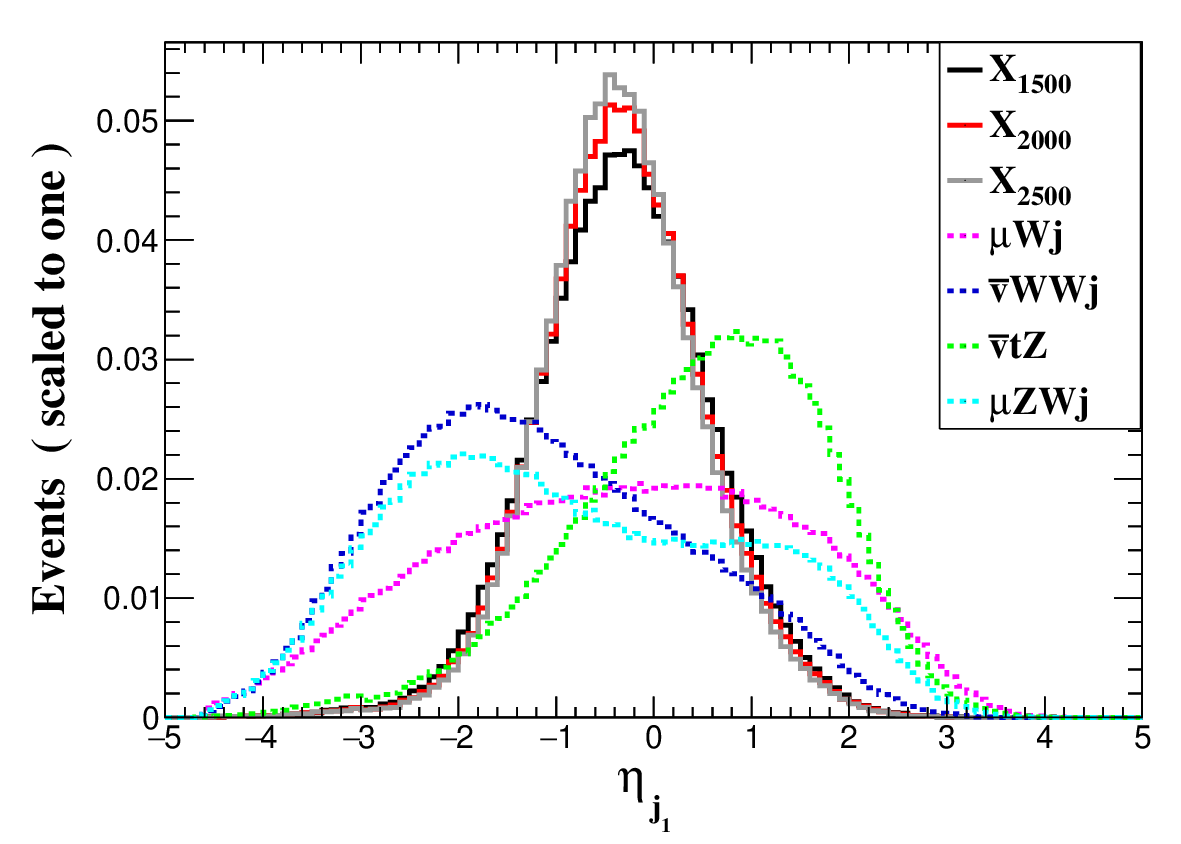}
  \includegraphics[width=0.45\textwidth]{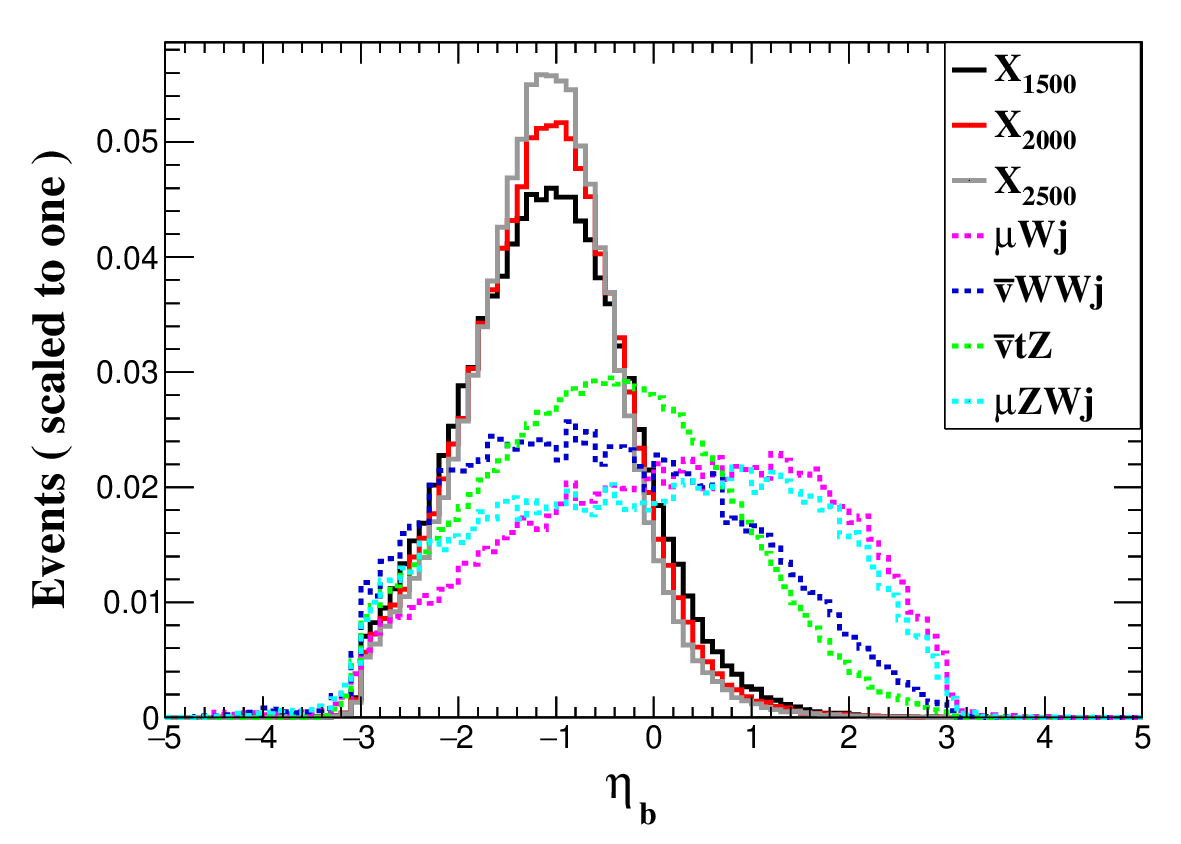}
  \includegraphics[width=0.45\textwidth]{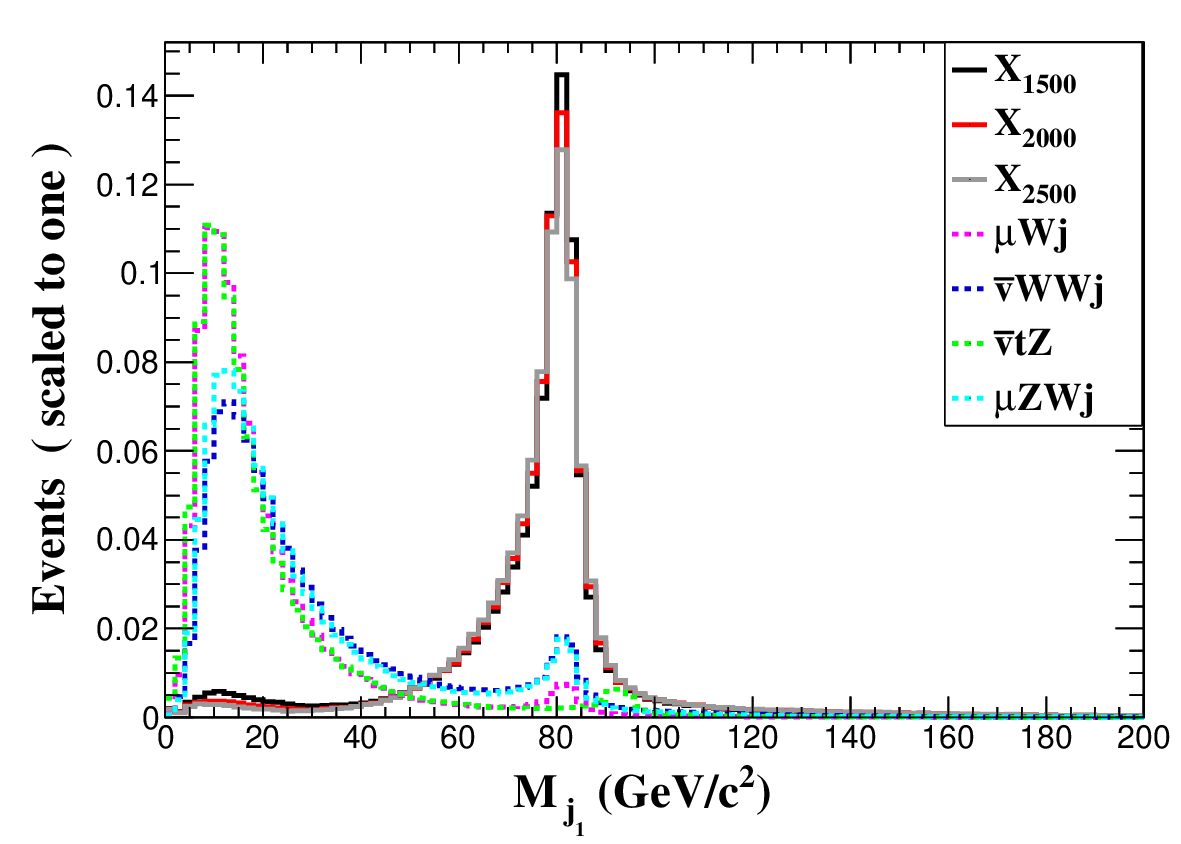}
  \includegraphics[width=0.45\textwidth]{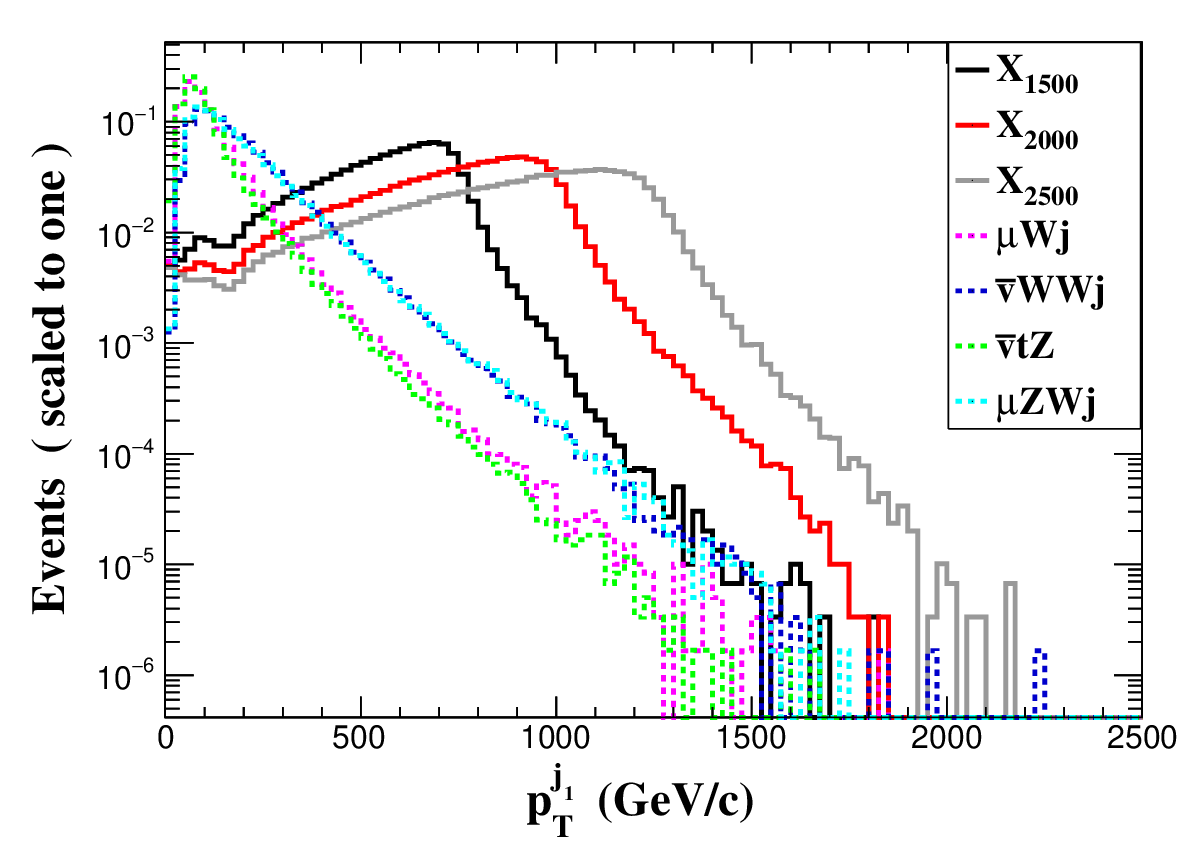}
  \includegraphics[width=0.45\textwidth]{figures/case1Ptb.eps}
\caption{Normalized distributions of representative kinematic observables for the SL2 signal with $m_X = 1500$, $2000$ and $2500\ \mathrm{GeV}$ and the dominant backgrounds at $\sqrt{s}=5.29\ \mathrm{TeV}$.}
  \label{fig10}
\end{figure}

This distinct decay topology leads to the characteristic kinematic distributions shown in Figure~\ref{fig10}.
The leading fat-jet, originating from the directly produced $W$ boson, is preferentially
emitted into the backward region, motivating the requirement $-1.6 < \eta_{j_1} < 0.4$.
Since this jet corresponds to a genuine hadronic $W$ decay, its 
invariant mass exhibits a pronounced peak around the $W$-boson mass.
In contrast, the leading jet in the dominant SM backgrounds (such as $\mu W j$
with leptonic $W$ decay) typically originates from initial- or final-state 
radiation, thereby populating the low-mass region. Therefore,
the mass window $60\ \mathrm{GeV}<M_{j_1}<100\ \mathrm{GeV}$ 
efficiently identifies boosted hadronic $W$ bosons from the signal while rejecting these non-resonant backgrounds.

The middle-right panel further shows that the leading fat-jet possesses
a significantly harder transverse momentum spectrum than the backgrounds
because it directly inherits the large Lorentz boost of the heavy VLX.
This motivates the requirement $p_T^{j_1}>600~\mathrm{GeV}$, which 
constitutes one of the most powerful selections in this channel.

The heavy mass of the VLX imparts a significant boost to the intermediate top quark,
forcing its subsequent $b$-jet to possess a highly energetic $p_T$ spectrum, as shown in
the bottom panel, and a narrow, backward-peaked pseudorapidity distribution, as demonstrated in 
the top-right panel. These features motivate the requirements $-2.4<\eta_b<0$ and $p_T^b>380\ \mathrm{GeV}$.

\begin{table}[htbp]
\centering
  \vspace{0.1cm}
  \begin{tabular}
  {cccccccc}
  \toprule[1pt]
  \multirow{2}{*}{Cuts} & \multicolumn{3}{c}{Signal(fb)} & \multicolumn{4}{c}{Backgrounds(fb)} \\ 
  \cline{2-4} \cline{5-8}
  & $X_{1500}$ & $X_{2000}$ & $X_{2500}$ & $\mu Wj$ & $\bar{\nu}WWj$ & $\bar{\nu}Zt$ & $\mu ZWj$ \\ 
  \cline{1-8} \midrule[0.8pt]
  Basic cuts & 31.90 & 13.20 & 4.66 & 9941 & 61.07 & 32.04 & 5.35 \\
  $N_j\ge1$ & 31.62 & 13.03 & 4.62 & 9940 & 61.05 & 31.91 & 5.35 \\
  $N_b=1$ & 15.82 & 6.73 & 2.52 & 409.50 & 2.41 & 16.89 & 0.21 \\
  $-1.6<\eta_{j_1}<0.4$ & 11.30 & 5.15 & 2.02 & 143.70 & 0.94 & 6.26 & 0.07 \\
  $-2.4<\eta_b<0$ & 9.54 & 4.65 & 1.89 & 55.38 & 0.50 & 3.89 & 0.03 \\
  $60\ \mathrm{GeV}<M_{j_1}<100\ \mathrm{GeV}$ & 7.98 & 3.88 & 1.57 & 4.00 & 0.07 & 0.33 & 0.004 \\
  $p_{T}^{j_1}>600\ \mathrm{GeV}$ & 5.10 & 3.58 & 1.54 & 0.25 & 0.004 & 0.012 & 0.0004 \\
  $p_{T}^{b}>380\ \mathrm{GeV}$ & 2.85 & 2.72 & 1.29 & 0.08 & 0.001 & 0.004 & 0.0001 \\
  \cline{1-8} \midrule[0.8pt]
  Efficiency & 9\% & 21\% & 28\% & 0.0008\% & 0.002\% & 0.01\% & 0.003\% \\
\bottomrule[1pt]
\end{tabular}
\caption{Cut flow for the SL2 signal ($m_X = 1500$, $2000$ and $2500\ \mathrm{GeV}$) and the backgrounds at $\sqrt{s}=5.29\ \mathrm{TeV}$.}
\label{table10}
\end{table}

Table~\ref{table10} details the cut flow and ensuing efficiencies. The dominant $\mu W j$ background experiences a
massive suppression from $9941$ fb down to $0.08$ fb (a total efficiency of $0.0008\%$). 
This robust background rejection relies primarily on the $b$-jet requirement to remove
light-flavor processes, followed by the $W$ fat-jet mass window and hard $p_T$ requirements 
to filter out the remaining non-boosted topologies.
Furthermore, because heavier VLXs yield highly energetic final states that easily pass 
the discussed kinematic criteria, the signal efficiency increases from $9\%$
for $m_X=1500\ \mathrm{GeV}$ to $28\%$ for $m_X=2500\ \mathrm{GeV}$.

The corresponding cut flow tables at $\sqrt{s}=6.48$ and $9.16~\mathrm{TeV}$ are presented in Appendix as Tables~\ref{table11} and~\ref{table12}, respectively.

\section{Exclusion and Discovery Reach}

The expected sensitivity of a $\mu p $ collider to the VLX signal
is evaluated using the Asimov significance variables~\cite{Cowan:2010js},
\begin{equation}
\mathcal{Z}_{\mathrm{exc}} = \sqrt{2\left[s - b\ln\left(1 + \frac{s}{b}\right)\right],}
\end{equation}
\begin{equation}
\mathcal{Z}_{\mathrm{dis}} = \sqrt{2\left[(s + b)\ln\left(1 + \frac{s}{b}\right) - s\right],}
\end{equation}
where $s$ and $b$ denote the expected signal and background event yields.
The exclusion and discovery reaches are defined by 
$\mathcal{Z}_{\mathrm{exc}} = 2$ and $\mathcal{Z}_{\mathrm{dis}} = 5$,
respectively.

\begin{figure}[htbp]
  \centering
  \includegraphics[width=0.45\textwidth, height=5.7cm]{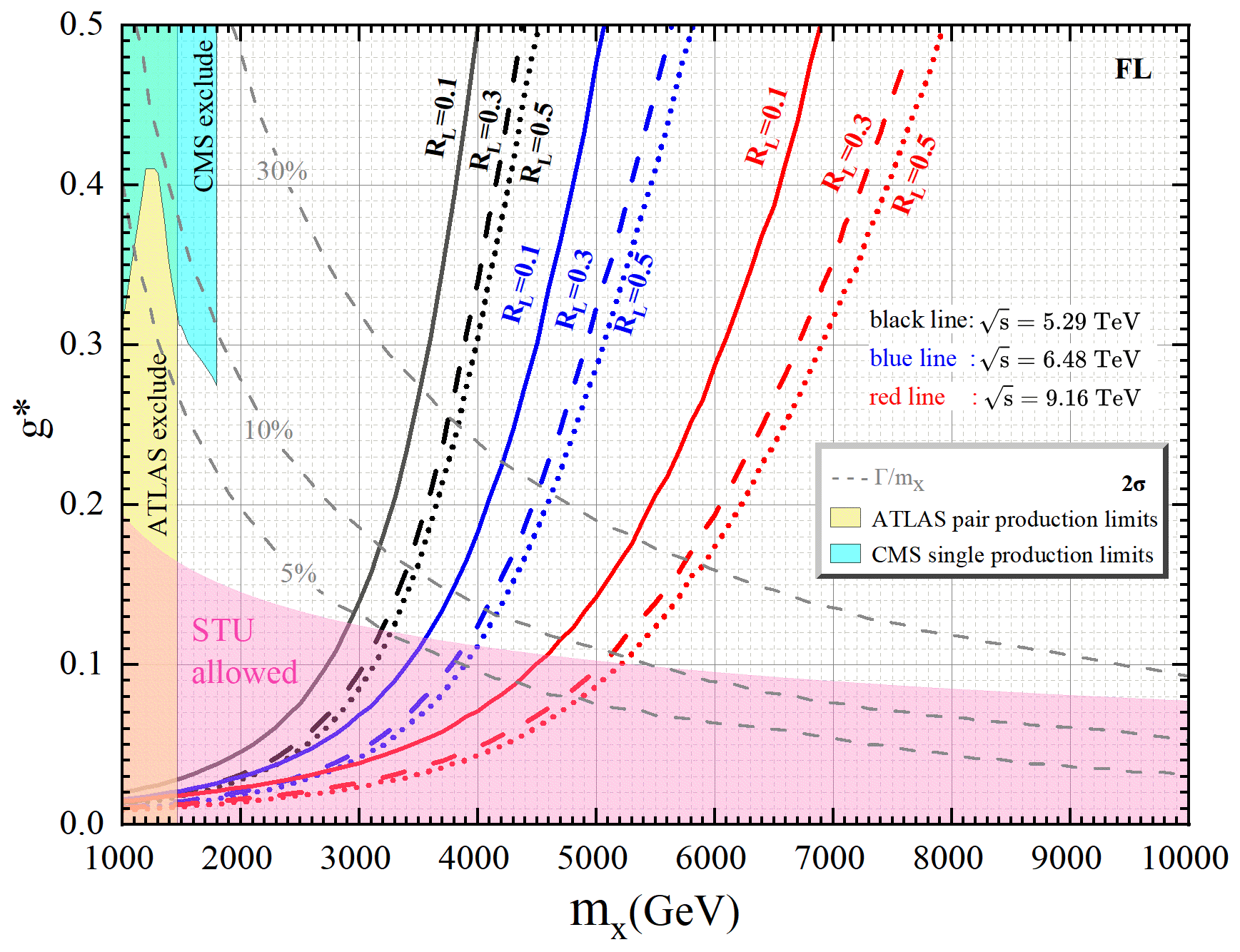}
  \includegraphics[width=0.45\textwidth, height=5.7cm]{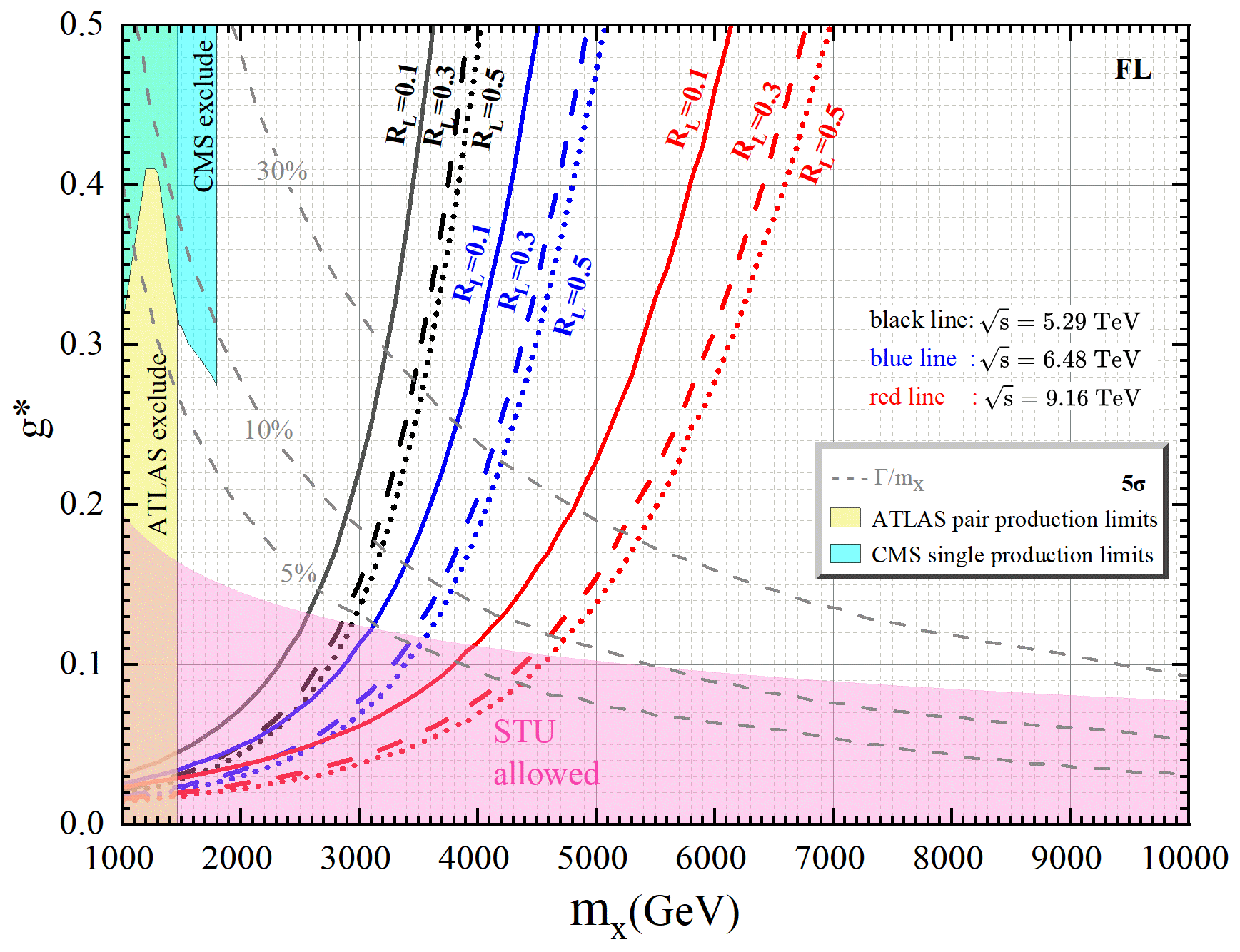}
  \caption{Expected $2\sigma$ exclusion limits and $5\sigma$ discovery reaches in the $g^*{-}m_X$ plane for the FL mode at $\sqrt{s}=5.29$, $6.48$ and $9.16$ TeV for different values of $R_L$.
  The black, blue and red curves correspond to the three collider energies,
  while different line styles denote different values of $R_L$.
  Contours of width-to-mass ratio ($\Gamma_X/m_X$)
  are overlaid as dashed lines.
  The shaded regions indicate the existing constraints from EWPOs 
  and current LHC searches.
  }
  \label{fig11}
\end{figure}

\begin{figure}[htbp]
  \centering
  \includegraphics[width=0.45\textwidth, height=5.7cm]{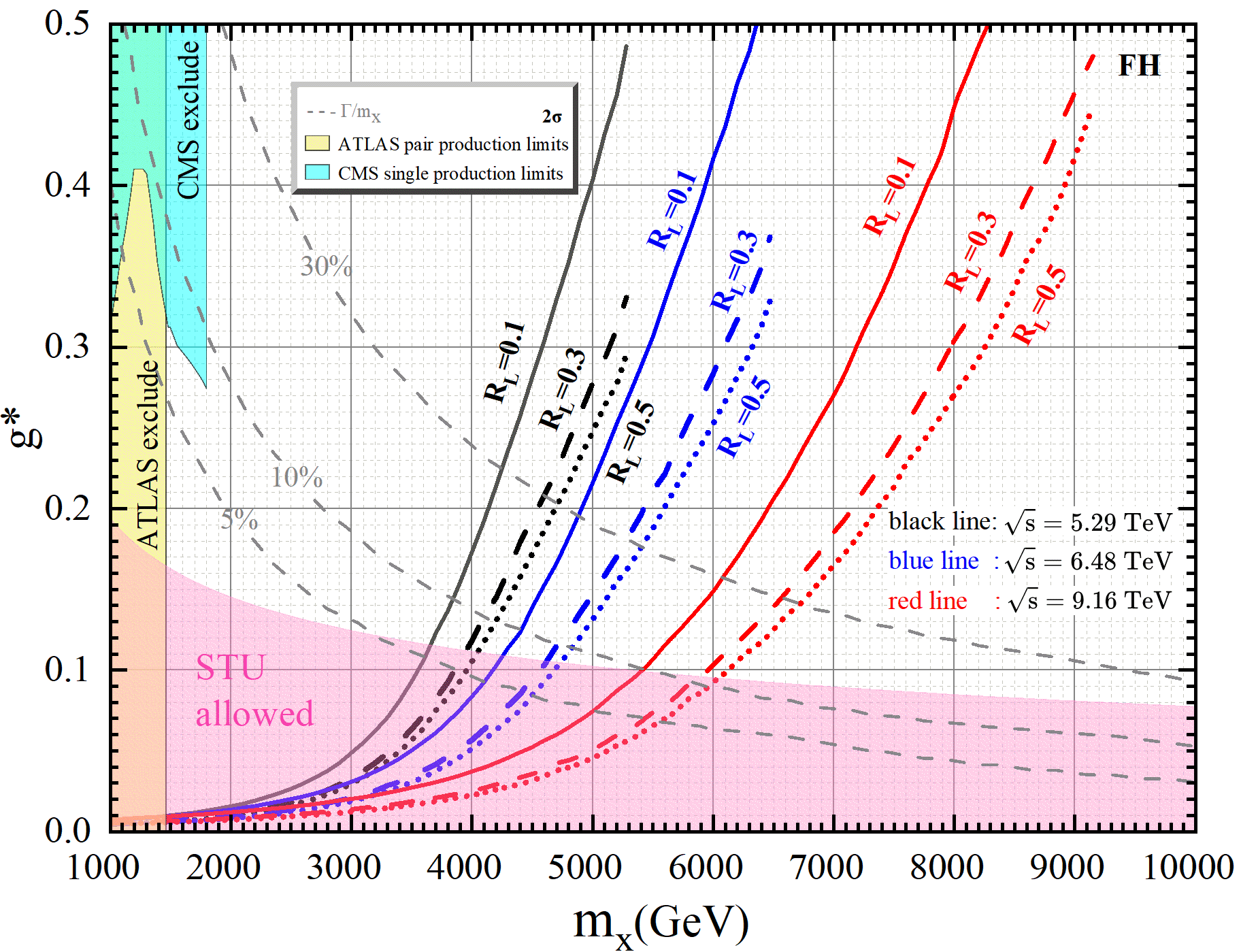}
  \includegraphics[width=0.45\textwidth, height=5.7cm]{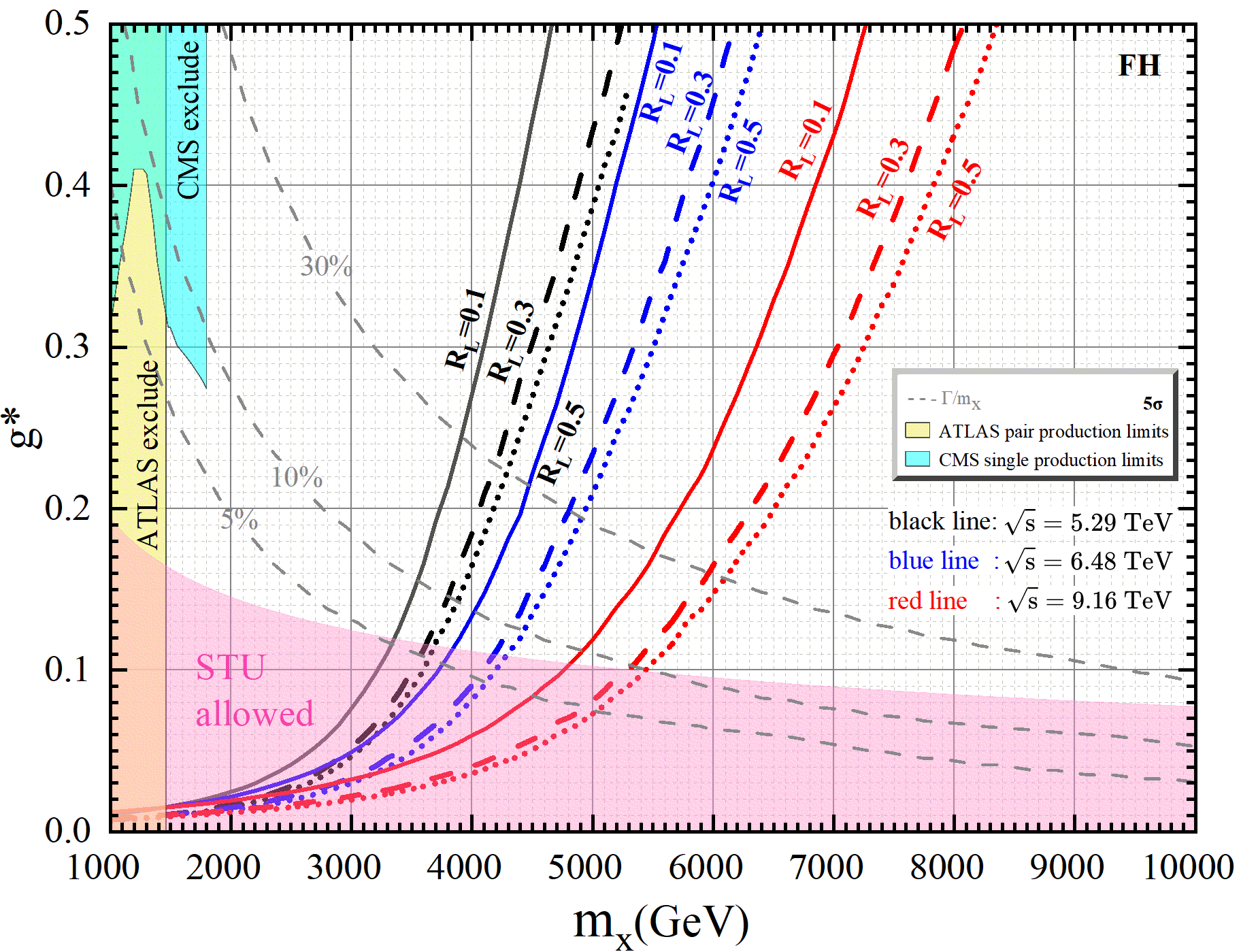}
  \caption{Same as Fig.\ref{fig11}, but for the FH mode.
  }
  \label{fig12}
\end{figure}

\begin{figure}[htbp]
  \centering
  \includegraphics[width=0.45\textwidth, height=5.7cm]{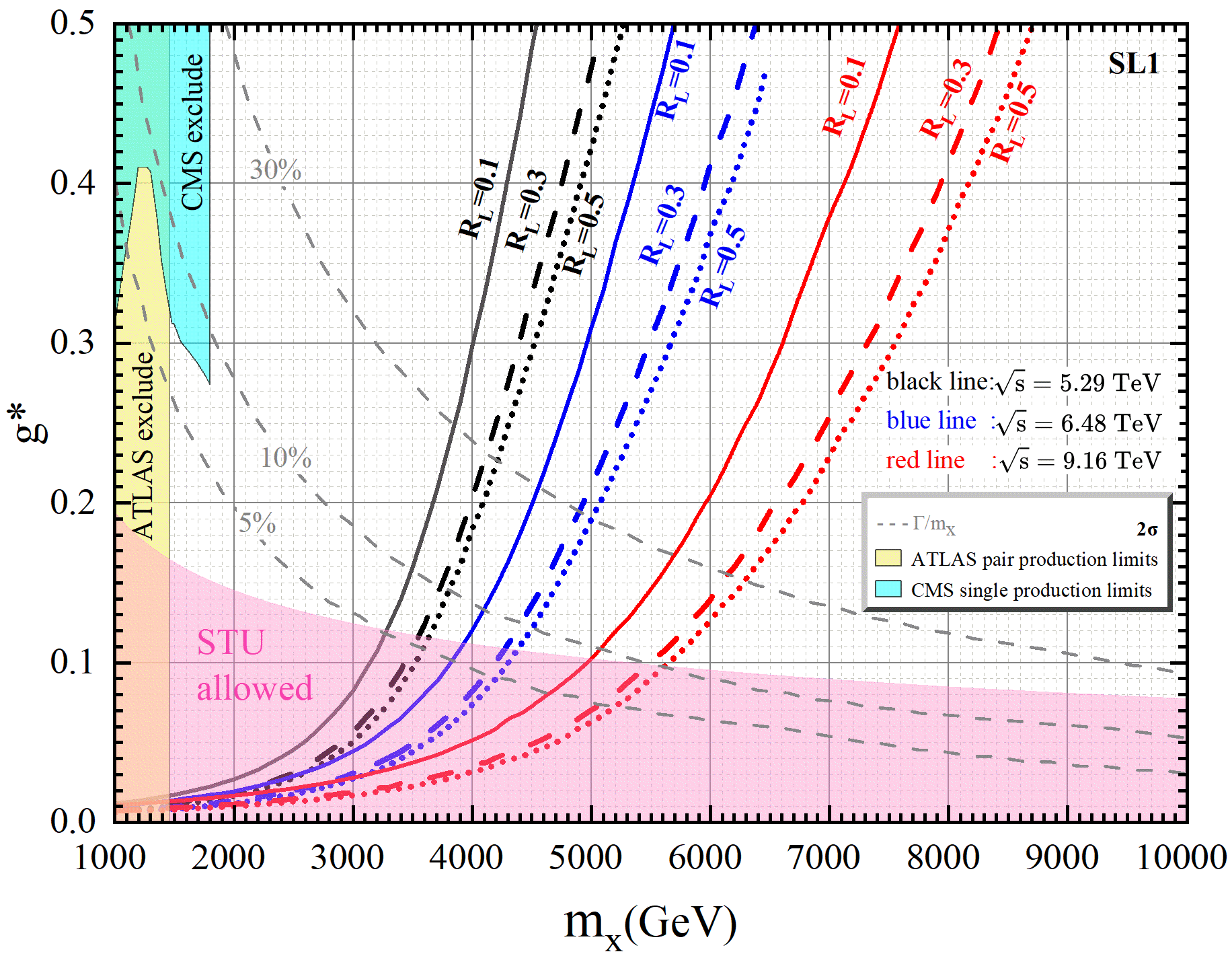}
  \includegraphics[width=0.45\textwidth, height=5.7cm]{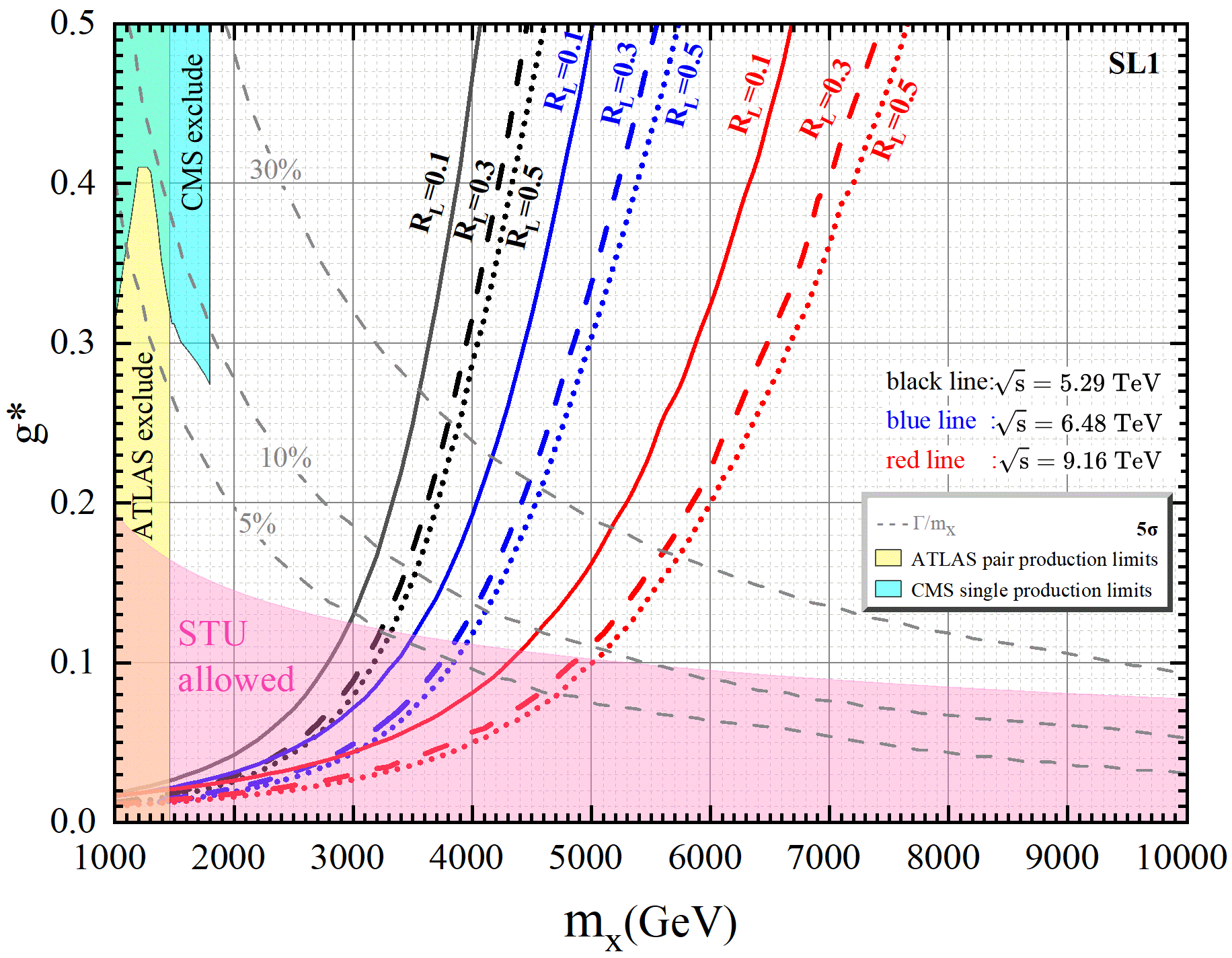}
    \caption{Same as Fig.\ref{fig11}, but for the SL1 mode.
  }
  \label{fig13}
\end{figure}

\begin{figure}[htbp]
  \centering
  \includegraphics[width=0.45\textwidth, height=5.7cm]{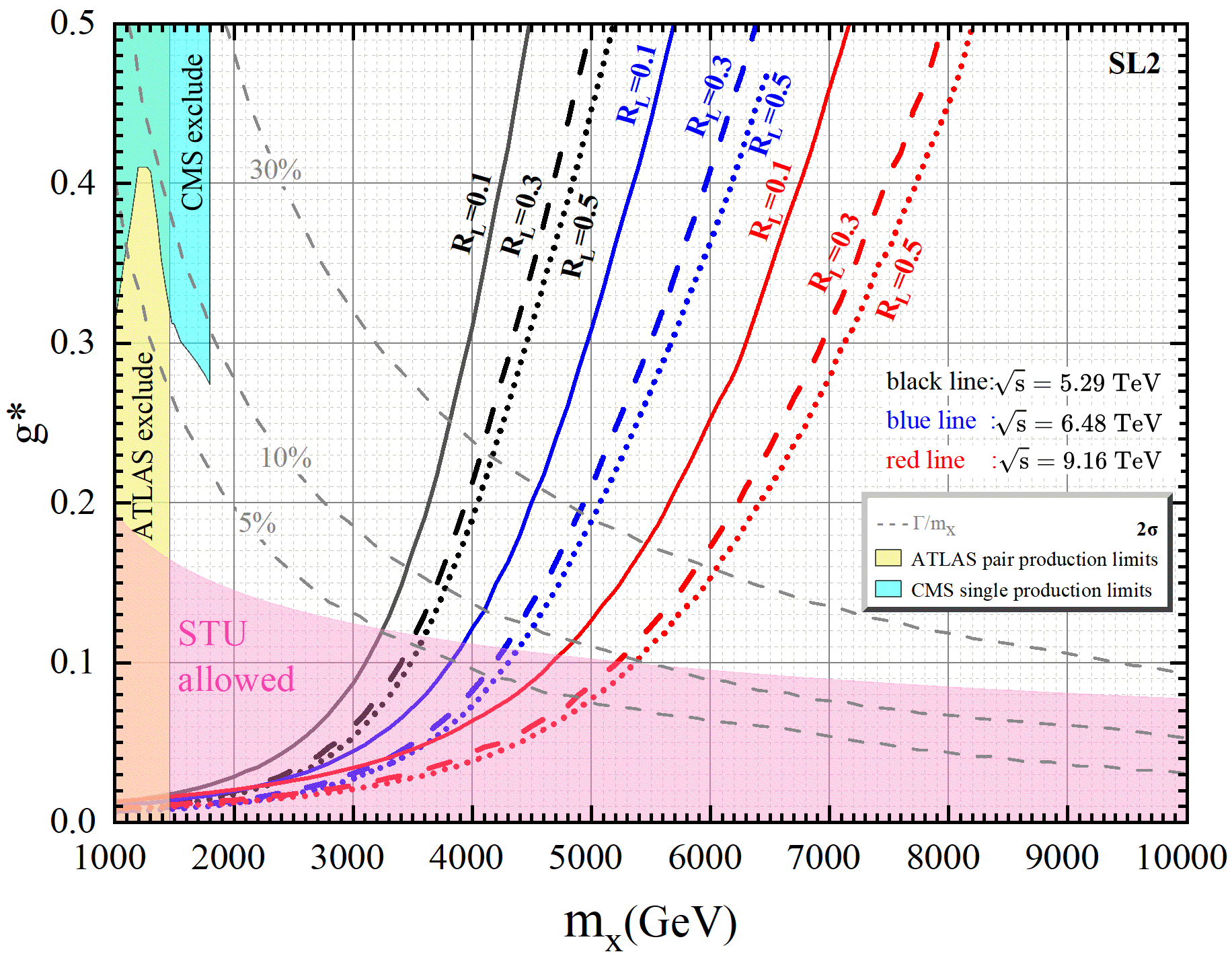}
  \includegraphics[width=0.45\textwidth, height=5.7cm]{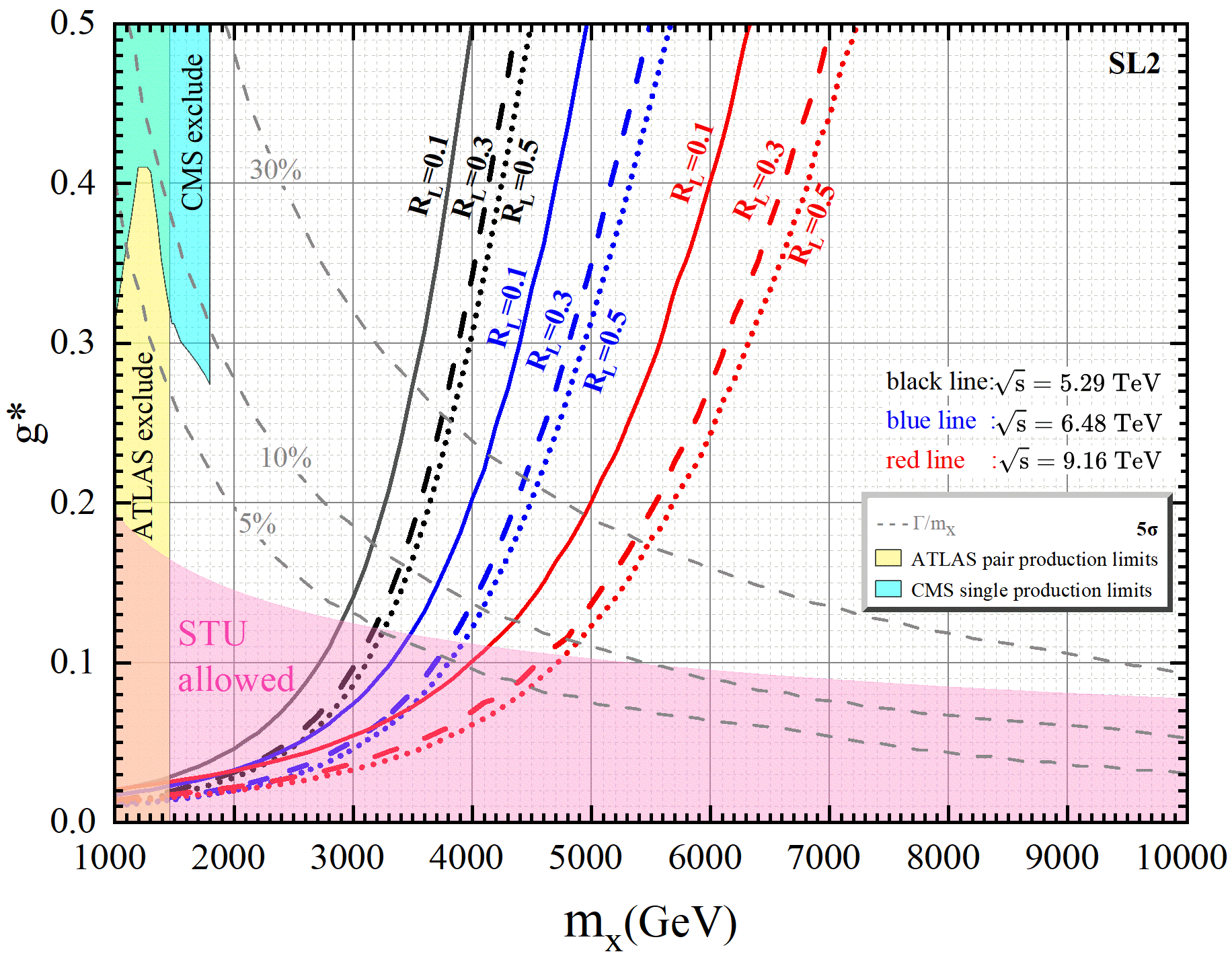}
      \caption{Same as Fig.\ref{fig11}, but for the SL2 mode.
  }
  \label{fig14}
\end{figure}

In Figures~\ref{fig11}-\ref{fig14}, we show the expected sensitivities in the four decay modes.
The solid curves correspond to different collider energies,
while different line styles indicate the dependence on the parameter $R_L$.
The regions above the curves represent the parameter space that can be excluded or discovered.
The pink shaded region denotes the allowed parameter space from EWPOs~\cite{Cao:2022mif},
whereas the yellow and cyan regions indicate the existing limits from ATLAS pair-production~\cite{ATLAS:2022tla} and 
CMS single-production ~\cite{Benbrik:2024fku} searches (we adopt the relation of $g^*=\sqrt{2}\kappa$),
respectively.

The grey dashed contours represent the width-to-mass ratio,
$\Gamma_X/m_X$, which is used to assess the validity of the Narrow-Width
Approximation (NWA), in the spirit of Refs.~\cite{Barducci:2013zaa,Moretti:2016gkr,Moretti:2017qby,Prager:2017hnt,Prager:2017owg,Carvalho:2018jkq}. Since the NWA neglects off-shell effects and the
interference between the resonant VLX contribution and
the corresponding SM background amplitude, its validity requires
a sufficiently small $\Gamma_X/m_X$~\cite{Deandrea:2021vje,Carvalho:2018jkq,Berdine:2007uv,Moretti:2016gkr}.
The contours demonstrate that most of the parameter space considered in this analysis remains
with the NWA regime.

\begin{table}[htbp]
\centering
  \vspace{0.1cm}
  \begin{tabular}
  {cccccccc}
  \toprule[1pt]
  \multirow{2}{*}{$\sqrt{s}$ (TeV)} & \multirow{2}{*}{$R_L$} & \multicolumn{2}{c}{$2\sigma$} && \multicolumn{2}{c}{$5\sigma$} \\ 
  \cline{3-4} \cline{6-7}
  & & $g^*$ & $m_X$ (GeV) && $g^*$ & $m_X$ (GeV) \\ 
  \cline{1-6} \midrule[0.8pt]
  \multirow{3}{*}{5.29} 
  & 0.1 & $[0.019,0.50]$ & $[1000,4000]$ && $[0.031,0.50]$ & $[1000,3620]$ \\
  & 0.3 & $[0.014,0.50]$ & $[1000,4390]$ && $[0.022,0.50]$ & $[1000,3930]$ \\
  & 0.5 & $[0.012,0.50]$ & $[1000,4500]$ && $[0.019,0.50]$ & $[1000,4020]$ \\
  \midrule[0.8pt]
  \multirow{3}{*}{6.48}
  & 0.1 & $[0.015,0.50]$ & $[1000,5090]$ && $[0.026,0.50]$ & $[1000,4500]$ \\
  & 0.3 & $[0.011,0.50]$ & $[1000,5640]$ && $[0.018,0.50]$ & $[1000,4960]$ \\
  & 0.5 & $[0.009,0.50]$ & $[1000,5800]$ && $[0.016,0.50]$ & $[1000,5090]$ \\
  \midrule[0.8pt]
  \multirow{3}{*}{9.16}
  & 0.1 & $[0.015,0.50]$ & $[1000,6900]$ && $[0.023,0.50]$ & $[1000,6150]$ \\
  & 0.3 & $[0.010,0.50]$ & $[1000,7700]$ && $[0.016,0.50]$ & $[1000,6800]$ \\
  & 0.5 & $[0.008,0.50]$ & $[1000,7940]$ && $[0.014,0.50]$ & $[1000,7000]$ \\
  \bottomrule[1pt]
  \end{tabular}
  \caption{
  Expected $2\sigma$ exclusion limits and $5\sigma$ discovery reaches for the
  FL mode at different $\sqrt{s}$ and $R_L$ values.}

\label{table13}
\end{table}

\begin{table}[htbp]
\centering
  \vspace{0.1cm}
  \begin{tabular}
  {cccccccc}
  \toprule[1pt]
  \multirow{2}{*}{$\sqrt{s}$ (TeV)} & \multirow{2}{*}{$R_L$} & \multicolumn{2}{c}{$2\sigma$} && \multicolumn{2}{c}{$5\sigma$} \\ 
  \cline{3-4} \cline{6-7}
  & & $g^*$ & $m_X$ (GeV) && $g^*$ & $m_X$ (GeV) \\ 
  \cline{1-6} \midrule[0.8pt]
  \multirow{3}{*}{5.29} 
  & 0.1 & $[0.008,0.49]$ & $[1000,5290]$ && $[0.014,0.50]$ & $[1000,4670]$ \\
  & 0.3 & $[0.006,0.32]$ & $[1000,5290]$ && $[0.009,0.50]$ & $[1000,5220]$ \\
  & 0.5 & $[0.005,0.29]$ & $[1000,5290]$ && $[0.008,0.46]$ & $[1000,5290]$ \\
  \midrule[0.8pt]
  \multirow{3}{*}{6.48}
  & 0.1 & $[0.009,0.50]$ & $[1000,6370]$ && $[0.014,0.50]$ & $[1000,5520]$ \\
  & 0.3 & $[0.006,0.37]$ & $[1000,6480]$ && $[0.009,0.50]$ & $[1000,6190]$ \\
  & 0.5 & $[0.005,0.32]$ & $[1000,6480]$ && $[0.008,0.50]$ & $[1000,6400]$ \\
  \midrule[0.8pt]
  \multirow{3}{*}{9.16}
  & 0.1 & $[0.009,0.50]$ & $[1000,8300]$ && $[0.016,0.50]$ & $[1000,7270]$ \\
  & 0.3 & $[0.007,0.48]$ & $[1000,9160]$ && $[0.011,0.50]$ & $[1000,8100]$ \\
  & 0.5 & $[0.006,0.45]$ & $[1000,9160]$ && $[0.009,0.50]$ & $[1000,8350]$ \\
  \bottomrule[1pt]
  \end{tabular}
    \caption{
  Expected $2\sigma$ exclusion limits and $5\sigma$ discovery reaches for the
  FH mode at different $\sqrt{s}$ and $R_L$ values.}
\label{table14}
\end{table}

\begin{table}[htbp]
\centering
  \vspace{0.1cm}
  \begin{tabular}
  {cccccccc}
  \toprule[1pt]
  \multirow{2}{*}{$\sqrt{s}$ (TeV)} & \multirow{2}{*}{$R_L$} & \multicolumn{2}{c}{$2\sigma$} && \multicolumn{2}{c}{$5\sigma$} \\ 
  \cline{3-4} \cline{6-7}
  & & $g^*$ & $m_X$ (GeV) && $g^*$ & $m_X$ (GeV) \\ 
  \cline{1-6} \midrule[0.8pt]
  \multirow{3}{*}{5.29} 
  & 0.1 & $[0.013,0.50]$ & $[1000,4530]$ && $[0.021,0.50]$ & $[1000,4080]$ \\
  & 0.3 & $[0.009,0.50]$ & $[1000,5120]$ && $[0.014,0.50]$ & $[1000,4480]$ \\
  & 0.5 & $[0.008,0.50]$ & $[1000,5280]$ && $[0.012,0.50]$ & $[1000,4610]$ \\
  \midrule[0.8pt]
  \multirow{3}{*}{6.48}
  & 0.1 & $[0.011,0.50]$ & $[1000,5700]$ && $[0.017,0.50]$ & $[1000,5010]$ \\
  & 0.3 & $[0.007,0.50]$ & $[1000,6400]$ && $[0.012,0.50]$ & $[1000,5560]$ \\
  & 0.5 & $[0.006,0.47]$ & $[1000,6480]$ && $[0.011,0.50]$ & $[1000,5730]$ \\
  \midrule[0.8pt]
  \multirow{3}{*}{9.16}
  & 0.1 & $[0.012,0.50]$ & $[1000,7600]$ && $[0.019,0.50]$ & $[1000,6690]$ \\
  & 0.3 & $[0.008,0.50]$ & $[1000,8450]$ && $[0.013,0.50]$ & $[1000,7420]$ \\
  & 0.5 & $[0.007,0.50]$ & $[1000,8700]$ && $[0.012,0.50]$ & $[1000,7650]$ \\
  \bottomrule[1pt]
  \end{tabular}
    \caption{
  Expected $2\sigma$ exclusion limits and $5\sigma$ discovery reaches for the
  SL1 mode at different $\sqrt{s}$ and $R_L$ values.}
\label{table15}
\end{table}

\begin{table}[htbp]
\centering
  \vspace{0.1cm}
  \begin{tabular}
  {cccccccc}
  \toprule[1pt]
  \multirow{2}{*}{$\sqrt{s}$ (TeV)} & \multirow{2}{*}{$R_L$} & \multicolumn{2}{c}{$2\sigma$} && \multicolumn{2}{c}{$5\sigma$} \\ 
  \cline{3-4} \cline{6-7}
  & & $g^*$ & $m_X$ (GeV) && $g^*$ & $m_X$ (GeV) \\ 
  \cline{1-6} \midrule[0.8pt]
  \multirow{3}{*}{5.29} 
  & 0.1 & $[0.013,0.50]$ & $[1000,4490]$ && $[0.021,0.50]$ & $[1000,4000]$ \\
  & 0.3 & $[0.009,0.50]$ & $[1000,5000]$ && $[0.014,0.50]$ & $[1000,4390]$ \\
  & 0.5 & $[0.008,0.50]$ & $[1000,5200]$ && $[0.012,0.50]$ & $[1000,4500]$ \\
  \midrule[0.8pt]
  \multirow{3}{*}{6.48}
  & 0.1 & $[0.011,0.50]$ & $[1000,5700]$ && $[0.017,0.50]$ & $[1000,4970]$ \\
  & 0.3 & $[0.007,0.50]$ & $[1000,6400]$ && $[0.012,0.50]$ & $[1000,5500]$ \\
  & 0.5 & $[0.006,0.46]$ & $[1000,6480]$ && $[0.011,0.50]$ & $[1000,5720]$ \\
  \midrule[0.8pt]
  \multirow{3}{*}{9.16}
  & 0.1 & $[0.012,0.50]$ & $[1000,7180]$ && $[0.019,0.50]$ & $[1000,6310]$ \\
  & 0.3 & $[0.008,0.50]$ & $[1000,7950]$ && $[0.013,0.50]$ & $[1000,7010]$ \\
  & 0.5 & $[0.007,0.50]$ & $[1000,8210]$ && $[0.012,0.50]$ & $[1000,7230]$ \\
  \bottomrule[1pt]
  \end{tabular}
    \caption{
  Expected $2\sigma$ exclusion limits and $5\sigma$ discovery reaches for the
  SL2 mode at different $\sqrt{s}$ and $R_L$ values.}
\label{table16}
\end{table}

As summarized in Tables~\ref{table13}-\ref{table16}, the sensitivity improves
with increasing CoM energy and with larger $R_L$.
For example, for the FL mode, the $2\sigma$ exclusion reach extends from
$m_X\simeq 4.0$ TeV at $\sqrt{s}=5.29$ TeV to about $6.9$ TeV at 
$\sqrt{s}=9.16$ TeV for $R_L=0.1$. The corresponding $5\sigma$ discovery
reach increases from about $3.6$ TeV to $6.1$ TeV.

Among the four modes, the FH mode provides the strongest sensitivity.
At $\sqrt{s}=9.16$ TeV and $R_L=0.1$, it probes VLX masses up to 
approximately $8.3$ TeV at the $2\sigma$ level
and $7.3$ TeV at the $5\sigma$ level, with
couplings down to $g^* \sim 0.009$ and $0.016$, respectively.
This enhanced sensitivity mainly originates from the larger
BR of fully hadronic $W$-boson decays and the efficient
reconstruction of boosted hadronic objects.

The SL modes provide intermediate
sensitivities, reaching
$m_X \simeq 7.6$ TeV (SL1) and $7.2$ TeV (SL2) at $\sqrt{s}=9.16$ TeV and $R_L=0.1$
for $2\sigma$ exclusion.
The FL mode has the cleanest signature but suffers
from a smaller BR, resulting in the weakest mass
reach among the four modes.

\begin{figure}[htbp]
  \centering
  \includegraphics[width=0.45\textwidth, height=5.3cm]{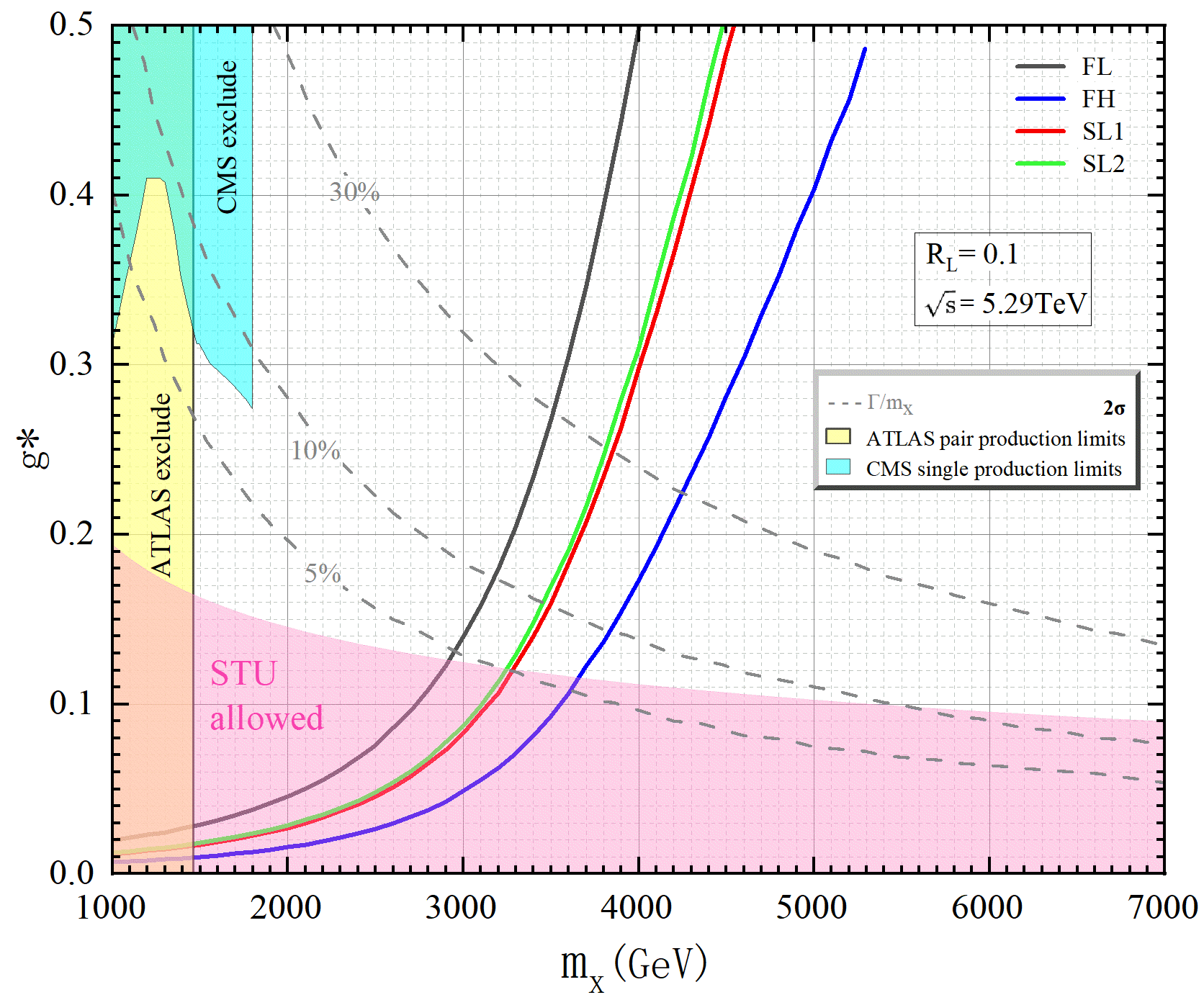}
  \includegraphics[width=0.45\textwidth, height=5.3cm]{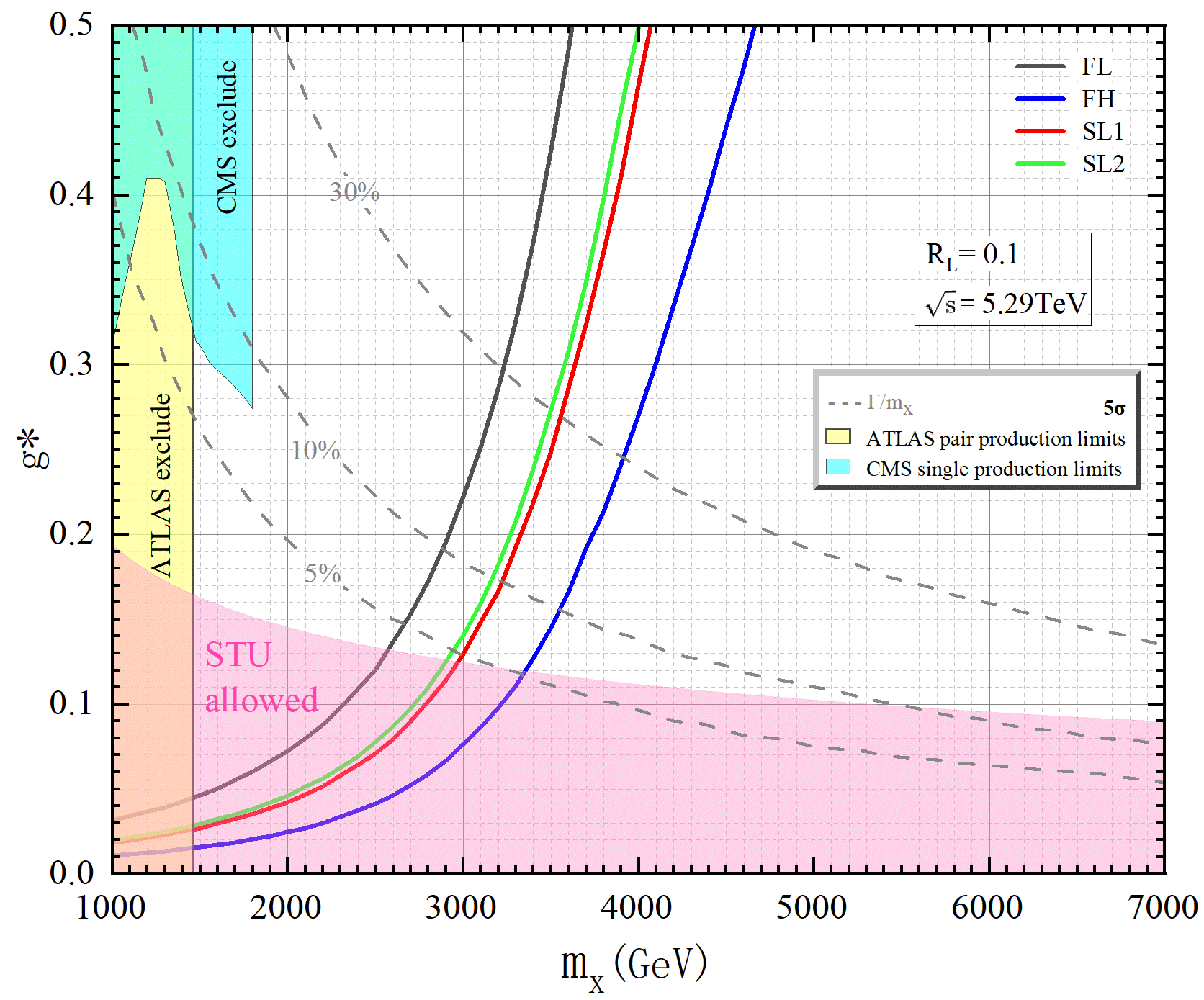} \\
  \includegraphics[width=0.45\textwidth, height=5.3cm]{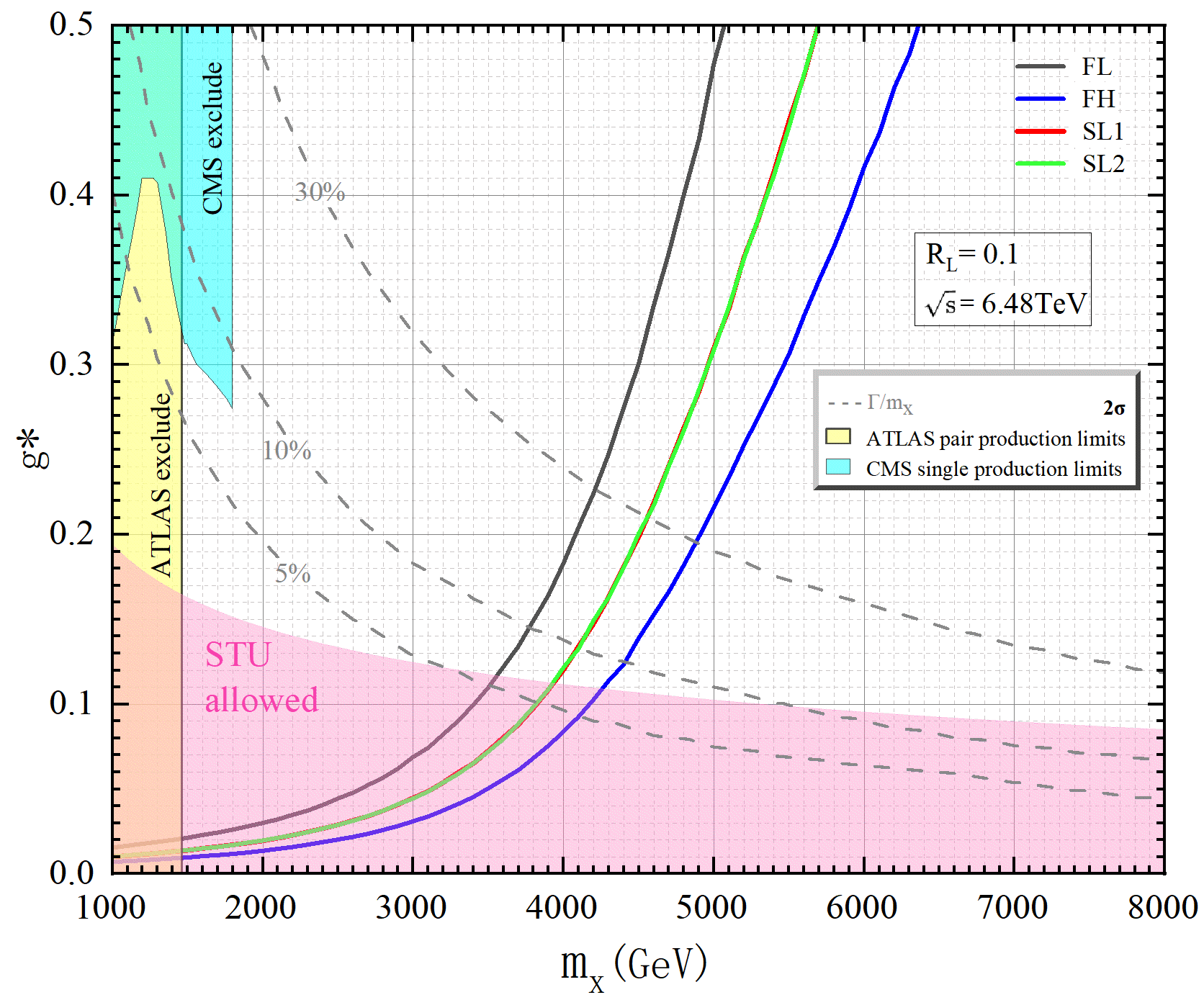}
  \includegraphics[width=0.45\textwidth, height=5.3cm]{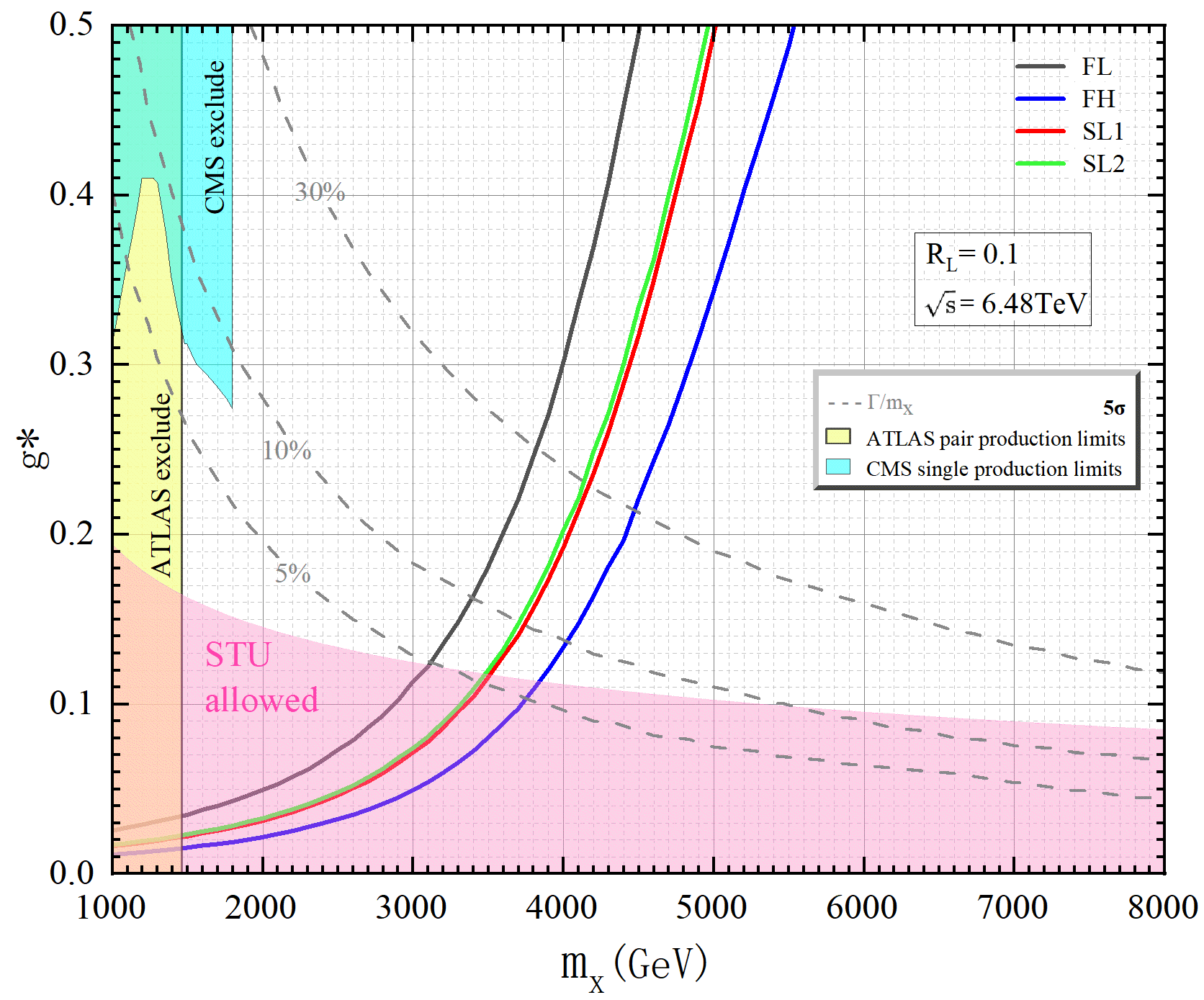} \\
  \includegraphics[width=0.45\textwidth, height=5.3cm]{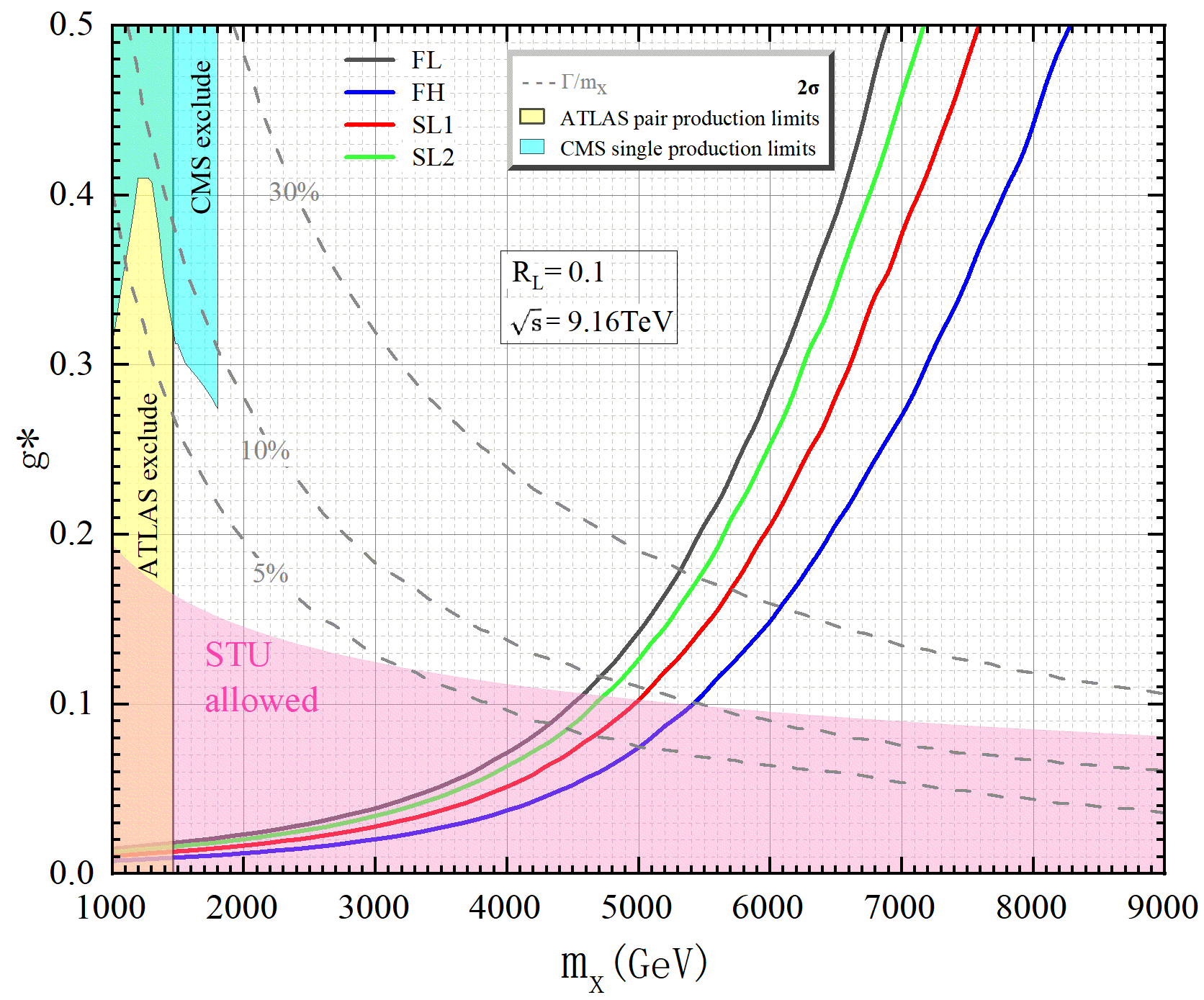}
  \includegraphics[width=0.45\textwidth, height=5.3cm]{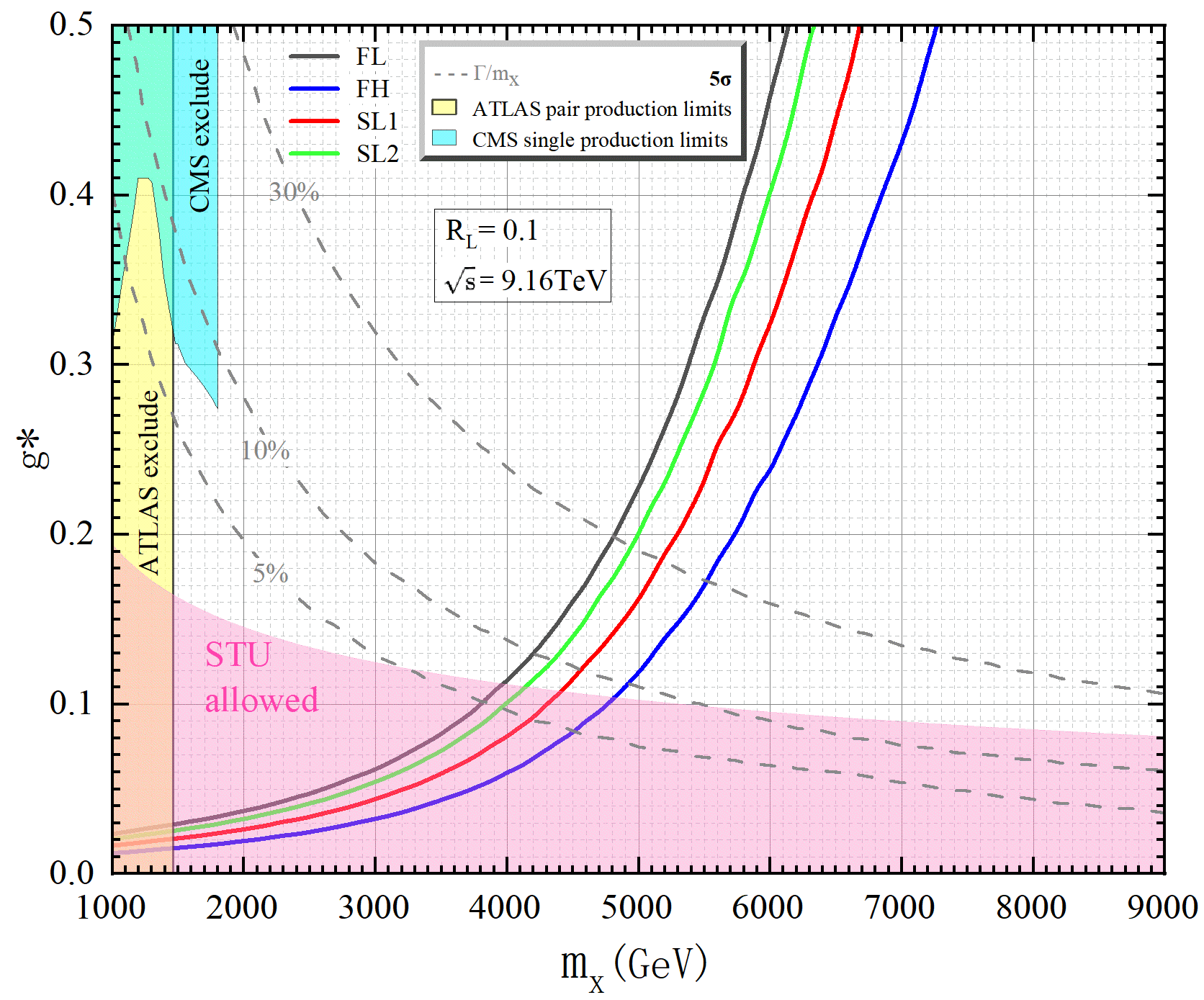}
  \caption{Comparison of the expected $2\sigma$ exclusion limits
  and $5\sigma$ discovery reaches for the four modes in the $g^*{-}m_X$ plane at
$\sqrt{s} = 5.29$, $6.48$ and $9.16$~TeV with $R_L = 0.1$.
The solid lines correspond to the FL, FH, SL1, and SL2 modes.
Contours of width-to-mass ratio ($\Gamma_X/m_X$)
  are overlaid as dashed lines.
  The shaded regions indicate the existing constraints from EWPOs and current LHC searches.
}
  \label{fig15}
\end{figure}

The mode comparison is presented in Figure~\ref{fig15}. 
For a fixed $R_L=0.1$, the exclusion and 
discovery capabilities of the four modes at different collider energies are summarized.
The results clearly demonstrate that the future $\mu p$
collider can probe VLX masses well beyond the current LHC and future High-luminosity LHC (HL-LHC) reaches,
extending the sensitivity to the multi-TeV region.

\begin{table}[htbp]
\centering
  \vspace{0.1cm}
  \begin{tabular}
  {cccccccc}
  \toprule[1pt]
  \multirow{2}{*}{Mode} & \multirow{2}{*}{$\sqrt{s}$ (TeV)} & \multicolumn{2}{c}{$2\sigma$} && \multicolumn{2}{c}{$5\sigma$} \\ 
  \cline{3-4} \cline{6-7}
  & & $g^*$ & $m_X$ (GeV) && $g^*$ & $m_X$ (GeV) \\ 
  \cline{1-6} \midrule[0.8pt]
  \multirow{3}{*}{FL} 
  & 5.29 & $[0.019, 0.50]$ & $[1000, 4000]$ && $[0.031, 0.50]$ & $[1000, 3620]$ \\
  & 6.48 & $[0.015, 0.50]$ & $[1000, 5090]$ && $[0.026, 0.50]$ & $[1000, 4500]$ \\
  & 9.16 & $[0.015, 0.50]$ & $[1000, 6900]$ && $[0.023, 0.50]$ & $[1000, 6150]$ \\
  \midrule[0.8pt]
  \multirow{3}{*}{FH} 
  & 5.29 & $[0.008, 0.49]$ & $[1000, 5290]$ && $[0.014, 0.50]$ & $[1000, 4670]$ \\
  & 6.48 & $[0.009, 0.50]$ & $[1000, 6370]$ && $[0.014, 0.50]$ & $[1000, 5520]$ \\
  & 9.16 & $[0.009, 0.50]$ & $[1000, 8300]$ && $[0.016, 0.50]$ & $[1000, 7270]$ \\
  \midrule[0.8pt]
  \multirow{3}{*}{SL1} 
  & 5.29 & $[0.013, 0.50]$ & $[1000, 4530]$ && $[0.021, 0.50]$ & $[1000, 4080]$ \\
  & 6.48 & $[0.011, 0.50]$ & $[1000, 5700]$ && $[0.017, 0.50]$ & $[1000, 5010]$ \\
  & 9.16 & $[0.012, 0.50]$ & $[1000, 7600]$ && $[0.019, 0.50]$ & $[1000, 6690]$ \\
  \midrule[0.8pt]
  \multirow{3}{*}{SL2} 
  & 5.29 & $[0.013, 0.50]$ & $[1000, 4490]$ && $[0.021, 0.50]$ & $[1000, 4000]$ \\
  & 6.48 & $[0.011, 0.50]$ & $[1000, 5700]$ && $[0.017, 0.50]$ & $[1000, 4970]$ \\
  & 9.16 & $[0.012, 0.50]$ & $[1000, 7180]$ && $[0.019, 0.50]$ & $[1000, 6310]$ \\
  \bottomrule[1pt]
  \end{tabular}
  \caption{
  Comparison of the expected $2\sigma$ exclusion limits and $5\sigma$ discovery reaches among the FL, FH, SL1, and SL2
  modes at fixed $R_L=0.1$.}

\label{table17}
\end{table}

A direct comparison of the sensitivities among the four modes
for a fixed value of $R_L=0.1$ is provided in Table~\ref{table17}.
Several interesting features can be observed. First, the sensitivity
gain from increasing the collider energy is more significant in
the high-mass region, where the enhanced partonic CoM
energy compensates for the rapidly decreasing quark parton luminosity in the PDFs.
This behavior allows the $\mu p$ collider to extend the VLX mass coverage
into the multi-TeV regime.
Second, the relative performance of different decay mode is mainly
determined by the interplay between BRs, signal efficiencies,
and background rejection capabilities. 
Although the FL mode benefits from a cleaner leptonic signature,
its smaller BR limits the overall event yield.
In contrast, the FH mode benefits from the dominant hadronic decay modes
of the $W$ boson and efficient boosted-object reconstruction,
leading to the best sensitivity. The SL modes provide a
complementary search strategy by combining the advantages of 
leptonic identification and sizable hadronic
BRs.
Their sensitivities lies between the FH and FL modes,
demonstrating the importance of exploring multiple decay topologies
at future $\mu p$ colliders.

\section{Summary}

We have presented a comprehensive phenomenological study of the single VLX production
at future $\mu p$ colliders 
through the process
$\mu^+ p \to \bar{\nu}_\mu X \to \bar{\nu}_\mu t W^+$.
Working within a simplified effective model,
we have investigated four complementary decay channels,
namely the FL, FH, SL1, and SL2 modes,
at CoM energies of
$\sqrt{s}=5.29$, $6.48$, and $9.16\ \mathrm{TeV}$.
Dedicated event selections were optimized
for each channel using
the characteristic kinematic properties
of the signal, including boosted-object
reconstruction with fat-jet techniques for 
hadronic decays.

The projected sensitivities were evaluated
using the Asimov significance and compared with
the existing constraints from
EWPOs and current 
LHC searches.
The discovery potential improves significantly
with increasing collider energy and 
larger values of the generation-mixing
parameter $R_L$.
Among the four modes, the FH mode
provides the best overall sensitivity,
benefiting from the large
hadronic BRs and the
efficient reconstruction of boosted hadronic objects.
The SL1 and SL2 modes yield comparable intermediate 
sensitivities,
whereas the FL mode yield gives the
weakest reach because of its smaller BR,
despite its clean experimental signature. As an example, for the benchmark scenario with $R_L =0.1$,
the FH mode achieves a projected $2\sigma$
exclusion reach of 
$m_X\simeq 8.3\ \mathrm{TeV}$ with $g^* = 0.009 $ 
and a $5\sigma$ discovery reach
of $m_X \simeq 7.3 \ \mathrm{TeV}$ with $g^* = 0.016 $
at $\sqrt{s}=9.16\ \mathrm{TeV}$.

Overall, our results demonstrate that 
future $\mu p$ colliders can substantially
extend the search for exotic VLQs 
beyond the current (HL-)LHC limits,
providing excellent sensitivity to the high-mass
and weak-coupling regions of parameter space.

\begin{acknowledgments}
SM is supported in part through the NExT Institute and STFC Consolidated Grant ST/X000583/1. 
The work of all others is supported by the Henan Provincial International Science, Technology Cooperation Incubation Project 262102520027 and the High Performance Computing Platform of Henan Normal University.
\end{acknowledgments}

\bibliography{ref}  

\appendix
\section{Additional Material}   
\label{app:extra}

\begin{table}[htbp]
  \centering
  \vspace{0.1cm}
  \begin{tabular}
  {cccccccc}
  \toprule[1pt]
  \multirow{2}{*}{Cuts} & \multicolumn{3}{c}{Signal(fb)} && \multicolumn{3}{c}{Backgrounds(fb)} \\ 
  \cline{2-4} \cline{6-8}
  & $X_{1500}$ & $X_{2000}$ & $X_{2500}$ && $\mu Wj$ & $\bar{\nu}WWj$ & $\mu ZWj$ \\ 
  \cline{1-8} \midrule[0.8pt]
  Basic cuts & 16.39 & 8.34 & 3.89 && 3911 & 14.71 & 2.49 \\
  $N_l = 2$ & 9.12 & 3.64 & 1.35 && 2573 & 11.60 & 1.91 \\
  $N_b = 1$ & 6.03 & 2.40 & 0.87 && 72.88 & 0.27 & 0.05 \\
  $-3.0 < \eta_{l_1} < 0.6$ & 4.60 & 1.78 & 0.63 && 3.90 & 0.12 & 0.0034 \\
  $-4.0 < \eta_{l_2} < 0.5$ & 3.90 & 1.54 & 0.55 && 0.56 & 0.06 & 0.0004 \\
  $E_T > 500\ \mathrm{GeV}$ & 2.86 & 1.34 & 0.51 && 0.03 & 0.01 & 0.0002 \\
  $M_{b l_1 l_2} > 500\ \mathrm{GeV}$ & 2.57 & 1.24 & 0.47 && 0.02 & 0.0036 & 0.00012 \\
   \cline{1-8} \midrule[0.8pt]
  Efficiency & 16\% & 15\% & 12\% &&0.0005\% & 0.025\% & 0.005\% \\
  \bottomrule[1pt]
  \end{tabular}
  \caption{Cut flow for the FL signal ($m_X = 1500$, $2000$ and $2500\ \mathrm{GeV}$) and the backgrounds at $\sqrt{s}=6.48\ \mathrm{TeV}$.}
  \label{table2}
\end{table}

\begin{table}[htbp]
  \centering
  \vspace{0.1cm}
  \begin{tabular}
  {cccccccc}
  \toprule[1pt]
  \multirow{2}{*}{Cuts} & \multicolumn{3}{c}{Signal(fb)} && \multicolumn{3}{c}{Backgrounds(fb)} \\ 
  \cline{2-4} \cline{6-8}
  & $X_{1500}$ & $X_{2000}$ & $X_{2500}$ && $\mu Wj$ & $\bar{\nu}WWj$ & $\mu ZWj$ \\ 
  \cline{1-8} \midrule[0.8pt]
  Basic cuts & 28.39 & 18.13 & 11.15 && 4829 & 26.20 & 4.03 \\
  $N_l = 2$ & 15.62 & 7.78 & 3.74 && 2473 & 20.39 & 2.72 \\
  $N_b = 1$ & 10.39 & 5.12 & 2.42 && 79.41 & 0.54 & 0.08 \\
  $-3.0 < \eta_{l_1} < 1.0$ & 7.77 & 3.77 & 1.70 && 3.01 & 0.24 & 0.0043 \\
  $-4.0 < \eta_{l_2} < 1.0$ & 6.71 & 3.28 & 1.50 && 1.0 & 0.12 & 0.00088 \\
  $E_T > 500\ \mathrm{GeV}$ & 5.74 & 3.10 & 1.44 && 0.03 & 0.04 & 0.00035 \\
  $M_{b l_1 l_2} > 500\ \mathrm{GeV}$ & 4.93 & 2.78 & 1.34 && 0.02 & 0.01 & 0.00017 \\
   \cline{1-8} \midrule[0.8pt]
  Efficiency & 17\% & 15\% & 12\% &&0.0005\% & 0.051\% & 0.004\% \\
  \bottomrule[1pt]
  \end{tabular}
    \caption{Cut flow for the FL signal ($m_X = 1500$, $2000$ and $2500\ \mathrm{GeV}$) and the backgrounds at $\sqrt{s}=9.16\ \mathrm{TeV}$.}
  \label{table3}
\end{table}

\begin{table}[htbp]
\centering
  \vspace{0.1cm}
  \begin{tabular}
  {ccccccccc}
  \toprule[1pt]
  \multirow{2}{*}{Cuts} & \multicolumn{3}{c}{Signal(fb)} && \multicolumn{4}{c}{Backgrounds(fb)} \\ 
  \cline{2-4} \cline{6-9}
  & $X_{1500}$ & $X_{2000}$ & $X_{2500}$ && $\bar{\nu}tZ$ & $\bar{\nu} WWj$ & $\bar{\nu} Zjj$ & $\bar{\nu}Wjj$ \\
  \cline{1-9} \midrule[0.8pt]
  Basic cuts & 147.47 & 75.10 & 34.94 && 146 & 132 & 1455 & 3932 \\
  $N_j\ge2$ & 127.83 & 62.22 & 28.14 && 145.27 & 132.33 & 1453.48 & 3924.84 \\
  $N_b=1$ & 71.51 & 35.24 & 16.17 && 74.66 & 8.44 & 84.20 & 226.60 \\
  $-2.4<\eta_{b}<0.4$ & 62.70 & 31.74 & 14.79 && 47.23 & 4.60 & 44.55 & 118.50 \\
  $H_T>900\ \mathrm{GeV}$ & 50.33 & 29.05 & 14.10 && 1.87 & 0.42 & 1.90 & 5.17 \\
  $70\ \mathrm{GeV}<M_{j_1}<90\ \mathrm{GeV}$ & 34.72 & 19.65 & 9.40 && 0.29 & 0.16 & 0.34 & 1.41 \\
  $p_{T}^{j_1}>600\ \mathrm{GeV}$ & 21.75 & 16.70 & 8.59 && 0.05 & 0.04 & 0.11 & 0.40 \\
  $p_{T}^{b}>450\ \mathrm{GeV}$ & 18.45 & 15.29 & 8.06 && 0.03 & 0.01 & 0.03 & 0.08 \\
  \cline{1-9} \midrule[0.8pt]
  Efficiency & 12\% & 20\% & 23\% && 0.02\% & 0.01\% & 0.002\% & 0.002\% \\
\bottomrule[1pt]
\end{tabular}
  \caption{Cut flow for the FH signal ($m_X = 1500$, $2000$ and $2500\ \mathrm{GeV}$) and the backgrounds at $\sqrt{s}=6.48\ \mathrm{TeV}$.}
\label{table5}
\end{table}

\begin{table}[htbp]
\centering
  \vspace{0.1cm}
  \begin{tabular}
  {ccccccccc}
  \toprule[1pt]
  \multirow{2}{*}{Cuts} & \multicolumn{3}{c}{Signal(fb)} && \multicolumn{4}{c}{Backgrounds(fb)} \\ 
  \cline{2-4} \cline{6-9}
  & $X_{1500}$ & $X_{2000}$ & $X_{2500}$ && $\bar{\nu}tZ$ & $\bar{\nu} WWj$ & $\bar{\nu} Zjj$ & $\bar{\nu}Wjj$ \\
  \cline{1-9} \midrule[0.8pt]
  Basic cuts & 255.47 & 163.12 & 100.31 && 281.49 & 236.04 & 2534 & 6531 \\
  $N_j\ge2$ & 226.22 & 139.56 & 84.01 && 279.88 & 235.97 & 2530.83 & 6520.51 \\
  $N_b=1$ & 128.95 & 80.96 & 49.54 && 141.07 & 15.14 & 151 & 384.10 \\
  $-2.2<\eta_{b}<0.8$ & 110.66 & 72.11 & 44.97 && 85.95 & 8.17 & 80.78 & 199 \\
  $H_T>800\ \mathrm{GeV}$ & 95.92 & 67.90 & 43.63 && 7.38 & 1.48 & 7.71 & 18.63 \\
  $70\ \mathrm{GeV}<M_{j_1}<90\ \mathrm{GeV}$ & 64.29 & 45.03 & 28.49 && 1.25 & 0.55 & 1.33 & 4.66 \\
  $p_{T}^{j_1}>600\ \mathrm{GeV}$ & 37.94 & 37.38 & 25.83 && 0.17 & 0.14 & 0.29 & 1.00 \\
  $p_{T}^{b}>500\ \mathrm{GeV}$ & 30.10 & 33.58 & 23.99 && 0.10 & 0.04 & 0.07 & 0.33 \\
  \cline{1-9} \midrule[0.8pt]
  Efficiency & 12\% & 21\% & 24\% && 0.03\% & 0.015\% & 0.0025\% & 0.004\% \\
\bottomrule[1pt]
\end{tabular}
  \caption{Cut flow for the FH signal ($m_X = 1500$, $2000$ and $2500\ \mathrm{GeV}$) and the backgrounds at $\sqrt{s}=9.16\ \mathrm{TeV}$.}
\label{table6}
\end{table}

\begin{table}[htbp]
\centering
  \vspace{0.1cm}
  \begin{tabular}
  {cccccccc}
  \toprule[1pt]
  \multirow{2}{*}{Cuts} & \multicolumn{3}{c}{Signal(fb)} & \multicolumn{4}{c}{Backgrounds(fb)} \\ 
  \cline{2-4} \cline{5-8}
  & $X_{1500}$ & $X_{2000}$ & $X_{2500}$ & $\mu Wj$ & $\bar{\nu}WWj$ & $\bar{\nu}Zt$ & $\mu ZWj$ \\ 
  \cline{1-8} \midrule[0.8pt]
  Basic cuts & 49.10 & 25.02 & 11.60 & 11743 & 88.10 & 48.70 & 7.46 \\
  $N_b=1$ & 30.60 & 15.10 & 6.94 & 701 & 5.06 & 18.03 & 0.43 \\
  $N_l=1$ & 26.65 & 13.10 & 6.02 & 494.90 & 4.10 & 12.58 & 0.34 \\
  $-2.6<\eta_b<0$ & 21.87 & 11.16 & 5.30 & 176.30 & 2.03 & 7.33 & 0.13 \\
  $-1.4<\eta_l<0.4$ & 13.23 & 7.09 & 3.55 & 3.82 & 0.69 & 3.78 & 0.004 \\
  $p_{T}^{b}>400\ \mathrm{GeV}$ & 5.72 & 4.72 & 2.71 & 0.08 & 0.02 & 0.03 & 0.0002 \\
  $\not\!\!\mathrm{H}_\mathrm{T}>500\ \mathrm{GeV}$ & 5.36 & 4.64 & 2.69 & 0.04 & 0.01 & 0.005 & 0.0001 \\
  \cline{1-8} \midrule[0.8pt]
  Efficiency & 11\% & 19\% & 23\% & 0.0003\% & 0.01\% & 0.01\% & 0.001\% \\
\bottomrule[1pt]
\end{tabular}
  \caption{Cut flow for the SL1 signal ($m_X = 1500$, $2000$ and $2500\ \mathrm{GeV}$) and the backgrounds at $\sqrt{s}=6.48\ \mathrm{TeV}$.}
\label{table8}
\end{table}

\begin{table}[htbp]
\centering
  \vspace{0.1cm}
  \begin{tabular}
  {cccccccc}
  \toprule[1pt]
  \multirow{2}{*}{Cuts} & \multicolumn{3}{c}{Signal(fb)} & \multicolumn{4}{c}{Backgrounds(fb)} \\ 
  \cline{2-4} \cline{5-8}
  & $X_{1500}$ & $X_{2000}$ & $X_{2500}$ & $\mu Wj$ & $\bar{\nu}WWj$ & $\bar{\nu}Zt$ & $\mu ZWj$ \\ 
  \cline{1-8} \midrule[0.8pt]
  Basic cuts & 85.12 & 54.33 & 33.44 & 14483 & 156.60 & 93.87 & 12.10 \\
  $N_b=1$ & 52.95 & 32.80 & 19.85 & 852.20 & 9.48 & 33.44 & 0.69 \\
  $N_l=1$ & 45.90 & 28.39 & 17.13 & 498 & 7.60 & 22.98 & 0.50 \\
  $-2.6<\eta_b<1.0$ & 42.50 & 26.90 & 16.43 & 249.90 & 4.90 & 16.80 & 0.27 \\
  $-1.4<\eta_l<0.6$ & 23.50 & 15.43 & 9.91 & 6.83 & 1.50 & 7.92 & 0.007 \\
  $p_{T}^{b}>400\ \mathrm{GeV}$ & 10.08 & 10.23 & 7.55 & 0.12 & 0.06 & 0.11 & 0.0004 \\
  $\not\!\!\mathrm{H}_\mathrm{T}>500\ \mathrm{GeV}$ & 9.38 & 10.05 & 7.49 & 0.07 & 0.03 & 0.02 & 0.0002 \\
  \cline{1-8} \midrule[0.8pt]
  Efficiency & 11\% & 19\% & 22\% & 0.0005\% & 0.02\% & 0.02\% & 0.002\% \\
\bottomrule[1pt]
\end{tabular}
  \caption{Cut flow for the SL1 signal ($m_X = 1500$, $2000$ and $2500\ \mathrm{GeV}$) and the backgrounds at $\sqrt{s}=9.16\ \mathrm{TeV}$.}
\label{table9}
\end{table}

\begin{table}[htbp]
\centering
  \vspace{0.1cm}
  \begin{tabular}
  {cccccccc}
  \toprule[1pt]
  \multirow{2}{*}{Cuts} & \multicolumn{3}{c}{Signal(fb)} & \multicolumn{4}{c}{Backgrounds(fb)} \\ 
  \cline{2-4} \cline{5-8}
  & $X_{1500}$ & $X_{2000}$ & $X_{2500}$ & $\mu Wj$ & $\bar{\nu}WWj$ & $\bar{\nu}Zt$ & $\mu ZWj$ \\ 
  \cline{1-8} \midrule[0.8pt]
  Basic cuts & 49.15 & 25.02 & 11.65 & 11743 & 88.14 & 48.68 & 7.46 \\
  $N_j\ge1$ & 48.77 & 24.83 & 11.56 & 11741 & 88.12 & 48.48 & 7.46 \\
  $N_b=1$ & 24.46 & 12.90 & 6.40 & 486.70 & 3.61 & 25.28 & 0.30 \\
  $-1.4<\eta_{j_1}<1.0$ & 19.45 & 10.83 & 5.50 & 198.20 & 1.52 & 11.62 & 0.11 \\
  $-1.8<\eta_{b}<0.2$ & 14.11 & 8.66 & 4.68 & 64.45 & 0.65 & 6.00 & 0.04 \\
  $60\ \mathrm{GeV}<M_{j_1}<100\ \mathrm{GeV}$ & 11.74 & 7.21 & 3.88 & 5.54 & 0.09 & 0.57 & 0.006 \\
  $p_{T}^{j_1}>600\ \mathrm{GeV}$ & 7.29 & 6.60 & 3.80 & 0.18 & 0.01 & 0.03 & 0.0009 \\
  $p_{T}^{b}>380\ \mathrm{GeV}$ & 4.12 & 5.00 & 3.21 & 0.06 & 0.004 & 0.01 & 0.0003 \\
  \cline{1-8} \midrule[0.8pt]
  Efficiency & 9\% & 20\% & 28\% & 0.0005\% & 0.004\% & 0.02\% & 0.004\% \\
\bottomrule[1pt]
\end{tabular}
  \caption{Cut flow for the SL2 signal ($m_X = 1500$, $2000$ and $2500\ \mathrm{GeV}$) and the backgrounds at $\sqrt{s}=6.48\ \mathrm{TeV}$.}
\label{table11}
\end{table}

\begin{table}[htbp]
\centering
  \vspace{0.1cm}
  \begin{tabular}
  {cccccccc}
  \toprule[1pt]
  \multirow{2}{*}{Cuts} & \multicolumn{3}{c}{Signal(fb)} & \multicolumn{4}{c}{Backgrounds(fb)} \\ 
  \cline{2-4} \cline{5-8}
  & $X_{1500}$ & $X_{2000}$ & $X_{2500}$ & $\mu Wj$ & $\bar{\nu}WWj$ & $\bar{\nu}Zt$ & $\mu ZWj$ \\ 
  \cline{1-8} \midrule[0.8pt]
  Basic cuts & 85.12 & 54.33 & 33.44 & 14483 & 156.60 & 93.87 & 12.09 \\
  $N_j\ge1$ & 84.54 & 53.99 & 33.23 & 14481 & 156.57 & 93.54 & 12.09 \\
  $N_b=1$ & 42.88 & 28.38 & 18.58 & 586.30 & 6.69 & 48.00 & 0.49 \\
  $-1.0<\eta_{j_1}<2.0$ & 36.24 & 24.97 & 16.76 & 286.80 & 3.13 & 28.34 & 0.21 \\
  $-1.8<\eta_{b}<0.6$ & 27.10 & 20.15 & 14.28 & 104.40 & 1.53 & 15.31 & 0.08 \\
  $60\ \mathrm{GeV}<M_{j_1}<100\ \mathrm{GeV}$ & 22.14 & 16.58 & 11.74 & 12.30 & 0.29 & 1.69 & 0.02 \\
  $p_{T}^{j_1}>550\ \mathrm{GeV}$ & 14.38 & 14.64 & 11.19 & 1.23 & 0.04 & 0.15 & 0.003 \\
  $p_{T}^{b}>380\ \mathrm{GeV}$ & 7.62 & 10.79 & 9.28 & 0.34 & 0.01 & 0.06 & 0.0009 \\
  \cline{1-8} \midrule[0.8pt]
  Efficiency & 9\% & 20\% & 28\% & 0.002\% & 0.007\% & 0.068\% & 0.007\% \\
\bottomrule[1pt]
\end{tabular}
  \caption{Cut flow for the SL2 signal ($m_X = 1500$, $2000$ and $2500\ \mathrm{GeV}$) and the backgrounds at $\sqrt{s}=9.16\ \mathrm{TeV}$.}
\label{table12}
\end{table}

\end{document}